\documentclass[prb,twocolumn,superscriptaddress,floatfix,letterpaper]{revtex4}
\pdfoutput=1

\newcommand{\frc}[2]{{{#1}/{#2}}}

\begin{document}

\title{A theory of 2+1D bosonic topological orders}

\author{Xiao-Gang Wen} \affiliation{Department of Physics, Massachusetts
Institute of Technology, Cambridge, Massachusetts 02139, USA}
\affiliation{Perimeter Institute for Theoretical Physics, Waterloo, Ontario,
N2L 2Y5 Canada}

\begin{abstract} 
In primary school, we were told that there are four phases of matter: solid,
liquid, gas, and plasma.  In college, we learned that there are much more than
four phases of matter, such as hundreds of crystal phases, liquid crystal
phases, ferromagnet, anti-ferromagnet, superfluid, \etc.  Those phases of
matter are so rich, it is amazing that they can be understood systematically by
the symmetry breaking theory of Landau.  However, there are even more
interesting phases of matter that are beyond Landau symmetry breaking theory.
In this paper, we review new ``topological'' phenomena, such as topological
degeneracy, that reveal the existence of those new zero-temperature phases --
topologically ordered phases. Microscopically, topologically orders are
originated from the patterns of long-range entanglement in the ground states.
As a truly new type of order and a truly new kind of phenomena, topological
order and long-range entanglement require a new language and a new mathematical
framework, such as unitary fusion category and modular tensor category to
describe them.  In this paper, we will describe a simple mathematical framework
based on measurable quantities of topological orders $(S,T,c)$ proposed around
1989. The framework allows us to systematically describe all 2+1D
bosonic topological orders (\ie topological orders in local bosonic/spin/qubit systems).

\end{abstract}

\maketitle

{\small \setcounter{tocdepth}{2} \tableofcontents }

\section{Introduction}

\subsection{Orders, phase transitions, and symmetries} \label{deforder}

Condensed matter physics is a branch of science that study various properties
of all kinds of materials, such as mechanical properties, hydrodynamic
properties, electric properties, magnetic properties, optical properties,
thermal properties,  \etc.  Since there are so many different kinds of
materials with vastly varying properties, not surprisingly, condensed matter
physics is a very rich field.  Usually for each kind of material, we need a
different theory (or model) to explain its properties. So there are many
different theories and models to explain various properties of different
materials.  

However, after seeing many different type of theories/models for condensed
matter systems, a common theme among those theories start to emerge.  The
common theme is the \emph{principle of emergence},\index{principle of
emergence} which states that the properties of a material are mainly determined
by how particles are organized in the material.  Different organizations of
particles lead to different materials and/or different phases of matter, which
in turn leads to different properties of materials.

Typically, one may think that the properties of a material should be determined
by the components that form the material.  However, this simple intuition is
incorrect, since all the materials are made of same three components:
electrons, protons and neutrons.  So we cannot use the richness of the
components to understand the richness of the materials. The various properties
of different materials originate from various ways in which the particles are
organized.  The organizations of the particles are called orders.  The orders
(the organizations of particles) determine the physics properties of a
material. 

Therefore, according to the principle of emergence, the key to understand a
material is to understand how electrons, protons and neutrons are organized in
the material.  However, to develop a theory for all possible organizations of
particles, we need to have a more precise description/definition of
organizations of particles.

First, we need to find a way to determine if two organizations of particles
should be regard as the same (or more precisely, belong to the same class or
belong to the same phase) or not. Here, we need to rely on the phenomena of
phase transition.  If we can deform the system (such as changing temperature,
magnetic field, or other parameters of the system) in such a way that the state
of the system before the deformation and the state of the system after the
deformation are smoothly connected without any phase
transition,\footnote{``Without any phase transition'' means that 
all local quantities
change smoothly under the deformation.} then we say the two states before and
after the deformation belong to the same phase and the particles in the two
states are regarded to have the same organization.  If there is no way to
deform the system to connect two states in a smooth way, then, the two states
belong to two different phases and the particles in the two states are regarded
to have to two different organizations.

We note that our definition of organizations is a definition of an equivalent
class.  Two states that can be connected without a phase transition are defined
to be equivalent. The equivalent class defined in this way is called the
universality class. Two states with different organizations can also be said to
belong to different universality classes.  We introduce a formal name ``order''
to refer to the ``organization'' defined above.

Based on a deep insight into phase and phase transition, Landau developed a
general theory of orders as well as transitions between different phases of
matter\cite{L3726,GL5064,LanL58}.  Landau 
points out that the reason that different phases (or orders) are different is
because they have different symmetries.  A phase transition is simply a
transition that changes the symmetry.  Introducing order parameters that
transform non-trivially under the symmetry transformations, Ginzburg and Landau
developed Ginzburg-Landau theory, which became the standard theory for phase
and phase transition.\cite{GL5064} For example, in Ginzburg-Landau theory, the
order parameter can be used to characterize different symmetry breaking phase:
if the order parameter is zero, then we are in a symmetric phase;  if the order
parameter is non-zero, then we are in a symmetry break phase.  The symmetry
breaking phase transition is the process in which the order parameter change
from zero to non-zero. 

Landau's theory is very successful.\index{symmetry breaking} Using Landau's
theory and the related group theory for symmetries, we can classify all of the
230 different kinds of crystals that can exist in three dimensions.  By
determining how symmetry changes across a continuous phase transition, we can
obtain the critical properties of the phase transition.  The symmetry breaking
also provides the origin of many gapless excitations, such as phonons, spin
waves, etc., which determine the low-energy properties of many
systems.\cite{N6080,G6154} Many of the properties of those excitations,
including their gaplessness, are directly determined by the
symmetry.\index{Landau's symmetry-breaking theory} 

As Landau's symmetry-breaking theory has such a broad and fundamental impact on
our understanding of matter, it became a corner-stone of condensed matter
theory. The picture painted by Landau's theory is so satisfactory that one
starts to have a feeling that we understand, at least in principle, all kinds
of orders that matter can have.  One starts to have a feeling of seeing the
beginning of the end of the condensed matter theory. 

However, through the researches in last 25 years, a different picture starts to
emerge. It appears that what we have seen is just the end of beginning.  There
is a whole new world ahead of us waiting to be explored.  A peek into the new
world is offered by the discovery of fractional quantum Hall (FQH)
effect.\cite{TSG8259} Another peek is offered by the discovery of high $T_c$
superconductors.\cite{BM8689} Both phenomena are completely beyond the paradigm
of Landau's symmetry breaking theory. Rapid and exciting developments in FQH
effect and in high $T_c$ superconductivity resulted in many new ideas and new
concepts.  Looking back at those new developments, it becomes more and more
clear that, in last 25 years, we were actually witnessing an emergence of a new
theme in condensed matter physics. The new theme is associated with new kinds
of orders, new states of matter and new class of materials beyond Landau's
symmetry breaking theory.  This is an exciting time for condensed matter
physics.  The new paradigm may even have an impact in our understanding of
fundamental questions of nature -- the emergence of elementary particles and
the four fundamental
interactions.\cite{W0202,W0303a,LWqed,W1301,YBX1451,YX14124784}

\subsection{The discovery of topological order}

After the discovery of high $T_c$ superconductors in 1986,\cite{BM8689} some
theorists believed that quantum spin liquids play a key role in understanding
high $T_c$ superconductors\cite{A8796} and started to construct and study
various spin liquids.\cite{BZA8773,AM8874,RK8876,AZH8845,DFM8826} Despite the
success of Landau symmetry-breaking theory in describing all kind of states,
the theory cannot explain and does not even allow the existence of spin
liquids. This leads many theorists to doubt the very existence of spin liquids.
In 1987, a special kind of spin liquids -- chiral spin
state\cite{KL8795,WWZcsp} -- was introduced in an attempt to explain high
temperature superconductivity.  In contrast to many other proposed spin liquids
at that time, the chiral spin liquid was shown to correspond to a stable
zero-temperature phase and is more likely to exist.\footnote{Recently,  chiral
spin liquid is shown to exist in Heisenberg model on Kagome lattice with
$J_1$-$J_2$-$J_3$ coupling.\cite{HC14072740,GS14121571}}  At first, not believing Landau
symmetry-breaking theory fails to describe spin liquids, people still wanted to
use the symmetry breaking theory to characterize the chiral spin state. They
identified the chiral spin state as a state that breaks the time reversal and
parity symmetries, but not the spin rotation and translation
symmetries.\cite{WWZcsp} However, it was quickly realized that there are many
different chiral spin states (with different spinon statistics and spin Hall conductances) that have exactly the same symmetry, so symmetry
alone is not enough to characterize different chiral spin states. This means
that the chiral spin states contain a new kind of order that is beyond symmetry
description.\cite{Wtop} This new kind of order was named\cite{Wrig} topological
order.\footnote{The name ``topological order'' is motivated by the
low energy effective theory of the chiral spin states, which is a topological
quantum field theory.\cite{W8951}.} 

But experiments soon indicated that high-temperature superconductors do not
break the time reversal and parity symmetries and chiral spin states do not
describe high-temperature superconductors.\cite{LSL9239} Thus the concept of
topological order became a concept with no experimental realization. 

Although the concept of topological order is introduced in a theoretical study,
about a state that is not known to exist in nature, this does not prevent
topological order to become a useful concept.  As we will see later that the
concept of topological order contains inherent self consistency and stability.
If we believe in nature's richness, all nice concepts should be realized one
way or another. The concept of topological order is not an exception.

Long before the discovery of high $T_c$ superconductors, Tsui, Stormer, and
Gossard discovered FQH effect,\cite{TSG8259} such as the filling fraction
$\nu=1/m$ Laughlin state\cite{L8395}
\begin{align}
 \Psi_{\nu=1/m}(\{z_i\})=\prod (z_i-z_j)^m \ee^{-\frac14 \sum |z_i|^2}
\end{align}
where $z_i=x_i+\ii y_i$.
People realized that the FQH states are new states of matter.  However,
influenced by the previous success of Landau's symmetry breaking theory, people
still want to use order parameters and long range correlations to describe the
FQH states.\cite{GM8752,ZHK8982,R8986}  But, if we concentrate on physical
measurable quantities,
we will see that all those different FQH states have exactly the same symmetry
and conclude that we cannot use Landau symmetry-breaking theory and symmetry
breaking order parameters to describe different orders in FQH states.  So the
order parameters and long range correlations of local operators are not the
correct  way to describe the internal structures of FQH states.  In fact, just
like chiral spin states, FQH states also contain new kind of orders beyond
Landau's symmetry breaking theory.  Different FQH states are also described by
different topological orders.\cite{WNtop} Thus the concept of topological order
does have experimental realizations in FQH systems.

In addition to the Laughlin states, more exotic non-abelian FQH states were
proposed in 1991 by two independent works.  \Ref{Wnab} pointed out that the FQH
states described by wave functions
\begin{align}
 \Psi_{\nu=n/m}(\{z_i\})&=[\chi_n(\{z_i\})]^m, 
\nonumber\\
\text { or }
 \Psi_{\nu=n/(m+n)}(\{z_i\})&=\chi_1(\{z_i\})[\chi_n(\{z_i\})]^m 
\end{align}
have excitations with non-abelian statistics, where $\chi_n$ is the fermion
wave function of $n$-filled Landau levels.  The edge of the above FQH states
are described by $U(1)^{nm}/SU(m)_n$ or $U(1)^{nm+1}/SU(m)_n$ Kac-Moody current
algebra.\cite{Wcll,Wedgerev,Wtoprev,BW9215}  Those results were obtained by
deriving their low energy effective $SU(m)$ level $n$ Chern-Simons theory or
$U(1)\times SU(m)$ level $n$ Chern-Simons theory.  In the same year,
\Ref{MR9162} conjectured that the FQH state described by $p$-wave paired wave
function\cite{GWW9105,GWW9267}
\begin{align}
 \Psi_{\nu=1/2}=
\cA\Big[ \frac{1}{z_1-z_2} \frac{1}{z_3-z_4} \cdots\Big]
\ee^{-\frac14 \sum |z_i|^2}
\prod (z_i-z_j)^2  .
\end{align}
has excitations with non-abelian statistics.  Its edge states were studied
numerically in \Ref{Wnabhalf} and were found to be described by a $c=1$
chiral-boson conformal field theory (CFT) plus a $c=1/2$ Majorana fermion CFT.
Such a result about the edge states supports the conjecture that the  $p$-wave
paired FQH state is non-abelian, since the edge for abelian FQH states always
have integer chiral central charge $c$.\cite{WZ9290,Wedgerev,Wtoprev} A few
years later, the non-abelian statistics in $p$-wave paired wave function was
also confirmed by its low energy effective $SO(5)$ Chern-Simons
theory.\cite{W9927}

It is interesting to point out that
long before the discovery of  FQH states, Onnes discovered superconductor in
1911.\cite{O1122} The Ginzburg-Landau theory for symmetry breaking phases is
largely developed to explain  superconductivity.  However, the superconducting
order, that motivates the Ginzburg-Landau theory for symmetry breaking, itself
is not a symmetry breaking order.  Superconducting order (in real life with
dynamical $U(1)$ gauge field) is an order that is beyond Landau symmetry
breaking theory.  Superconducting order (in real life) is an topological order
(or more precisely a $Z_2$ topological order).\cite{W9141,HOS0497} It is quite
amazing that the experimental discovery of superconducting order did not lead
to a theory of topological order, but instead, lead to a theory of symmetry
breaking order, that fails to describe superconducting order itself.

\section{What is topological order?} \label{deftop}

\subsection{Topological ground state degeneracy}

The above description of topological order is highly incomplete and highly
unsatisfactory. This is because the characterization of topological order is
through specifying what it is not: topological order is a kind of orders that
cannot be described by symmetry breaking.  But what \emph{is} the topological
order? 

To appreciate the difficulty of describing topological order, let me tell a
story about a tribe. The tribe uses a language that contains only four words
for counting: one, two, three, and many-many.  It is very hard for a tribe
member to describe a naturally occurring phenomenon -- a large herd of deers.
He can only describe the number of deers in the herd by what it is not -- the
number is not one, nor two, nor three.

Similarly, the possible organizations of many particles in naturally occurring
states can be very rich, much richer than those described by symmetry breaking.
To describe the new orders (such as the topological orders), we need to
introduce new tools and new languages.  The richness of nature is not bounded
by the known theoretical formalism.  The Landau's symmetry breaking theory
corresponds to ``one'', ``two'', ``three'' which describes a small class of
orders.  Many other orders also exist in nature, but we do not know how to
describe them. Therefore, we introduced terms like ``spin liquid'', ``non-Fermi
liquid'', ``exotic order'', ``preformed pair'', ``dynamical stripe'', \etc.
Just like the term ``many many'' in the above story, those terms mainly
describe what it is not than what it is.

The symmetry breaking theory is the only language that we know to describe
phases and orders.  But topological order, by definition, cannot be described
by the symmetry breaking theory.  If we abandon the only language that we know,
how can we say any thing? Where do we start to understand the topological
order?  So the development of topological order theory is mainly trying to come
up with a proper way to name/label topological orders. We hope the name/label
to carry information that allows us to derive all the universal properties of
the corresponding topological order from its name/label.

To make progress, let us point out that, in physics, to define and to introduce
a concept is to design an experiment (a laboratory one or a numerical one).  We
need to identify measurable quantities such that the measurement of those
quantities facilitate the definition of the concept.  So in physics, once you
design an experiment, you define a concept. And only after you design an
experiment, do you define a concept.

So what experiments or what measurable quantities define the concept of
topological order? It was noted that a $\nu=1/m$ Laughlin FQH state has $m$
fold degenerate ground states on torus and a non-degenerate ground state on
sphere.\cite{YHL8319,H8305,S8469,TW8497,NTW8572,HR8529,AS8559,H8595} However,
the different degeneracies was regarded as finite size and/or group theoretical
effects without thermodynamical implications.  

In \Ref{Wtop,Wrig,WNtop}, it was shown that the ground state
degeneracy of a chiral spin state or a FQH state 
is stable against \emph{any local
perturbations}, including random perturbations that break all the
symmetries.\cite{Wrig,WNtop} Thus the topology-dependent ground state
degeneracies are a robust or universal property with important thermodynamical
implications: the topology-dependent and topologically robust degeneracies can
be used to define a phase (or a universality class) of a thermodynamical system
(\ie a system with a large size).  So the topology-dependent ground state
degeneracies is just what we are looking for:  the measurable quantities (in a
numerical experiment) that can be used to (partially) define topological order
in chiral spin states and FQH states.\cite{Wtop,WNtop} Such kind of universal
properties are also call \emph{topological invariants}, since they are robust
against any local perturbations. 

\begin{figure}[tb] \centerline{ \includegraphics[scale=0.6]{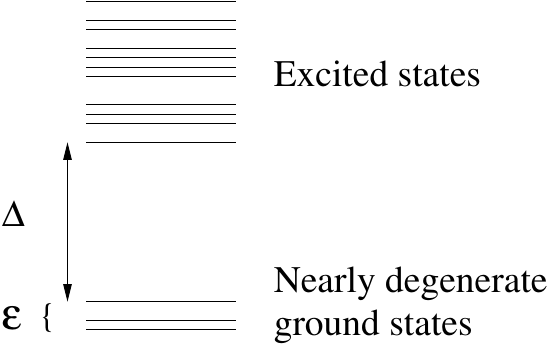} } 
\caption{
The energy levels of a topologically ordered state with a finite size.  The
splitting $\eps$ of the nearly degenerate ground states approaches to zero in
the large-system limit.  } 
\label{gdeg} 
\end{figure}

We would like to remark that the ground state degeneracy discussed above is
only an approximate degeneracy for a finite system, \ie there is a small
energy splitting $\eps$ between different degenerate ground states.  The energy
gap to other excited states is given by $\Del$ (see Fig. \ref{gdeg}).  It was
shown in \Ref{Wrig} and \Ref{WNtop} that, for chiral spin states and FQH
states, $\eps$ is exponentially small: $\eps \sim e^{-L/\xi}$ while $\Del$ is
finite in the limit where the system size approaches infinite: $L\to \infty$.

The topology-dependent ground state degeneracy 
is an amazing phenomenon.  In both FQH and chiral spin states, the correlation
of \emph{any local operators} are short ranged. This seems to imply that FQH
and chiral spin states are ``short sighted'' and they cannot know the topology
of space which is a global and long-distance property.  However, the fact that
ground state degeneracy does depend on the topology of space implies that FQH
and chiral spin states are not ``short sighted'' and they do find a way to know
the global and long-distance structure of space.  So, despite the short-ranged
correlations of all the local operators, the FQH and chiral spin states must
contain certain hidden long-range structure.  The  robustness of the ground
state degeneracy suggests that the hidden long-range structure in
FQH/chiral-spin states is also robust and universal.  A term topological order
was introduced to describe such a ``robust hidden long range
structure''.\cite{Wrig}

More recently, such a ``robust hidden long range structure'' was identified to
be the long-range entanglement defined by local unitary
transformations.\cite{CGW1038,ZW1490,SM1403} Thus  topological order is nothing
but the pattern of long range entanglement.  Different patterns of
long-range entanglement (or different topological orders) correspond to
different quantum phases.  Chiral spin liquids,\cite{KL8795,WWZcsp}
integral/fractional quantum Hall states\cite{KDP8094,TSG8259,L8395}, $Z_2$ spin
liquids,\cite{RS9173,Wsrvb,MS0181} non-Abelian FQH
states,\cite{MR9162,Wnab,WES8776,RMM0899} \etc are examples of topologically
ordered or long-range entangled phases.

\subsection{Topological order and phase transitions}

In section \ref{deforder}, we define a quantum phase as a region bounded by
lines of singularity in the ground state energy (or some other local
quantities).  In section \ref{deftop}, we define a topologically ordered
phase as a region characterized by a certain ground state degeneracy.  Are
these two definition self consistent?  As one topologically ordered  phase
changes into another topologically ordered  phase, the ground state degeneracy
may change from one value to another value.  So why a change in the ground
state degeneracy corresponds to a singularity in some averages of local
quantities?

\begin{figure}[tb] \centerline{ \includegraphics[scale=0.6]{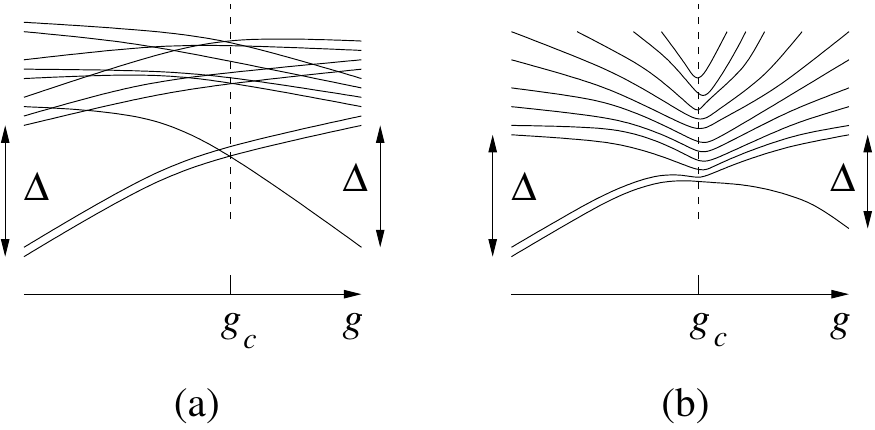} }
\caption{ Two types of phase transitions between two gapped states.  (a) Energy
levels of a Hamiltonian $H_a(g)$ as functions of the coupling constant $g$. The
topologically order state for $g<g_c$ changes into a trivial state for $g>g_c$
via a first order phase transition.  (b) Energy levels of a Hamiltonian
$H_b(g)$ as functions of the coupling constant $g$. The topologically order
state for $g<g_c$ changes into a trivial state for $g>g_c$ via a continuous
phase transition.  In this case, the ground state of $H_b(g_c)$ is a quantum
critical state.  } \label{toptran} \end{figure}

We note that the ground state degeneracy that characterize topological order is
robust again any perturbations. So a small change in the Hamiltonian will not
change the ground state degeneracy. However, a large change of Hamiltonian can
cause a change in ground state degeneracy.  The ground state degeneracy can
change in two different ways as described by Fig. \ref{toptran}.

In Fig. \ref{toptran}a, the ground state degeneracy changes due to an level
crossing. The ground state energy has a discontinuous first order derivative at
the crossing point. The corresponding phase transition is a first order phase
transition.

\begin{figure}[tb] \centerline{ \includegraphics[scale=0.6]{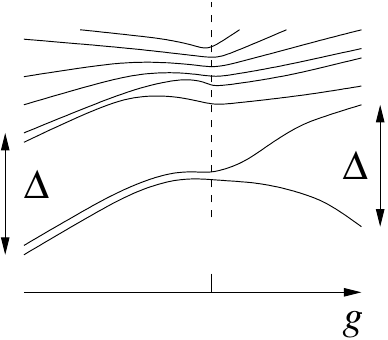} } \caption{
An impossible $g$ dependence of energy levels.  } \label{iltran} \end{figure}

Since the ground state degeneracy is robust against any perturbations, this
means that the degenerate ground state wave functions are locally
indistinguishable.  The ground states cannot be split without losing the
locally indistinguishable property.  The only way to lose locally
indistinguishable property is to develop a long range correlation \ie closing
the energy gap of other excitations.  So, the situation described by Fig.
\ref{iltran} cannot happen.  The possible situation is described by Fig.
\ref{toptran}b, where the energy gap of the excitations closes as $g\to g_c$
and reopens as $g$ passes $g_c$. The closing of the energy gap allow the ground
state degeneracy to change. The closing of the energy gap at $g_c$ cause a
singularity in the ground state energy (or some other local quantities).  Such
a phase transition is a continuous phase transition.

We see that the change of ground state degeneracy of topologically ordered
state and singularity in ground state energy always happen at the same place.
Thus the topological ground state degeneracy characterize a phase and a change
of the topological ground state degeneracy marks a phase transition.

\subsection{Topological invariants -- Towards a complete characterization of
topological orders} 
\label{topinv}

Soon after the introduction of topological order through topologically robust
and topology-dependent ground state degeneracies, it was realized that the
topology-dependent degeneracies are not enough to characterize all different
topological orders [note that, as discussed in section \ref{deforder}, orders
are defined through phase transitions]. Certain different topological orders can
have exactly the same set of ground state degeneracies for all compact spaces.

To obtain a complete topological invariant that can fully characterize
topological orders, in \Ref{Wrig,KW9327}, it was conjectured that \emph{the
non-Abelian geometric phases\cite{WZ8411} (both the $U(1)$ part and the
non-Abelian part) of degenerate ground states generated by the automorphism of
Riemann surfaces can completely characterize different topological
orders}.\cite{Wrig} 

We note that an automorphism of a Riemann surface change the Hamiltonian $H$
defined on the surface to another Hamiltonian $H'$ which is defined on the same
surface.  If we smoothly deform the Hamiltonian $H$ to $H'$, plus the
automorphism transformation at the end, we will get a family of Hamiltonians
that form a ``closed loop''.\cite{Wrig,KW9327} We can use such a loop-like
deformation path of Hamiltonians, with their degenerate ground states, to
define a non-Abelian geometric phase.\cite{WZ8411}  Thus, for every
automorphism of Riemann surfaces, we can produce a non-Abelian geometric phase
which is a unitary matrix.  

Such a unitary matrix is uniquely determined by the automorphism (up to a path
dependent over all $U(1)$ phase).  To understand such a result, let us assume
that the unitary matrix is not uniquely determined by the automorphism, \ie a
small change of deformation path leads to a different unitary matrix beyond the
different  over all $U(1)$ phase. This will mean that the small change of
deformation path causes  different phase shifts for different degenerate ground
states.  Since the small change of deformation path are local perturbations,
the different phase shifts for different degenerate ground states will implies
that the degenerate ground states are locally distinguishable, which contradict
the robustness of the degeneracy against any local perturbations and the
locally indistinguishable property of the degenerate ground states.

As a result, the above unitary matrices form a projective representation of
automorphism group of Riemann surfaces.\cite{Wrig,KW1458,MW1418} The
automorphism group $G_\text{Aut}$ contain a connected subgroup
$G^0_\text{Aut}$.  $G_\text{Aut}/G^0_\text{Aut}$ is the mapping class group
(MCG).  We note that the non-Abelian geometric phases for the automorphisms in
$G^0_\text{Aut}$ are all pure $U(1)$ phases, since the loops that correspond to
the automorphisms in $G^0_\text{Aut}$ are all contractible to a trivial point.
Thus the non-Abelian geometric phase also generate a projective representation
of MCG.

We see that the non-Abelian geometric phases contain a universal non-Abelian
part\cite{Wrig,KW9327} and a path dependent Abelian part\cite{Wrig,W1221}. The
non-Abelian part carries information about the  projective representation of
MCG.  For torus, the MCG is $SL(2,\Z)$, which is generate by a 90$^\circ$
rotation and a Dehn twist.  For such two generators of MCG, the associated
non-Abelian geometric phases is denoted by $S$ and $T$, which are unitary
matrices.  $S$ and $T$ generate a projective representation of MCG $SL(2,\Z)$
for torus.  

The Abelian part of the non-Abelian geometric phases is also important: it is
related to the gravitational Chern-Simons
term\cite{KF9732,HLP1242,KW1458,BR150204126} and carries information about the
chiral central charge $c$ for the gapless edge
excitations.\cite{Wedgerev,Wtoprev}  The chiral central charge $c$ can be
measured directly via the thermal Hall conductivity
$K_H=c\frac{\pi}{6}\frac{k_B^2}{\hbar}T$ of the sample.\cite{KF9732,HLP1242}

It is believed that $(S,T,c)$ form a complete and one-to-one description of
2+1D topological orders, which is consistent with the previous conjecture in
\Ref{Wrig}.  So $(S,T,c)$ are the new words, like ``four'', ``five'', ``six'',
``ten'', ``eleven'', ``twelve'', \etc in our tribe story, that we are looking
for, to describe/label topological orders.  Since $(S,T,c)$ may completely
describe 2+1D topological orders, we may be able develop a theory of 2+1D
topological order based  $(S,T,c)$.  

\subsection{Wave-function-overlap approach to obtain $S,T$}

We like to mention that in addition to use non-Abelian geometric phases to
obtain $(S,T)$ matrices, there are several other ways to obtain
them.\cite{ZGT1251,CV1223,ZMP1233,TZQ1251} In particular, one can use wave
function overlap to extract $(S,T)$ matrices directly from the degenerate
ground states wave functions on torus, provided that the system have
translation symmetry.\cite{HW1339,MW1418,HMW1457,MW1427} (The non-Abelian-geometric-phase approach can obtain $(S,T)$ matrices even from systems without
translation symmetry.) It was argued that for a system on a $d$-dimensional
torus $T^d$ of volume $V$ with the set of topologically degenerate ground
states $\{|\psi_i\>\}_{i=1}^N$, the overlaps of the degenerate ground states
have the following form \begin{equation} \label{eq:overlap} \<\psi_i|\hat
W|\psi_j\> = e^{-f V + o(1/V)} M^W_{ij}, \end{equation} where $\hat W$ are
transformations of the wave functions induced by the MCG transformations of the
space $T^d \to T^d$, $f$ is a non-universal constant, and $M^W$ is an
\emph{universal} unitary matrix.  

We know that a  MCG transformation $\hat W$ maps the space $T^d$ to itself:
$T^d \to T^d$. It transforms a ground state wave function $|\psi_j\>$ on space
$T^d$ to another wave function $\hat W|\psi_j\>$ on the same space $T^d$.
Since the MCG transformation $\hat W$ is not a symmetry of the Hamiltonian, the
new wave function $\hat W|\psi_j\>$ is not longer a ground state of the
Hamiltonian.  So the overlap of  $\hat W|\psi_j\>$ with a ground state
$|\psi_i\>$ is exponentially small in large volume limit.  $\<\psi_i|\hat
W|\psi_j\> \sim e^{-f V}$.  It seems that such an overlap contains no useful
universal information about topological order.  What was discovered in
\Ref{MW1418} is that if we separate out the volume dependent exponential factor,
the volume-independent constant factor $M^W$ contains useful universal
information about topological order.

We note that the volume-independent constant factor $M^W$ is a unitary matrix.
In contrast to non-Abelian geometric phases, such a unitary matrix has no
$U(1)$ phase ambiguity.  Those unitary matrices (from different  MCG
transformations $\hat W$) form a representation of the MCG of the space $T^d$,
$\texttt{MCG}(T^d)=SL(d,\Z)$, which is robust against any perturbations.  For
2+1D cases, the MCG of the torus $T^2$ is generate by 90$^\circ$ rotation $\hat
S$ and Dehn twist $\hat T$.  The corresponding unitary matrices $\hat S\to
M^S\equiv S$ and $\hat T\to M^T\equiv T$ generate a unitary representation of
$SL(2,\Z)$ (instead of a projective representation as for the case of
non-Abelian geometric phases).  As a result, we can use a unitary representation
of MCG, $(S,T)$, plus the chiral central charge $c$ to characterize all the
2+1D topological orders.

We also like to point out that we can always choose a so called
\emph{excitation basis} for the degenerate ground state (see Section
\ref{MTC}). In such a basis, $T$ is diagonal and $S_{1i}$ are real and
positive.  It is $(S,T)$ in such a basis, plus the chiral central charge $c$,
that may fully characterize all the 2+1D topological orders.

\subsection{The current systematic theories of topological orders}

We like to remark that topological order (\ie long-range entanglement) is truly
a new phenomena. They require new mathematical language to describe them.  Some
early researches suggest that tensor category
theory\cite{FNS0428,LWstrnet,CGW1038,GWW1017,KK1251,GWW1332} and simple current
algebra\cite{MR9162,BW9215,WW9455,LWW1024} (or pattern of zeros
\cite{WW0808,WW0809,BW0932,SL0604,BKW0608,SY0802,BH0802,BH0802a,BH0882}) may be
part of the new  mathematical language.  Using tensor category theory, we have
developed a systematic and quantitative theory that classify topological orders
with gappable edge for 2+1D interacting boson and fermion
systems.\cite{LWstrnet,CGW1038,GWW1017,GWW1332} 

For 2+1D topological orders (with gapped or gapless edge) that have only
Abelian statistics, we have a more complete and simpler result: we find that we
can use integer $K$-matrices to classify all of them.\cite{WZ9290}  So the
integer matrices $K$ are also the new words, like ``4'', ``5'', ``6'', \etc in
our tribe story, that can be used to describe/label a subset of topological
orders -- Abelian topological orders.  Such a $K$-label completely determine
the low energy universal properties of the corresponding topological order. For
example, the low energy effective theory for the topological order labeled by
$K$ is given by the following $U(1)$ Chern-Simons
theory\cite{BW9045,R9002,FK9169,WZ9290,BM0535,KS1193,Wtoprev} 
\begin{align} 
\label{csK}
{\cal L}= \frac{K_{IJ}}{4\pi} a_{I\mu} \prt_\nu a_{J\la}\eps^{\mu\nu\la} .
\end{align} 
Such an effective theory or the topological order labeled by $K$
can be realized by a concrete physical system -- a multi-layer FQH state:
\begin{align} 
\label{wavK} 
\prod_{I;i<j} (z_i^I-z_j^I)^{K_{II}} \prod_{I<J;i,j}
(z_i^I-z_j^J)^{K_{IJ}} \ee^{-\frac14 \sum_{i,I} |z_i^I|^2}, 
\end{align}
where $z_i^I=x_i^I+\ii y_i^I$ is the coordinate of the $i^\text{th}$ particle
in $I^\text{th}$ layer.

Certainly, the topological order described by $K$ are also described by
$(S,T,c)$.  The non-Abelian geometric phases for some canonical choices of path
are calculated for the bosonic $K$ topological order described by
\eqn{wavK}:\cite{W1221}
\begin{align}
 \t T_{\v\al\v\bt}&=
\ee^{\frac{\ii\pi}{12}  (\v N^TK \v N-\v K_d^T\v N )} \ee^{\ii \pi
(\v\al^TK\v\al-\frac{c}{12})}\del_{\v\al\v\bt}
\nonumber\\
 \t S_{\v\al\v\bt}&=
(-)^{(\v N^TK \v N-\v K_d^T\v N )/2} \frac{\ee^{ -\ii 2\pi \v\bt^T K\v\al }} {
\sqrt{|\det(K)|} },
\end{align}
 where $c$ is the difference in the numbers of
positive and negative eigenvalues of $K$, $\v
K_d^T=(K^{11},K^{22},\cdots,K^{\ka\ka})$ is two times the spin vector of the
Abelian FQH state, and $\v N^T=(N_{1},N_{2},\cdots,N_{\ka})$ are the numbers of
the bosonic ``electrons'' in each layer.\cite{Wtoprev} We see that the $U(1)$
factors depend on the number of electrons and are not universal.  But we can
isolate the universal  non-Abelian part by taking the limit $\v N\to 0$, and
find
\begin{align}
 \label{STK} T_{\v\al\v\bt}&= \ee^{-\ii 2\pi \frac{c}{24}}
\ee^{\ii \pi \v\al^TK\v\al}\del_{\v\al\v\bt}
\nonumber\\
 S_{\v\al\v\bt}&= \frac{\ee^{ -\ii 2\pi \v\bt^T K\v\al }} {
\sqrt{|\det(K)|} } = S_{\v\al\v\bt} . 
\end{align}
 We see that the  universal
non-Abelian part of the non-Abelian geometric phases determines $S$, $T$, and
$c$ mod 24.

%

\section{Topological excitations}

We have seen that we can use  unitary representation of MCG and the chiral
central charge, $(S,T,c)$,  to characterize/label/name all the 2+1D topological
orders. It is possible that $(S,T,c)$ is a full characterization of 2+1D
topological orders, in the sense that all other universal properties of
topological orders can be determined from the data $(S,T,c)$. In this
section, we will discuss some other universal properties of 2+1D topological
orders, and see how those universal properties are determined by the data
$(S,T,c)$.

\subsection{Local excitations and topological excitations}

Topologically ordered states in 2+1D are characterized by their unusual
particle-like excitations which may carry fractional/non-Abelian statistics.
To understand and to classify particle-like excitations in topologically
ordered states, it is important to understand the notions of local
quasiparticle excitations and topological quasiparticle excitations.  

\begin{figure}[tb] 
\centering \includegraphics[scale=0.5]{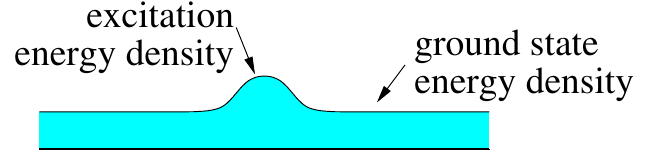} 
\caption{The energy density distribution of a quasiparticle.} 
\label{exceng} 
\end{figure}

First we define the notion of ``particle-like'' excitations.  Consider a gapped
system with translation symmetry.  The ground state has a uniform energy
density.  If we have a state with an excitation, we can measure the energy
distribution of the state over the space.  If for some local area, the energy
density is higher than ground state, while for the rest area the energy density
is the same as ground state, one may say there is a ``particle-like''
excitation, or a quasiparticle, in this area (see Figure \ref{exceng}).

  Quasiparticles defined like this can be divided into two types.  The first
type can be created or annihilated by local operators, such as a spin flip.
So, the first type of particle-like excitation is called local quasiparticle
excitations.  The second type cannot be  created or annihilated by any finite
number of local operators (in the infinite system size limit).  In other words,
the higher local energy density cannot be created or removed by \emph{any}
local operators in that area.  The second type of particle-like excitation is
called topological quasiparticle excitations.

From the notions of local quasiparticles and topological quasiparticles, we can
further introduce the notion \emph{topological} quasiparticle type, or simply,
quasiparticle type.  We say that local quasiparticles are of the trivial type,
while topological quasiparticles are of nontrivial types. Two topological
quasiparticles are of the same type if and only if they differ by local
quasiparticles.  In other words, we can turn one topological quasiparticle into
the other one of the same type by applying some local operators.  

\subsection{Fusion space and internal degrees of freedom for the
quasiparticles} \label{Fspace}

The quasiparticles have locational degrees of freedom, as well as
internal degrees of freedom. 

To understand the notion of
internal degrees of freedom, 
let us
discuss another way to define quasiparticles:\\
 \emph{Consider a gapped local
Hamiltonian qubit system defined by a local Hamiltonian $H_0$ in $d$
dimensional space $M^d$ without boundary.  A collection of quasiparticle
excitations labeled by $i$ and located at $\v x_i$ can be produced as
\emph{gapped} ground states of $H_0+\del H$ where $\del H$ is non-zero only
near $\v x_i$'s.  By choosing different $\del H$ we can create (or trap) all
kinds of quasiparticles.  We will use $i_i$ to label the type of the
quasiparticle at $\v x_i$.}
The gapped ground states of $H_0+\del H$ may have a degeneracy $D(
M^d;i_1,i_2,\cdots )$ which depends on the quasiparticle types $i_i$ and the
topology of the space $M^d$. The degeneracy is not exact, but becomes exact in
the large space and large particle separation limit.  We will use
$\cV(M^d;i_1,i_2,\cdots )$ to denote the space of the degenerate ground states.
If the Hamiltonian $H_0+\del H$ is not gapped, we will say $D(
M^d;i_1,i_2,\cdots )=0$ ({i.e.,}\ $\mathcal V(M^d;i_1,i_2,\cdots )$ has zero
dimension).  If $H_0+\del H$ is gapped, but if $\del H$ also creates
quasiparticles away from $\v x_i$'s (indicated by the bump in the energy
density away from $\v x_i$'s), we will also say $D( M^d;i_1,i_2,\cdots )=0$.
(In this case quasiparticles at $\v x_i$'s do not fuse to trivial
quasiparticles.) So, if $D( M^d;i_1,i_2,\cdots )>0$, $\del H$ only
creates/traps quasiparticles at $\v x_i$'s.

If we choose the space to be a $d$-dimensional sphere $M^d=S^d$, then the
number of the degenerate ground states, $D( S^d;i_1,i_2,\cdots )$ represents
the total number of internal degrees of freedom for the quasiparticles
$i_1,i_2,\cdots )$.
To obtain the  number of internal degrees of freedom for type-$i$ quasiparticle,
we consider the dimension $D( S^d;i,i,\cdots, i)$
of the fusion space on $n$ type-$i$ particles on $S^d$.
In large $n$ limit $D( S^d;i,i,\cdots, i)$ has a form
\begin{align}
 \ln D( S^d;i,i,\cdots, i) =n (\ln d_i +o(1/n) ).
\end{align}
Here $d_i$ is called the quantum dimension of the type-$i$ particle, which
describe the internal degrees of freedom the particle.  For example, a spin-0
particle has a quantum dimension $d=1$, while a spin-1  particle has a quantum
dimension $d=3$.
For particles with abelian statistics, their quantum dimensions are always 
equal to $1$. For particles with non-abelian statistics,
the quantum dimensions $d>1$, but in general the quantum dimensions $d$ may not be integers.

\subsection{Simple type and composite type} \label{types}

Even after quotient out the local quasiparticle excitations, topological
quasiparticle type still have two kinds: \emph{simple type} and \emph{composite
type}.  

We can also use the traping Hamiltonian $H_0+\del H$ and the
associated fusion space $\cV(M^d;i_1,i_2,\cdots )$
to understand the notion of simple type and composite type. 

If the degeneracy $D(M^d;i_1,i_2,\cdots)$ (the dimension of
$\cV(M^d;i_1,i_2,\cdots )$) cannot not be lifted by any small local
perturbation near $\v x_1$, then the particle type $i_1$ at $\v x_1$ is said to
be simple. Otherwise, the  particle type $i_1$ at $\v x_1$ is said to be
composite.  

When $i_1$ is composite, the space of the degenerate ground states
$\cV(M^d;i_1,i_2,i_3,\cdots)$ has a direct sum decomposition:
\begin{align}
&\quad\ \cV(M^d;i_1,i_2,i_3,\cdots)
\nonumber \\
 &=
\cV(M^d;j_1,i_2,i_3,\cdots)\oplus \cV(M^d;k_1,i_2,i_3,\cdots)
\nonumber \\
& \ \ \ \ \ \ \
\oplus \cV(M^d;l_1,i_2,i_3,\cdots)\oplus \cdots
\end{align}
 where $j_1$,
$k_1$, $l_1$, {\it etc.} are simple types.  To see the above result, we note
that when $i_1$ is composite the ground state degeneracy can be split by adding
some small perturbations near $\v x_1$.  After splitting, the original
degenerate ground states become groups of degenerate  states, each group of
degenerate  states span the space $\mathcal V(M^d;j_1,i_2,i_3,\cdots)$ or
$\mathcal V(M^d;k_1,i_2,i_3,\cdots)$ {\it etc.} which correspond to simple
quasiparticle types at $\v x_1$.  The above decomposition allows us to denote
the composite type $i_1$ as
\begin{align}
 i_1=j_1\oplus k_1\oplus l_1\oplus
\cdots. 
\end{align}

The degeneracy $D(M^d;i_1,i_2,\cdots)$ for simple particle types $i_i$ is a
universal property ({i.e.,}\ a topological invariant) of the topologically
ordered state.  In this paper, when we said particle/topological type, we
usually mean simple type. The number of simple types (including the trivial
type) is also a topological invariant of the topological order. Such a number
is referred as the rank of the topological order.

We have claimed that $(S,T,c)$ can determine all other topological invariants
of a topological order, including its rank. Indeed, the dimension of the $S$ or
$T$ matrices is the rank of the topological order.


\subsection{Fusion of quasiparticles}

When we fuse two simple types of topological particles $i$
and $j$ together, it may become a topological particle of a composite type:
\begin{align} i\otimes j=l=k_1\oplus k_2 \oplus \cdots,
\end{align}
 where
$i,j,k_i$ are simple types and $l$ is a composite type.  Here, we will use an
integer tensor $N^{ij}_k$ to describe the quasiparticle fusion, where $i,j,k$
label simple types.  Such an integer tensor $N^{ij}_k$ is referred as the
fusion coefficients of the topological order, which is a universal property of
the topologically ordered state.

When $N^{ij}_k=0$, the fusion of $i$ and $j$ does not contain $k$.  When
$N^{ij}_k=1$, the fusion of $i$ and $j$ contain one $k$: $i\otimes b=k \oplus
k_1  \oplus k_2 \oplus \cdots$.  When $N^{ij}_k=2$, the fusion of $i$ and $j$
contain two $k$'s: $i\otimes j =k \oplus k \oplus k_1  \oplus k_2 \oplus
\cdots$.  This way, we can denote that fusion of simple types as
\begin{align}
i\otimes j=\oplus_k N^{ij}_k k . 
\end{align}
 In physics, the quasiparticle
types always refer to simple types. The fusion rules $N^{ij}_k$ is a universal
property of the topologically ordered state.  The degeneracy
$D(M^d;i_1,i_2,\cdots)$ is determined completely by the fusion rules
$N^{ij}_k$.

Let us then consider the fusion of 3 simple quasiparticles $i,j,k$.  We may
first fuse $i,j$, and then with $k$, $(i\otimes j)\otimes k=(\oplus_{m}
N^{ij}_m m)\otimes k=\oplus_l (\sum_m N^{ij}_m N^{mk}_l) l$. We may also first
fuse $j,k$ and then with $i$, $i\otimes (j\otimes k)=i\otimes (\oplus_{m}
N^{jk}_m m)=\oplus_l (\sum_m N^{im}_l N^{jk}_m)l$.  The two ways of fusion
should produce the same result and this requires that
\begin{align}
\label{NN=NN} \sum_m N^{ij}_m N^{mk}_l=\sum_m N^{im}_l N^{jk}_m.  
\end{align}
Note that here, we do not require $N^{ij}_k = N^{ji}_k$. 

The fusion coefficients $N^{ij}_k$ are also topological invariants of the
topological order.  $(S,T,c)$ can determine such topological invariants.  In
fact, $S$ alone can determine $N^{ij}_k$:
\begin{align}
 N^{ij}_k
=\sum_{l=1}^{n} \frac{ S_{li} S_{lj} (S_{lk})^*}{ S_{l1} }
\end{align}
 which is
the famous Verlinde formula.\cite{V8860}



The internal degrees of freedom (\ie the quantum dimension $d_i$) for the
type-$i$ simple particle can be calculated directly from $N^{ij}_k$.  In fact
$d_i$ is the largest eigenvalue of the matrix $N_i$, whose elements are
$(N_{i})_{kj} = N^{ij}_k$. We see that $S$ matrix determines the  internal
degrees of freedom of the simple particles.

\subsection{Quasiparticle intrinsic spin}

For 2+1D topological orders, the quasiparticles can also braid. We also need
data to describe the braiding of the quasiparticles in addition to the fusion
rules We will discuss the braiding in this and next subsections.

If we twist the quasiparticle at $\v x_1$ by rotating $\del H$ at $\v x_1$ by
360$^\circ$ (note that $\del H$ at $\v x_1$ has no rotational symmetry), all
the degenerate ground states in $\mathcal V(M^d;i_1,i_2,i_3,\cdots)$ will
acquire the same geometric phase $\ee^{\ii\theta_{i_1}}$ provided that the
quasiparticle type $i_1$ is a simple type.  This is because when  $i_1$ is a
simple type, no local perturbations near $\v x_1$ can split the degeneracy.
Thus the  degenerate ground states are locally indistinguishable near $\v x_1$.
As a result, the 360$^\circ$ rotation cause the same phase shift
$\ee^{\ii\theta_{i_1}}$ for all the degenerate ground states.  We will call
$s_i=\frac{\theta_{i}}{2\pi}$ mod 1 the intrinsic spin (or simply spin) of the
simple type $i$, which is another universal property of the topologically
ordered state. 
$(S,T,c)$ can determine the topological invariants $s_i$ as well.  In fact,
$s_i$ mod 1 are given by the eigenvalues or the diagonal elements of $T$ and
$c$:
\begin{align}
 \ee^{\ii 2\pi s_i } = \ee^{\ii 2\pi \frac{c}{24} } T_{ii}
\end{align} (note that $T$ is diagonal in the excitation basis).

\subsection{Quasiparticle mutual statistics}

If we move the quasiparticle $i_2$ at $\v x_2$ around the quasiparticle $i_1$
at $\v x_1$, we will generate a non-Abelian geometric phase -- a unitary
transformation acting on the degenerate ground states in $\mathcal
V(M^d;i_1,i_2,i_3,\cdots)$.  Such a unitary transformation not only depends on
the types $i_1$ and $i_2$, but also depends on the quasiparticles at other
places.  So, here we will consider three quasiparticles of simple types $i$,
$j$, $\bar k$ on a 2D sphere $S^2$.  The ground state degenerate space is
$\cV(S^2;i,j,\bar k)$.  For some choices of $i$, $j$, $\bar k$,
$D(S^2;i,j,\bar k) \geq 1$, which is the dimension of $\cV(S^2;i,j,\bar k)$.
Now, we move the quasiparticle $j$ around the quasiparticle $i$.  All the
degenerate ground states in $\cV(S^2;i,j,\bar k)$ will acquire the same
geometric phase 
\begin{align}
\label{thijk}
\ee^{\ii \th_{ij}^{(k)}}= \dfrac 
{\ee^{\ii 2\pi s_{ k}}} 
{\ee^{\ii 2\pi s_i}\ee^{\ii 2\pi s_j}} 
.  
\end{align}
This is because, in $\mathcal
V(S^2;i,j,\bar k)$, the quasiparticles $i$ and $j$ fuse into $k$ (the
anti-quasiparticle of $\bar k$).  Moving quasiparticle $j$ around the
quasiparticle $i$ plus rotating $i$ and $j$ respectively by 360$^\circ$ is like
rotating $k$ by 360$^\circ$, \ie
$\ee^{\ii \th_{ij}^{(k)}}  
\ee^{\ii 2\pi s_i}\ee^{\ii 2\pi s_j} = \ee^{\ii 2\pi s_{ k}} $.
This leads to \eqn{thijk}.
We see that the quasiparticle mutual statistics is determined
by the quasiparticle spin $s_i$ and the  quasiparticle fusion rules $N^{ij}_k$.
For this reason, we call the set of data $(N^{ij}_k,s_i)$ quasiparticle
statistics. 

\begin{figure}[tb] 
\centering \includegraphics[scale=0.5]{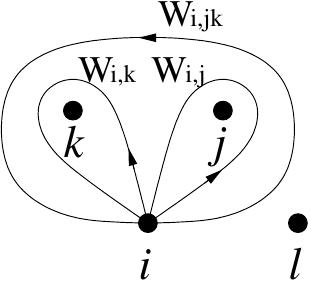} 
\caption{The braiding procedure to derive \eqn{vafaT}.} 
\label{vafa} 
\end{figure}

In fact, in order for data $(N^{ij}_k,s_i)$ to describe a valid quasiparticle
statistics, they must satisfy certain
conditions\cite{AM8841,V8821,Em0207007,E2009}.  Let us consider the fusion
space $\cV(S^2;i,j,k,l)$.  Let $W_{i,j}$ be the non-abelian geometric phase
(\ie the unitary matrix acting $\cV(S^2;i,j,k,l)$) generated by moving particle
$i$ around particle $j$, $W_{i,k}$ by moving particle $i$ around particle $k$,
and $W_{i,jk}$ by moving particle $i$ around both particle $j$ and $k$ (see Fig
\ref{vafa}).  We see that $W_{i,jk}=W_{i,k}W_{i,j}$, or
\begin{align}
 \det(W_{i,jk})=\det(W_{i,k})\det(W_{i,j})
\end{align}
We note that
\begin{align}
 \det(W_{i,j}) &=\prod_r
\Big( \frac {\ee^{\ii 2\pi s_{ r}}} {\ee^{\ii 2\pi s_i}\ee^{\ii 2\pi s_j}} 
\Big)^{N^{ij}_r N^{rk}_{\bar l}},
\nonumber\\
 \det(W_{i,k}) &=\prod_r
\Big( \frac {\ee^{\ii 2\pi s_{ r}}} {\ee^{\ii 2\pi s_i}\ee^{\ii 2\pi s_k}} 
\Big)^{N^{ik}_r N^{rj}_{\bar l}},
\nonumber\\
 \det(W_{i,jk}) &=\prod_r
\Big( \frac {\ee^{\ii 2\pi s_{\bar l}}} {\ee^{\ii 2\pi s_i}\ee^{\ii 2\pi s_r}} 
\Big)^{N^{jk}_r N^{ri}_{\bar l}}.
\end{align}
This way, we obtain
\begin{align}
\label{vafaT}
&\ \ \ \ \ee^{\ii 2\pi \sum_r s_r( 
N^{ij}_r N^{kl}_{\bar r} 
+ 
N^{jk}_r N^{il}_{\bar r} 
+
N^{ik}_r N^{jl}_{\bar r})}
\nonumber\\
&=
\ee^{\ii 2\pi \sum_r 
s_i (N^{jk}_r N^{ri}_{\bar l}-N^{ij}_r N^{rk}_{\bar l}-N^{ik}_r N^{rj}_{\bar l})}\times
\nonumber\\
& \ \ \ \
\ee^{\ii 2\pi \sum_r (
-s_j N^{ij}_r N^{rk}_{\bar l}
-s_k  N^{ik}_r N^{rj}_{\bar l}
-s_l N^{jk}_r N^{ri}_{\bar l}
)}
\nonumber\\
& =
\ee^{-\ii 2\pi 
(s_i+s_j+s_k+s_l)
\sum_r 
N^{ij}_r N^{kl}_{\bar r}
},
\end{align}
where the properties \eqn{Ncnd} and \eqn{Nc1} are used.  The above is the
relation between $N^{ij}_k$ and $s_i$.  For a given $N^{ij}_k$, the relation
determines $s_i$ up to discrete choices. This implies $s_i$ to be rational, and
we refer the above condition as the rational condition \eqn{Vs}.

\section{A theory of 2+1D bosonic topological orders} 

\subsection{A theory of 2+1D topological orders based on $(S,T,c)$} \label{STc}

In this section, we would like to develop a theory of 2+1D topological orders
based on $(S,T,c)$.  We have seen that we can measure  $(S,T,c)$ for every 2+1D
topological orders (in particular using the wave function overlap
\eqn{eq:overlap}), and every 2+1D topological orders are described by $(S,T,c)$
where $S,T$ are unitary matrices and $c$ is a rational number.  However, not
every $(S,T,c)$ can describe existing topological orders in 2+1D.  So to
develop a theory of topological order based on $(S,T,c)$, we need to find the
conditions on $(S,T,c)$.  If we find enough conditions on $(S,T,c)$, then every
$(S,T,c)$ that satisfies those conditions will describe an existing topological
order.  This way, we will have a theory of  topological orders.

So here, we will follow \Ref{W8951,GKh9410089,RSW0777,Wang10} and list the
known conditions satisfied by a $(S,T,c)$ that corresponds to an existing 2+1D
topological order:\\[2mm]
 \centerline{\textbf{$(S,T,c)$ conditions:}} 
\begin{enumerate} 
\item $S$
is symmetric and unitary with $S_{11}>0$, and satisfies the Verlinde
formula:\cite{V8860} 
\begin{align} 
\label{Ver} 
N^{ij}_k =\sum_{l=1}^{n} \frac{
S_{li} S_{lj} (S_{lk})^*}{ S_{l1} } \in \N , 
\end{align}
where $i,j,\cdots=1,2,\cdots,N$.  $N^{ij}_k$ is called fusion coefficient,
which gives the fusion rule for quasi-particles.  
\item
 Let
\begin{align}
d_i=\frac{S_{1i}}{S_{11}}
\end{align}
 which is called quantum dimension. Then
$d_i\geq 1$ is the largest eigenvalue of the matrix $N_i$, whose elements are
$(N_{i})_{kj} = N^{ij}_k$.  
\item 
$T$ is unitary and diagonal:
\begin{align}
T_{ij}=\ee^{\ii 2\pi s_i} \ee^{-\ii 2\pi \frac{c}{24}} \del_{ij}. 
\end{align}
Here $s_i$ is called topological spin ($\th_i=2\pi s_i$ is called
statistical angle).  $c$ is the chiral central charge.  
\item
 $S$ and $T$
satisfy:
\begin{align}
 (ST)^3=S^2=C, \ \ \ C^2=1, \ \ \ C_{ij}=N_1^{ij}.
\end{align} Thus $S$ and $T$ generate a unitary representation of $SL(2,\Z)$.
\item $S$ and $T$ also satisfy [see eqn. (223) in \Ref{K062}]:
\begin{align}
\label{dexch}
S_{ij}=\frac{1}{D}\sum_k N^{ij}_k \ee^{\ii 2\pi (s_i+s_j-s_k)} d_k ,
\end{align} 
where $D=\sqrt{\sum_i d_i^2}$.  
\item 
$N^{ij}_k$ and $\ee^{2\pi
\ii s_i}$ also satisfy\cite{V8821,Em0207007,E2009} (see \eqn{vafaT})
\begin{align}
\label{Vs}
\sum_r V_{ijkl}^r s_r =0 \text{ mod }1
\end{align} 
where
\begin{align}
\ \ \ \ \ \ 
V_{ijkl}^r &=  
N^{ij}_r N^{kl}_{\bar r}+
N^{il}_r N^{jk}_{\bar r}+
N^{ik}_r N^{jl}_{\bar r}
\nonumber\\
&\ \ \ \
- ( \del_{ir}+ \del_{jr}+ \del_{kr}+ \del_{lr}) \sum_m N^{ij}_m N^{kl}_{\bar m}
.
\end{align}
The above also implies that
\begin{align} 
\label{hAn} 
\sum_j A_{ij}s_j & =0 \text{ mod } 1,\ \ \ 
A_{ij}=M_{ij}-\frac43 \sum_k M_{ik} \del_{ij},
\end{align}
 where $M_{ij}= 2 N_{j}^{i\bar i}
N_{i}^{ij} + N_{j}^{ii} N_{i}^{j \bar i}$.
\item Let 
\begin{equation}
 \label{nuga}
 \nu_i=\frac{1}{D^2} \sum_{jk} N_i^{jk}d_{j}d_{k}\ee^{\ii 4\pi(s_{j}-s_{k})}, 
\end{equation}
then\cite{RSW0777,Wang10} $\nu_i=0$ if $i\neq \bar i$, and $\nu_i=\pm 1$ if $i
= \bar i$.  
\end{enumerate} 
The above are the necessary conditions in order for $(S,T,c)$ to describe an
existing 2+1D topological order.  In other words, the  $(S,T,c)$'s for all the
topological orders are included in the solutions.

However, it is not clear if those conditions are sufficient.  So it is possible
that some solutions are ``fake'' $(S,T,c)$ that do not correspond to any valid
topological order.  It is also possible that a valid solution $(S,T,c)$ may
correspond to several topological orders.


To see if there are any ``fake'' $(S,T,c)$'s in our lists, \Ref{SW150801111}
tries to construct explicit many-body wave functions for those $(S,T,c)$'s in
the lists, using simple current algebra.\cite{BW9215,WW9455,LWW1024} We find
that all the $(S,T,c)$'s in our lists are valid and correspond to existing
topological orders. 

\subsection{A theory of 2+1D topological orders based on $(N^{ij}_k,s_i,c)$} 

\label{Nsc}

From the above conditions, we see that, instead of using $(S,T,c)$, we can also
use $(N^{ij}_k,s_i,c)$ to describe topological orders, since $(S,T,c)$ can be
expressed in terms of $(N^{ij}_k,s_i,c)$, and $(N^{ij}_k,s_i,c)$ can be
expressed in terms of $(S,T,c)$.  So we can develop a theory of topological
orders based on $(N^{ij}_k,s_i,c)$, instead of $(S,T,c)$.  Again not all
$(N^{ij}_k,s_i,c)$ describe existing 2+1D topological orders.  Here we list the
necessary conditions on $(N^{ij}_k,s_i,c)$:\\[2mm]
\centerline{\textbf{$(N^{ij}_k,s_i,c)$ conditions:}} 
\begin{enumerate} 
\item
$N^{ij}_k$ are non-negative integers that satisfy 
\begin{align} 
\label{Ncnd} &
N^{ij}_k=N_k^{ji}, \ \ N_j^{1i}=\del_{ij}, \ \ \sum_{k=1}^N N_1^{i k}N_1^{
kj}=\del_{ij},
\nonumber\\
 & \sum_{m=1}^{n} N_m^{ij}N_l^{m k} = \sum_{n=1}^{n}
N_l^{in}N_n^{j k} \text{ or }  N_ k N_i = N_i N_ k 
\end{align} 
where
$i,j,\cdots=1,2,\cdots,n$, and the matrix $N_i$ is given by $(N_{i})_{ kj} =
N^{ij}_k$.  In fact $ N_1^{ij}$ defines a charge conjugation $i\to \bar i$:
\begin{align} 
N_1^{ij}=\del_{\bar ij}.  
\end{align} 
We also refer $n$ as the
rank of the corresponding topological order.  
\item 
$N^{ij}_k$ and $s_i$ satisfy\cite{V8821,Em0207007,E2009}  (see \eqn{vafaT})
\begin{align}
\label{Vcnd}
\sum_r V_{ijkl}^r s_r =0 \text{ mod }1
\end{align} 
where
\begin{align}
\ \ \ \ \ \ 
V_{ijkl}^r &=  
N^{ij}_r N^{kl}_{\bar r}+
N^{il}_r N^{jk}_{\bar r}+
N^{ik}_r N^{jl}_{\bar r}
\nonumber\\
&\ \ \ \
- ( \del_{ir}+ \del_{jr}+ \del_{kr}+ \del_{lr}) \sum_m N^{ij}_m N^{kl}_{\bar m}
\end{align}
Those are the conditions that allows us to
show $s_i$ and $c$ to be rational
numbers.\cite{V8821,AM8841,BK01,Em0207007,E2009}.  
\item Let $d_i$ be the largest
eigenvalue of the matrix $N_i$.  Let 
\begin{align} 
\label{SNsss}
S_{ij}=\frac{1}{\sqrt{\sum_i
d_i^2}}\sum_k N^{ij}_k \ee^{2\pi \ii ( s_i+ s_j- s_k)} d_k .  
\end{align} 
Then,
$S$ is unitary and satisfies \cite{V8860}
\begin{align}
 \label{Vform} S_{11}>0,
\ \ \ N^{ij}_k =\sum_{l} \frac{ S_{li} S_{lj} (S_{lk})^*}{ S_{1l} } .
\end{align} 
\item Let 
\begin{align} 
T_{ij}=\ee^{2\pi \ii  s_i} \ee^{-2\pi \ii
\frac{c}{24}} \del_{ij}.  
\end{align} 
Then
\begin{align}
 (ST)^3=S^2=C, \ \ \
C^2=1. 
\end{align}
 In fact $C_{ij}=N_1^{ij}$.  
\item Let 
\begin{equation}
\label{nuga1} 
\nu_i= \frac{1}{D^2} \sum_{jk} N_i^{jk} d_{j}d_{k} \ee^{4\pi \ii
( s_{j}- s_{k})}.  
\end{equation} 
Then\cite{RSW0777,Wang10} $\nu_i=0$ if $i\neq
\bar i$, and $\nu_i=\pm 1$ if $i = \bar i$.  
\end{enumerate} 
The above conditions are necessary for $(N^{ij}_k,s_i,c)$ to describe an
existing 2+1D topological order.  If the above conditions are also sufficient,
then the above will represent a classifying theory of 2+1D topological orders.  

In section \ref{smptop}, we will solve the above conditions to obtain a list
2+1D topological orders.  We like to mention that solving the above conditions
is closely related to classifying modular tensor categories.  \Ref{RSW0777}
have classified all the 70 modular tensor categories with rank $N=1,2,3,4$,
using Galois group.  In this paper, we will try to solve the  above conditions
numerically for higher ranks.

\section{2+1D topological orders with low ranks and low quantum dimensions}
\label{smptop}


\def\arraystretch{1.25} \setlength\tabcolsep{3pt} 
\begin{table*}[tb] 
\caption{ A list of all 45 bosonic topological orders in 2+1D with rank
$N=1,2,3,4,5$ and with max$(N^{ij}_k)\leq 3$.  The entries in red are the
topological orders with max$(N^{ij}_k)=2$.  All other topological orders have
$N^{ij}_k=0,1$. There are no topological orders with max$(N^{ij}_k)=3$ and
$N\leq 5$.  The entries in blue are the composite topological orders that can
be obtained by stacking lower rank topological orders.  The first column is the
rank $N$ and the central charge $c$ (mod 8).  The second column is the
topological entanglement entropy $S_\text{top}=\log_2 D$, $D=\sqrt{\sum_i
d_i^2}$. The quantum dimensions of topological excitations in the third column
are expressed in terms of
$\zeta_n^m=\frac{\sin[\pi(m+1)/(n+2)]}{\sin[\pi/(n+2)]}$. The fourth column is
the spins of the corresponding topological excitations.} 
\label{toplst} 
\centering
\begin{tabular}{ |c|c|l|l||c|c|l|l| } \hline $N^B_c$ & $S_\text{top}$ &
$d_1,d_2,\cdots$ & $s_1,s_2,\cdots$ & $N^B_c$ & $S_\text{top}$ & $d_1,d_2,\cdots$
& $s_1,s_2,\cdots$ \\
 \hline $1^B_{ 1}$ & $0$ & $1$ & $0$ & & & & \\
 \hline $2^B_{
1}$ & $0.5$ & $1,1$ & $0, \frac{1}{4}$ & $2^B_{-1}$ & $0.5$ & $1,1$ &
$0,-\frac{1}{4}$ \\
 $2^B_{\frc{14}{5}}$ & $0.9276$ & $1,\zeta_3^1$ & $0,
\frac{2}{5}$ & $2^B_{-\frc{14}{5}}$ & $0.9276$ & $1,\zeta_3^1$ &
$0,-\frac{2}{5}$ \\
 \hline $3^B_{ 2}$ & $0.7924$ & $1,1,1$ & $0, \frac{1}{3},
\frac{1}{3}$ & $3^B_{-2}$ & $0.7924$ & $1,1,1$ & $0,-\frac{1}{3},-\frac{1}{3}$ \\
$3^B_{\frc{1}{2}}$ & $1$ & $1,1,\zeta_2^1$ & $0,\frac{1}{2}, \frac{1}{16}$ &
$3^B_{-\frc{1}{2}}$ & $1$ & $1,1,\zeta_2^1$ & $0,\frac{1}{2},-\frac{1}{16}$ \\
$3^B_{\frc{3}{2}}$ & $1$ & $1,1,\zeta_2^1$ & $0,\frac{1}{2}, \frac{3}{16}$ &
$3^B_{-\frc{3}{2}}$ & $1$ & $1,1,\zeta_2^1$ & $0,\frac{1}{2},-\frac{3}{16}$ \\
$3^B_{\frc{5}{2}}$ & $1$ & $1,1,\zeta_2^1$ & $0,\frac{1}{2}, \frac{5}{16}$ &
$3^B_{-\frc{5}{2}}$ & $1$ & $1,1,\zeta_2^1$ & $0,\frac{1}{2},-\frac{5}{16}$ \\
$3^B_{\frc{7}{2}}$ & $1$ & $1,1,\zeta_2^1$ & $0,\frac{1}{2}, \frac{7}{16}$ &
$3^B_{-\frc{7}{2}}$ & $1$ & $1,1,\zeta_2^1$ & $0,\frac{1}{2},-\frac{7}{16}$ \\
$3^B_{\frc{8}{7}}$ & $1.6082$ & $1,\zeta_5^1,\zeta_5^2$ & $0,-\frac{1}{7},
\frac{2}{7}$ & $3^B_{-\frc{8}{7}}$ & $1.6082$ & $1,\zeta_5^1,\zeta_5^2$ & $0,
\frac{1}{7},-\frac{2}{7}$ \\
 \hline $4^{B,a}_{ 0}$ & $1$ & $1,1,1,1$ & $0, 0,
0,\frac{1}{2}$ & \color{blue} $4^{B,b}_{ 0}$ &\color{blue}  $1$ &\color{blue}
$1,1,1,1$ &\color{blue}  $0, 0, \frac{1}{4},-\frac{1}{4}$ \\
 $4^B_{ 1}$ & $1$ &
$1,1,1,1$ & $0, \frac{1}{8}, \frac{1}{8},\frac{1}{2}$ & $4^B_{-1}$ & $1$ &
$1,1,1,1$ & $0,-\frac{1}{8},-\frac{1}{8},\frac{1}{2}$ \\
 \color{blue} $4^B_{ 2}$
&\color{blue}  $1$ &\color{blue}  $1,1,1,1$ &\color{blue}  $0, \frac{1}{4},
\frac{1}{4},\frac{1}{2}$ & \color{blue} $4^B_{-2}$ &\color{blue}  $1$
&\color{blue}  $1,1,1,1$ &\color{blue}
$0,-\frac{1}{4},-\frac{1}{4},\frac{1}{2}$ \\
 $4^B_{ 3}$ & $1$ & $1,1,1,1$ & $0,
\frac{3}{8}, \frac{3}{8},\frac{1}{2}$ & $4^B_{-3}$ & $1$ & $1,1,1,1$ &
$0,-\frac{3}{8},-\frac{3}{8},\frac{1}{2}$ \\
 $4^B_{4}$ & $1$ & $1,1,1,1$ &
$0,\frac{1}{2},\frac{1}{2},\frac{1}{2}$ & \color{blue} $4^B_{\frc{9}{5}}$
&\color{blue} $1.4276$ &\color{blue} $1,1,\zeta_3^1,\zeta_3^1$ &\color{blue}
$0,-\frac{1}{4}, \frac{3}{20}, \frac{2}{5}$ \\
 \color{blue} $4^B_{-\frc{9}{5}}$
&\color{blue} $1.4276$ &\color{blue} $1,1,\zeta_3^1,\zeta_3^1$ &\color{blue}
$0, \frac{1}{4},-\frac{3}{20},-\frac{2}{5}$ &\color{blue} $4^B_{ \frc{19}{5}}$ &\color{blue} $1.4276$ &\color{blue} $1,1,\zeta_3^1,\zeta_3^1$
&\color{blue} $0, \frac{1}{4},-\frac{7}{20}, \frac{2}{5}$ \\
 \color{blue}
$4^B_{-\frc{19}{5}}$ &\color{blue} $1.4276$ &\color{blue}
$1,1,\zeta_3^1,\zeta_3^1$ &\color{blue} $0,-\frac{1}{4},
\frac{7}{20},-\frac{2}{5}$ &\color{blue} \color{blue} $4^{B,c}_{ 0}$ &\color{blue}
$1.8552$ &\color{blue} $1,\zeta_3^1,\zeta_3^1,\zeta_3^1\zeta_3^1$ &\color{blue}
$0, \frac{2}{5},-\frac{2}{5}, 0$ \\
 \color{blue} $4^B_{\frc{12}{5}}$
&\color{blue} $1.8552$ &\color{blue} $1,\zeta_3^1,\zeta_3^1,\zeta_3^1\zeta_3^1$
&\color{blue} $0,-\frac{2}{5},-\frac{2}{5}, \frac{1}{5}$ &\color{blue}
\color{blue} $4^B_{-\frc{12}{5}}$ &\color{blue} $1.8552$ &\color{blue}
$1,\zeta_3^1,\zeta_3^1,\zeta_3^1\zeta_3^1$ &\color{blue} $0, \frac{2}{5},
\frac{2}{5},-\frac{1}{5}$ \\
 $4^B_{\frc{10}{3}}$ & $2.1328$ &
$1,\zeta_7^1,\zeta_7^2,\zeta_7^3$ & $0, \frac{1}{3}, \frac{2}{9},-\frac{1}{3}$
& $4^B_{-\frc{10}{3}}$ & $2.1328$ & $1,\zeta_7^1,\zeta_7^2,\zeta_7^3$ &
$0,-\frac{1}{3},-\frac{2}{9}, \frac{1}{3}$ \\
 \hline $5^B_{ 0}$ & $1.1609$ &
$1,1,1,1,1$ & $0, \frac{1}{5}, \frac{1}{5},-\frac{1}{5},-\frac{1}{5}$ & $5^B_{4}$
& $1.1609$ & $1,1,1,1,1$ & $0, \frac{2}{5},
\frac{2}{5},-\frac{2}{5},-\frac{2}{5}$ \\
 $5^{B,a}_{ 2}$ & $1.7924$ &
$1,1,\zeta_4^1,\zeta_4^1,2$ & $0, 0, \frac{1}{8},-\frac{3}{8}, \frac{1}{3}$ &
$5^{B,b}_{ 2}$ & $1.7924$ & $1,1,\zeta_4^1,\zeta_4^1,2$ & $0, 0,-\frac{1}{8},
\frac{3}{8}, \frac{1}{3}$ \\
 $5^{B,b}_{-2}$ & $1.7924$ & $1,1,\zeta_4^1,\zeta_4^1,2$
& $0, 0, \frac{1}{8},-\frac{3}{8},-\frac{1}{3}$ & $5^{B,a}_{-2}$ & $1.7924$ &
$1,1,\zeta_4^1,\zeta_4^1,2$ & $0, 0,-\frac{1}{8}, \frac{3}{8},-\frac{1}{3}$ \\
$5^B_{\frc{16}{11}}$ & $2.5573$ & $1,\zeta_9^1,\zeta_9^2,\zeta_9^3,\zeta_9^4$ &
$0,-\frac{2}{11}, \frac{2}{11}, \frac{1}{11},-\frac{5}{11}$ &
$5^B_{-\frc{16}{11}}$ & $2.5573$ & $1,\zeta_9^1,\zeta_9^2,\zeta_9^3,\zeta_9^4$ &
$0, \frac{2}{11},-\frac{2}{11},-\frac{1}{11}, \frac{5}{11}$ \\
 \color{red} $5^B_{\frc{18}{7}}$ &\color{red}  $2.5716$ &\color{red}
$1,\zeta_5^2,\zeta_5^2,\zeta_{12}^2,\zeta_{12}^4$ &\color{red}
$0,-\frac{1}{7},-\frac{1}{7}, \frac{1}{7}, \frac{3}{7}$ &\color{red}
$5^B_{-\frc{18}{7}}$ &\color{red}  $2.5716$ &\color{red}
$1,\zeta_5^2,\zeta_5^2,\zeta_{12}^2,\zeta_{12}^4$ &\color{red}  $0,
\frac{1}{7}, \frac{1}{7},-\frac{1}{7},-\frac{3}{7}$ \\
 \hline \end{tabular}
\end{table*}

\begin{table*}[tb] 
\caption{A list of all 50 bosonic rank $N=6$ topological
orders in 2+1D with max$(N^{ij}_k)\leq 2$.  
} 
\label{toplst6} 
\centering
\begin{tabular}{ |c|c|c|l|l|c| } 
\hline 
$N^B_c$ & $S_\text{top}$ & $D^2$ & $d_1,d_2,\cdots$ & $s_1,s_2,\cdots$ & comment \\
 \hline 
$6^B_{ 1}$ &
$1.2924$ & $6$ & $1,1,1,1,1,1$ & $0, \frac{1}{12}, \frac{1}{12},-\frac{1}{4},
\frac{1}{3}, \frac{1}{3}$ & $2^B_{-1}\boxtimes 3^B_{ 2}$\\
 $6^B_{-1}$ & $1.2924$ & $6$
& $1,1,1,1,1,1$ & $0,-\frac{1}{12},-\frac{1}{12},
\frac{1}{4},-\frac{1}{3},-\frac{1}{3}$ & $2^B_{ 1}\boxtimes 3^B_{-2}$\\
 $6^B_{ 3}$ &
$1.2924$ & $6$ & $1,1,1,1,1,1$ & $0, \frac{1}{4}, \frac{1}{3},
\frac{1}{3},-\frac{5}{12},-\frac{5}{12}$ & $2^B_{ 1}\boxtimes 3^B_{ 2}$\\
 $6^B_{-3}$ &
$1.2924$ & $6$ & $1,1,1,1,1,1$ & $0,-\frac{1}{4},-\frac{1}{3},-\frac{1}{3},
\frac{5}{12}, \frac{5}{12}$ & $2^B_{-1}\boxtimes 3^B_{-2}$\\
 \hline $6^B_{
\frc{1}{2}}$ & $1.5$ & $8$ & $1,1,1,1,\zeta_2^1,\zeta_2^1$ & $0,
\frac{1}{4},-\frac{1}{4},\frac{1}{2},-\frac{1}{16}, \frac{3}{16}$ & $2^B_{
1}\boxtimes 3^B_{-\frc{1}{2}}$\\
 $6^B_{-\frc{1}{2}}$ & $1.5$ & $8$ &
$1,1,1,1,\zeta_2^1,\zeta_2^1$ & $0, \frac{1}{4},-\frac{1}{4},\frac{1}{2},
\frac{1}{16},-\frac{3}{16}$ & $2^B_{ 1}\boxtimes 3^B_{-\frc{3}{2}}$\\
 $6^B_{
\frc{3}{2}}$ & $1.5$ & $8$ & $1,1,1,1,\zeta_2^1,\zeta_2^1$ & $0,
\frac{1}{4},-\frac{1}{4},\frac{1}{2}, \frac{1}{16}, \frac{5}{16}$ & $2^B_{
1}\boxtimes 3^B_{\frc{1}{2}}$\\
 $6^B_{-\frc{3}{2}}$ & $1.5$ & $8$ &
$1,1,1,1,\zeta_2^1,\zeta_2^1$ & $0,
\frac{1}{4},-\frac{1}{4},\frac{1}{2},-\frac{1}{16},-\frac{5}{16}$ & $2^B_{
1}\boxtimes 3^B_{-\frc{5}{2}}$\\
 $6^B_{\frc{5}{2}}$ & $1.5$ & $8$ &
$1,1,1,1,\zeta_2^1,\zeta_2^1$ & $0, \frac{1}{4},-\frac{1}{4},\frac{1}{2},
\frac{3}{16}, \frac{7}{16}$ & $2^B_{ 1}\boxtimes 3^B_{\frc{3}{2}}$\\
$6^B_{-\frc{5}{2}}$ & $1.5$ & $8$ & $1,1,1,1,\zeta_2^1,\zeta_2^1$ & $0,
\frac{1}{4},-\frac{1}{4},\frac{1}{2},-\frac{3}{16},-\frac{7}{16}$ & $2^B_{
1}\boxtimes 3^B_{-\frc{7}{2}}$\\
 $6^B_{\frc{7}{2}}$ & $1.5$ & $8$ &
$1,1,1,1,\zeta_2^1,\zeta_2^1$ & $0, \frac{1}{4},-\frac{1}{4},\frac{1}{2},
\frac{5}{16},-\frac{7}{16}$ & $2^B_{ 1}\boxtimes 3^B_{\frc{5}{2}}$\\
$6^B_{-\frc{7}{2}}$ & $1.5$ & $8$ & $1,1,1,1,\zeta_2^1,\zeta_2^1$ & $0,
\frac{1}{4},-\frac{1}{4},\frac{1}{2},-\frac{5}{16}, \frac{7}{16}$ & $2^B_{
1}\boxtimes 3^B_{\frc{7}{2}}$\\
 \hline $6^B_{\frc{4}{5}}$ & $1.7200$ & $10.854$ &
$1,1,1,\zeta_3^1,\zeta_3^1,\zeta_3^1$ & $0,-\frac{1}{3},-\frac{1}{3},
\frac{1}{15}, \frac{1}{15}, \frac{2}{5}$ & $2^B_{\frc{14}{5}}\boxtimes 3^B_{-2}$\\
$6^B_{-\frc{4}{5}}$ & $1.7200$ & $10.854$ &
$1,1,1,\zeta_3^1,\zeta_3^1,\zeta_3^1$ & $0, \frac{1}{3},
\frac{1}{3},-\frac{1}{15},-\frac{1}{15},-\frac{2}{5}$ &
$2^B_{-\frc{14}{5}}\boxtimes 3^B_{ 2}$\\
 $6^B_{\frc{16}{5}}$ & $1.7200$ & $10.854$ &
$1,1,1,\zeta_3^1,\zeta_3^1,\zeta_3^1$ & $0,-\frac{1}{3},-\frac{1}{3},
\frac{4}{15}, \frac{4}{15},-\frac{2}{5}$ & $2^B_{-\frc{14}{5}}\boxtimes 3^B_{-2}$\\
$6^B_{-\frc{16}{5}}$ & $1.7200$ & $10.854$ &
$1,1,1,\zeta_3^1,\zeta_3^1,\zeta_3^1$ & $0, \frac{1}{3},
\frac{1}{3},-\frac{4}{15},-\frac{4}{15}, \frac{2}{5}$ & $2^B_{
\frc{14}{5}}\boxtimes 3^B_{ 2}$\\
 \hline $6^B_{\frc{3}{10}}$ & $1.9276$ & $14.472$
& $1,1,\zeta_2^1,\zeta_3^1,\zeta_3^1,\zeta_2^1\zeta_3^1$ &
$0,\frac{1}{2},-\frac{5}{16},-\frac{1}{10}, \frac{2}{5}, \frac{7}{80}$ & $2^B_{
\frc{14}{5}}\boxtimes 3^B_{-\frc{5}{2}}$\\
 $6^B_{-\frc{3}{10}}$ & $1.9276$ &
$14.472$ & $1,1,\zeta_2^1,\zeta_3^1,\zeta_3^1,\zeta_2^1\zeta_3^1$ &
$0,\frac{1}{2}, \frac{5}{16}, \frac{1}{10},-\frac{2}{5},-\frac{7}{80}$ &
$2^B_{-\frc{14}{5}}\boxtimes 3^B_{\frc{5}{2}}$\\
 $6^B_{\frc{7}{10}}$ & $1.9276$ &
$14.472$ & $1,1,\zeta_2^1,\zeta_3^1,\zeta_3^1,\zeta_2^1\zeta_3^1$ &
$0,\frac{1}{2}, \frac{7}{16}, \frac{1}{10},-\frac{2}{5}, \frac{3}{80}$ &
$2^B_{-\frc{14}{5}}\boxtimes 3^B_{\frc{7}{2}}$\\
 $6^B_{-\frc{7}{10}}$ & $1.9276$ &
$14.472$ & $1,1,\zeta_2^1,\zeta_3^1,\zeta_3^1,\zeta_2^1\zeta_3^1$ &
$0,\frac{1}{2},-\frac{7}{16},-\frac{1}{10}, \frac{2}{5},-\frac{3}{80}$ & $2^B_{
\frc{14}{5}}\boxtimes 3^B_{-\frc{7}{2}}$\\
 $6^B_{\frc{13}{10}}$ & $1.9276$ &
$14.472$ & $1,1,\zeta_2^1,\zeta_3^1,\zeta_3^1,\zeta_2^1\zeta_3^1$ &
$0,\frac{1}{2},-\frac{3}{16},-\frac{1}{10}, \frac{2}{5}, \frac{17}{80}$ & $2^B_{
\frc{14}{5}}\boxtimes 3^B_{-\frc{3}{2}}$\\
 $6^B_{-\frc{13}{10}}$ & $1.9276$ &
$14.472$ & $1,1,\zeta_2^1,\zeta_3^1,\zeta_3^1,\zeta_2^1\zeta_3^1$ &
$0,\frac{1}{2}, \frac{3}{16}, \frac{1}{10},-\frac{2}{5},-\frac{17}{80}$ &
$2^B_{-\frc{14}{5}}\boxtimes 3^B_{\frc{3}{2}}$\\
 $6^B_{\frc{17}{10}}$ & $1.9276$ &
$14.472$ & $1,1,\zeta_2^1,\zeta_3^1,\zeta_3^1,\zeta_2^1\zeta_3^1$ &
$0,\frac{1}{2},-\frac{7}{16}, \frac{1}{10},-\frac{2}{5}, \frac{13}{80}$ &
$2^B_{-\frc{14}{5}}\boxtimes 3^B_{-\frc{7}{2}}$\\
 $6^B_{-\frc{17}{10}}$ & $1.9276$ &
$14.472$ & $1,1,\zeta_2^1,\zeta_3^1,\zeta_3^1,\zeta_2^1\zeta_3^1$ &
$0,\frac{1}{2}, \frac{7}{16},-\frac{1}{10}, \frac{2}{5},-\frac{13}{80}$ & $2^B_{
\frc{14}{5}}\boxtimes 3^B_{\frc{7}{2}}$\\
 $6^B_{\frc{23}{10}}$ & $1.9276$ &
$14.472$ & $1,1,\zeta_2^1,\zeta_3^1,\zeta_3^1,\zeta_2^1\zeta_3^1$ &
$0,\frac{1}{2},-\frac{1}{16},-\frac{1}{10}, \frac{2}{5}, \frac{27}{80}$ & $2^B_{
\frc{14}{5}}\boxtimes 3^B_{-\frc{1}{2}}$\\
 $6^B_{-\frc{23}{10}}$ & $1.9276$ &
$14.472$ & $1,1,\zeta_2^1,\zeta_3^1,\zeta_3^1,\zeta_2^1\zeta_3^1$ &
$0,\frac{1}{2}, \frac{1}{16}, \frac{1}{10},-\frac{2}{5},-\frac{27}{80}$ &
$2^B_{-\frc{14}{5}}\boxtimes 3^B_{\frc{1}{2}}$\\
 $6^B_{\frc{27}{10}}$ & $1.9276$ &
$14.472$ & $1,1,\zeta_2^1,\zeta_3^1,\zeta_3^1,\zeta_2^1\zeta_3^1$ &
$0,\frac{1}{2},-\frac{5}{16}, \frac{1}{10},-\frac{2}{5}, \frac{23}{80}$ &
$2^B_{-\frc{14}{5}}\boxtimes 3^B_{-\frc{5}{2}}$\\
 $6^B_{-\frc{27}{10}}$ & $1.9276$ &
$14.472$ & $1,1,\zeta_2^1,\zeta_3^1,\zeta_3^1,\zeta_2^1\zeta_3^1$ &
$0,\frac{1}{2}, \frac{5}{16},-\frac{1}{10}, \frac{2}{5},-\frac{23}{80}$ & $2^B_{
\frc{14}{5}}\boxtimes 3^B_{\frc{5}{2}}$\\
 $6^B_{\frc{33}{10}}$ & $1.9276$ &
$14.472$ & $1,1,\zeta_2^1,\zeta_3^1,\zeta_3^1,\zeta_2^1\zeta_3^1$ &
$0,\frac{1}{2}, \frac{1}{16},-\frac{1}{10}, \frac{2}{5}, \frac{37}{80}$ & $2^B_{
\frc{14}{5}}\boxtimes 3^B_{\frc{1}{2}}$\\
 $6^B_{-\frc{33}{10}}$ & $1.9276$ &
$14.472$ & $1,1,\zeta_2^1,\zeta_3^1,\zeta_3^1,\zeta_2^1\zeta_3^1$ &
$0,\frac{1}{2},-\frac{1}{16}, \frac{1}{10},-\frac{2}{5},-\frac{37}{80}$ &
$2^B_{-\frc{14}{5}}\boxtimes 3^B_{-\frc{1}{2}}$\\
 $6^B_{\frc{37}{10}}$ & $1.9276$ &
$14.472$ & $1,1,\zeta_2^1,\zeta_3^1,\zeta_3^1,\zeta_2^1\zeta_3^1$ &
$0,\frac{1}{2},-\frac{3}{16}, \frac{1}{10},-\frac{2}{5}, \frac{33}{80}$ &
$2^B_{-\frc{14}{5}}\boxtimes 3^B_{-\frc{3}{2}}$\\
 $6^B_{-\frc{37}{10}}$ & $1.9276$ &
$14.472$ & $1,1,\zeta_2^1,\zeta_3^1,\zeta_3^1,\zeta_2^1\zeta_3^1$ &
$0,\frac{1}{2}, \frac{3}{16},-\frac{1}{10}, \frac{2}{5},-\frac{33}{80}$ & $2^B_{
\frc{14}{5}}\boxtimes 3^B_{\frc{3}{2}}$\\
 \hline $6^B_{\frc{1}{7}}$ & $2.1082$ &
$18.591$ & $1,1,\zeta_5^1,\zeta_5^1,\zeta_5^2,\zeta_5^2$ &
$0,-\frac{1}{4},-\frac{1}{7},-\frac{11}{28}, \frac{1}{28}, \frac{2}{7}$ &
$2^B_{-1}\boxtimes 3^B_{\frc{8}{7}}$\\
 $6^B_{-\frc{1}{7}}$ & $2.1082$ & $18.591$ &
$1,1,\zeta_5^1,\zeta_5^1,\zeta_5^2,\zeta_5^2$ & $0, \frac{1}{4}, \frac{1}{7},
\frac{11}{28},-\frac{1}{28},-\frac{2}{7}$ & $2^B_{ 1}\boxtimes 3^B_{-\frc{8}{7}}$\\
$6^B_{\frc{15}{7}}$ & $2.1082$ & $18.591$ &
$1,1,\zeta_5^1,\zeta_5^1,\zeta_5^2,\zeta_5^2$ & $0, \frac{1}{4},
\frac{3}{28},-\frac{1}{7}, \frac{2}{7},-\frac{13}{28}$ & $2^B_{ 1}\boxtimes 3^B_{
\frc{8}{7}}$\\
 $6^B_{-\frc{15}{7}}$ & $2.1082$ & $18.591$ &
$1,1,\zeta_5^1,\zeta_5^1,\zeta_5^2,\zeta_5^2$ & $0,-\frac{1}{4},-\frac{3}{28},
\frac{1}{7},-\frac{2}{7}, \frac{13}{28}$ & $2^B_{-1}\boxtimes 3^B_{-\frc{8}{7}}$\\
\hline 
$6^{B,a}_{ 0}$ & $2.1609$ & $20$ & $1,1,2,2,\sqrt{5},\sqrt{5}$ & $0, 0,
\frac{1}{5},-\frac{1}{5}, 0,\frac{1}{2}$ & primitive \\
 $6^{B,b}_{ 0}$ & $2.1609$ & $20$ &
$1,1,2,2,\sqrt{5},\sqrt{5}$ & $0, 0, \frac{1}{5},-\frac{1}{5},
\frac{1}{4},-\frac{1}{4}$ & primitive \\
 $6^{B,a}_{4}$ & $2.1609$ & $20$ &
$1,1,2,2,\sqrt{5},\sqrt{5}$ & $0, 0, \frac{2}{5},-\frac{2}{5}, 0,\frac{1}{2}$ &
primitive \\
 $6^{B,b}_{4}$ & $2.1609$ & $20$ & $1,1,2,2,\sqrt{5},\sqrt{5}$ & $0, 0,
\frac{2}{5},-\frac{2}{5}, \frac{1}{4},-\frac{1}{4}$ & primitive \\
 \hline 
$6^B_{ \frc{58}{35}}$ & $2.5359$ & $33.632$ &
$1,\zeta_3^1,\zeta_5^1,\zeta_5^2,\zeta_3^1\zeta_5^1,\zeta_3^1\zeta_5^2$ & $0,
\frac{2}{5}, \frac{1}{7},-\frac{2}{7},-\frac{16}{35}, \frac{4}{35}$ & $2^B_{
\frc{14}{5}}\boxtimes 3^B_{-\frc{8}{7}}$\\
 $6^B_{-\frc{58}{35}}$ & $2.5359$ &
$33.632$ &
$1,\zeta_3^1,\zeta_5^1,\zeta_5^2,\zeta_3^1\zeta_5^1,\zeta_3^1\zeta_5^2$ &
$0,-\frac{2}{5},-\frac{1}{7}, \frac{2}{7}, \frac{16}{35},-\frac{4}{35}$ &
$2^B_{-\frc{14}{5}}\boxtimes 3^B_{\frc{8}{7}}$\\
 $6^B_{\frc{138}{35}}$ & $2.5359$
& $33.632$ &
$1,\zeta_3^1,\zeta_5^1,\zeta_5^2,\zeta_3^1\zeta_5^1,\zeta_3^1\zeta_5^2$ & $0,
\frac{2}{5},-\frac{1}{7}, \frac{2}{7}, \frac{9}{35},-\frac{11}{35}$ & $2^B_{
\frc{14}{5}}\boxtimes 3^B_{\frc{8}{7}}$\\
 $6^B_{-\frc{138}{35}}$ & $2.5359$ &
$33.632$ &
$1,\zeta_3^1,\zeta_5^1,\zeta_5^2,\zeta_3^1\zeta_5^1,\zeta_3^1\zeta_5^2$ &
$0,-\frac{2}{5}, \frac{1}{7},-\frac{2}{7},-\frac{9}{35}, \frac{11}{35}$ &
$2^B_{-\frc{14}{5}}\boxtimes 3^B_{-\frc{8}{7}}$\\
 \hline {$6^B_{\frc{46}{13}}$}
& {$2.9132$} & {$56.746$} &
{$1,\zeta_{11}^{1},\zeta_{11}^{2},\zeta_{11}^{3},\zeta_{11}^{4},\zeta_{11}^{5}$}
& {$0, \frac{4}{13}, \frac{2}{13},-\frac{6}{13},
\frac{6}{13},-\frac{1}{13}$} & primitive \\
 {$6^B_{-\frc{46}{13}}$} & {$2.9132$} &
{$56.746$} &
{$1,\zeta_{11}^{1},\zeta_{11}^{2},\zeta_{11}^{3},\zeta_{11}^{4},\zeta_{11}^{5}$}
& {$0,-\frac{4}{13},-\frac{2}{13}, \frac{6}{13},-\frac{6}{13},
\frac{1}{13}$} &  primitive\\
 \hline \blue{$6^B_{\frc{8}{3}}$} & \blue{$3.1107$} &
\blue{$74.617$} &
\blue{$1,\zeta_{16}^{2},\zeta_{16}^{2},\zeta_{16}^{2},\zeta_{16}^{4},\zeta_{16}^{6}$}
& \blue{$0, \frac{1}{9}, \frac{1}{9}, \frac{1}{9}, \frac{1}{3},-\frac{1}{3}$} &
\color{blue} primitive \\
 \blue{$6^B_{-\frc{8}{3}}$} & \blue{$3.1107$} & \blue{$74.617$} &
\blue{$1,\zeta_{16}^{2},\zeta_{16}^{2},\zeta_{16}^{2},\zeta_{16}^{4},\zeta_{16}^{6}$}
& \blue{$0,-\frac{1}{9},-\frac{1}{9},-\frac{1}{9},-\frac{1}{3}, \frac{1}{3}$} &
\color{blue} primitive \\
\hline 
\color{blue} $6^B_{2}$ &\color{blue}  $3.3263$ &\color{blue}  $100.61$ &\color{blue}  $1,\frac{3+\sqrt{21}}{2},\frac{3+\sqrt{21}}{2},\frac{3+\sqrt{21}}{2},\frac{5+\sqrt{21}}{2},\frac{7+\sqrt{21}}{2}$ &\color{blue}  $0,-\frac{1}{7},-\frac{2}{7}, \frac{3}{7}, 0, \frac{1}{3}$ &\color{blue} primitive \\
\color{blue} $6^B_{-2}$ &\color{blue}  $3.3263$ &\color{blue}  $100.61$ &\color{blue}  $1,\frac{3+\sqrt{21}}{2},\frac{3+\sqrt{21}}{2},\frac{3+\sqrt{21}}{2},\frac{5+\sqrt{21}}{2},\frac{7+\sqrt{21}}{2}$ &\color{blue}  $0, \frac{1}{7}, \frac{2}{7},-\frac{3}{7}, 0,-\frac{1}{3}$ & \color{blue}primitive \\
 \hline 
\end{tabular} 
\end{table*}

\begin{table*}[tb] 
\caption{ A list of all 24 bosonic rank $N=7$ topological
orders in 2+1D with max$(N^{ij}_k)\leq 1$.   Since $N=7$ is a prime number, all
those 24 topological order are primitive.  } \label{toplst7} \centering
\begin{tabular}{ |c|c|c|l|l| } 
\hline $N^B_c$ & $S_\text{top}$ & $D^2$ & $d_1,d_2,\cdots$ & $s_1,s_2,\cdots$  \\
 \hline 
$7^{B,a}_{ 2}$ & $1.4036$ & $7$ &
$1,1,1,1,1,1,1$ & $0, \frac{1}{7}, \frac{1}{7}, \frac{2}{7},
\frac{2}{7},-\frac{3}{7},-\frac{3}{7}$ \\
 $7^{B,a}_{-2}$ & $1.4036$ & $7$ &
$1,1,1,1,1,1,1$ & $0,-\frac{1}{7},-\frac{1}{7},-\frac{2}{7},-\frac{2}{7},
\frac{3}{7}, \frac{3}{7}$ \\
 \hline $7^B_{\frc{1}{4}}$ & $2.3857$ & $27.313$ &
$1,1,\zeta_6^1,\zeta_6^1,\zeta_6^2,\zeta_6^2,\zeta_6^3$ &
$0,\frac{1}{2},-\frac{5}{32},-\frac{5}{32}, \frac{1}{4},-\frac{1}{4},
\frac{7}{32}$ \\
 $7^B_{-\frc{1}{4}}$ & $2.3857$ & $27.313$ &
$1,1,\zeta_6^1,\zeta_6^1,\zeta_6^2,\zeta_6^2,\zeta_6^3$ & $0,\frac{1}{2},
\frac{5}{32}, \frac{5}{32}, \frac{1}{4},-\frac{1}{4},-\frac{7}{32}$ \\
 $7^B_{ \frc{3}{4}}$ & $2.3857$ & $27.313$ &
$1,1,\zeta_6^1,\zeta_6^1,\zeta_6^2,\zeta_6^2,\zeta_6^3$ & $0,\frac{1}{2},
\frac{9}{32}, \frac{9}{32}, \frac{1}{4},-\frac{1}{4},-\frac{3}{32}$ \\
$7^B_{-\frc{3}{4}}$ & $2.3857$ & $27.313$ &
$1,1,\zeta_6^1,\zeta_6^1,\zeta_6^2,\zeta_6^2,\zeta_6^3$ &
$0,\frac{1}{2},-\frac{9}{32},-\frac{9}{32}, \frac{1}{4},-\frac{1}{4},
\frac{3}{32}$ \\
 $7^B_{\frc{5}{4}}$ & $2.3857$ & $27.313$ &
$1,1,\zeta_6^1,\zeta_6^1,\zeta_6^2,\zeta_6^2,\zeta_6^3$ &
$0,\frac{1}{2},-\frac{1}{32},-\frac{1}{32}, \frac{1}{4},-\frac{1}{4},
\frac{11}{32}$ \\
 $7^B_{-\frc{5}{4}}$ & $2.3857$ & $27.313$ &
$1,1,\zeta_6^1,\zeta_6^1,\zeta_6^2,\zeta_6^2,\zeta_6^3$ & $0,\frac{1}{2},
\frac{1}{32}, \frac{1}{32}, \frac{1}{4},-\frac{1}{4},-\frac{11}{32}$ \\
 $7^B_{ \frc{7}{4}}$ & $2.3857$ & $27.313$ &
$1,1,\zeta_6^1,\zeta_6^1,\zeta_6^2,\zeta_6^2,\zeta_6^3$ & $0,\frac{1}{2},
\frac{13}{32}, \frac{13}{32}, \frac{1}{4},-\frac{1}{4}, \frac{1}{32}$ \\
$7^B_{-\frc{7}{4}}$ & $2.3857$ & $27.313$ &
$1,1,\zeta_6^1,\zeta_6^1,\zeta_6^2,\zeta_6^2,\zeta_6^3$ &
$0,\frac{1}{2},-\frac{13}{32},-\frac{13}{32},
\frac{1}{4},-\frac{1}{4},-\frac{1}{32}$ \\
 $7^B_{\frc{9}{4}}$ & $2.3857$ &
$27.313$ & $1,1,\zeta_6^1,\zeta_6^1,\zeta_6^2,\zeta_6^2,\zeta_6^3$ &
$0,\frac{1}{2}, \frac{3}{32}, \frac{3}{32}, \frac{1}{4},-\frac{1}{4},
\frac{15}{32}$ \\
 $7^B_{-\frc{9}{4}}$ & $2.3857$ & $27.313$ &
$1,1,\zeta_6^1,\zeta_6^1,\zeta_6^2,\zeta_6^2,\zeta_6^3$ &
$0,\frac{1}{2},-\frac{3}{32},-\frac{3}{32},
\frac{1}{4},-\frac{1}{4},-\frac{15}{32}$ \\
 $7^B_{\frc{11}{4}}$ & $2.3857$ &
$27.313$ & $1,1,\zeta_6^1,\zeta_6^1,\zeta_6^2,\zeta_6^2,\zeta_6^3$ &
$0,\frac{1}{2},-\frac{15}{32},-\frac{15}{32}, \frac{1}{4},-\frac{1}{4},
\frac{5}{32}$ \\
 $7^B_{-\frc{11}{4}}$ & $2.3857$ & $27.313$ &
$1,1,\zeta_6^1,\zeta_6^1,\zeta_6^2,\zeta_6^2,\zeta_6^3$ & $0,\frac{1}{2},
\frac{15}{32}, \frac{15}{32}, \frac{1}{4},-\frac{1}{4},-\frac{5}{32}$ \\
 $7^B_{ \frc{13}{4}}$ & $2.3857$ & $27.313$ &
$1,1,\zeta_6^1,\zeta_6^1,\zeta_6^2,\zeta_6^2,\zeta_6^3$ & $0,\frac{1}{2},
\frac{7}{32}, \frac{7}{32}, \frac{1}{4},-\frac{1}{4},-\frac{13}{32}$ \\
$7^B_{-\frc{13}{4}}$ & $2.3857$ & $27.313$ &
$1,1,\zeta_6^1,\zeta_6^1,\zeta_6^2,\zeta_6^2,\zeta_6^3$ &
$0,\frac{1}{2},-\frac{7}{32},-\frac{7}{32}, \frac{1}{4},-\frac{1}{4},
\frac{13}{32}$ \\
 $7^B_{\frc{15}{4}}$ & $2.3857$ & $27.313$ &
$1,1,\zeta_6^1,\zeta_6^1,\zeta_6^2,\zeta_6^2,\zeta_6^3$ &
$0,\frac{1}{2},-\frac{11}{32},-\frac{11}{32}, \frac{1}{4},-\frac{1}{4},
\frac{9}{32}$ \\
 $7^B_{-\frc{15}{4}}$ & $2.3857$ & $27.313$ &
$1,1,\zeta_6^1,\zeta_6^1,\zeta_6^2,\zeta_6^2,\zeta_6^3$ & $0,\frac{1}{2},
\frac{11}{32}, \frac{11}{32}, \frac{1}{4},-\frac{1}{4},-\frac{9}{32}$ \\
 \hline
$7^{B,b}_{ 2}$ & $2.4036$ & $28$ & $1,1,2,2,2,\sqrt{7},\sqrt{7}$ & $0, 0,
\frac{1}{7}, \frac{2}{7},-\frac{3}{7}, \frac{1}{8},-\frac{3}{8}$ \\
 $7^{B,c}_{ 2}$ &
$2.4036$ & $28$ & $1,1,2,2,2,\sqrt{7},\sqrt{7}$ & $0, 0, \frac{1}{7},
\frac{2}{7},-\frac{3}{7},-\frac{1}{8}, \frac{3}{8}$ \\
 $7^{B,c}_{-2}$ & $2.4036$ &
$28$ & $1,1,2,2,2,\sqrt{7},\sqrt{7}$ & $0, 0,-\frac{1}{7},-\frac{2}{7},
\frac{3}{7}, \frac{1}{8},-\frac{3}{8}$ \\
 $7^{B,b}_{-2}$ & $2.4036$ & $28$ &
$1,1,2,2,2,\sqrt{7},\sqrt{7}$ & $0, 0,-\frac{1}{7},-\frac{2}{7},
\frac{3}{7},-\frac{1}{8}, \frac{3}{8}$ \\
 \hline 
$7^B_{8/5}$ & $3.2194$ & $86.750$ & $1,\zeta_{13}^{1},\zeta_{13}^{2},\zeta_{13}^{3},\zeta_{13}^{4},\zeta_{13}^{5},\zeta_{13}^{6}$ & $0,-\frac{1}{5}, \frac{2}{15}, 0, \frac{2}{5}, \frac{1}{3},-\frac{1}{5}$ \\
$7^B_{-8/5}$ & $3.2194$ & $86.750$ & $1,\zeta_{13}^{1},\zeta_{13}^{2},\zeta_{13}^{3},\zeta_{13}^{4},\zeta_{13}^{5},\zeta_{13}^{6}$ & $0, \frac{1}{5},-\frac{2}{15}, 0,-\frac{2}{5},-\frac{1}{3}, \frac{1}{5}$ \\
 \hline 
\end{tabular} 
\end{table*}

\subsection{A numerical approach}

Here, we will assume the conditions in section \ref{Nsc} to be sufficient, and
treat them as a classifying theory of 2+1D topological orders.  In this
section, we will describe how to numerically solve those conditions to obtain a
list of simple 2+1D topological orders.  Our approach is similar to that used
in \Ref{GKh9410089}, where a list of fusion rings are obtain.  Here, we will
obtain a list of 2+1D bosonic topological orders.

We first numerically solve the condition (1) in the $(N^{ij}_k,s_i,c)$
conditions in section \ref{Nsc} to obtain $N^{ij}_k$. Then we will use Smith
normal form of integer matrices $V_{ijkl}^r$ and/or $\t M_{ij}$ to solve the
condition (2) to obtain a list of $s_i$.  We then use other conditions to
obtain a list of $(N^{ij}_k,s_i)$'s that satisfy all those conditions by direct
checking.  The central charge $c$ mod 8 is obtained from the condition (4).

To numerically solve the condition (1) in the \textbf{$(N^{ij}_k,s_i,c)$ conditions}
efficiently, it is important to find as many conditions on $N^{ij}_k$ as
possible.  We first set $l=1$ in \eqn{Ncnd} and find the following symmetry
condition on $N^{ij}_k$ 
\begin{align} 
\label{Nc1}
N_{\bar k}^{ij}= N_{\bar i}^{jk} = N_{\bar j}^{ki} = N_{\bar j}^{ik} .  
\end{align}
The second kind of conditions on $N^{ij}_k$ is that 
\begin{align}
\label{Nc2}
[N_i,N_j]=0.
\end{align}

To find more conditions on $N^{ij}_k$, we note that since $S$ is unitary, we
may rewrite \eqn{Vform} as 
\begin{align} 
\sum_k N^{ij}_kS_{lk} = \frac{ S_{li}
S_{lj} }{ S_{1l} } \text{ or } v_l N_i = d^i_l v_l , 
\end{align} 
where the row eigenvector $v_i$ is given by $(v_l)_j =S_{lj}$ and the
eigenvalues $d^i_l=S_{li}/S_{1l}$.  In other words there exist a symmetric
unitary matrix that satisfies 
\begin{align} 
S N_i S^\dag = D^i, 
\end{align}
where $D^i$ is a diagonal matrix given by $(D^i)_{ll}=d^i_l=S_{li}/S_{1l}$.  We
see that even though $N_i$ may not be hermitian, we still require that 
\begin{align}
\label{Nc3}
N_i \text{ can be diagonalized by a unitary matrix }
\end{align}
This is the third kind of conditions on $N^{ij}_k$.


To get more information, let $u_l$ be the common eigenvectors of a set of
$N_i$'s, $i \in I$ and $I\subset \{1,\cdots,N\}$.  We will try to calculate $S$
from such a subset of $N_i$'s.  Let $V$ be a linear combination of the set of
$N_i$'s, $V=\sum_{i\in I} f(x_i)N_i$.  Let $d^i_l$ be the eigenvalue of $N_i$
for the eigenvector $u_l$.  Let $\t l$ belong to the set of indices that label
eigenvectors that have non-degenerate eigenvalues for $V$.  In this case, the
corresponding eigenvector $u_{\t l}$ is unique up to a $U(1)$ phase factor.
Then those non-degenerate normalized eigenvectors with the first element being
positive satisfies 
\begin{align} 
\label{tvS} 
(u_{\t l})_j = S^*_{p(\t l)j},\ \
\ \ (u_{\t  l})_1 \neq 0 .  
\end{align} 
where $p$ is a permutation map $\t  l
\to  l$.  For those  $(u_{\t  l})_j$'s, we have 
\begin{align}
\sum_i (u_{\t  l_1})_i [(u_{\t  l_2})_i]^* =\del_{\t l_1\t l_2} , \ \ \ d^i_{\t
l}=\frac{S_{p(\t l)i}}{S_{p(\t l) 1}} 
\end{align} 
In other words 
$d^i_{\t l} = \Big( \frac{(u_{\t  l})_i}{(u_{\t  l})_1} \Big)^* $.  

To summarize, let $u_l$ be the common eigenvectors of a set of
$N_i$'s ($i \in I$) with eigenvalue $d^i_l$, 
then
\begin{align} 
\label{Ncnd1} 
(u_{\t  l})_1 \neq
0, \ \sum_j (u_{\t  l_1})_j [(u_{\t  l_2})_j]^* =\del_{\t l_1\t l_2} , \
d^i_{\t l} = \Big(\frac{(u_{\t  l})_i}{(u_{\t  l})_1} \Big)^*  
\end{align} 
for any $i \in I$ and $\t l$'s in the set of that label non-degenerate
eigenvalues.  If the above conditions are not satisfied, then
corresponding $N_i$ does not satisfy the necessary conditions to describe a
topological order.
 
Also, if the all the eigenvalues of $V$'s are non-degenerate, then $u_l$
determines $S$ upto a permutation of the rows (see \eqn{tvS}).  In this case,
we can determine the full $N^{ij}_k$ using \eqn{Ver}.

We wrote a program to numerically search for $N^{ij}_k$'s that satisfy the
condition \eqn{Nc1}, \eqn{Nc2}, \eqn{Ncnd1}, and \eqn{Ver} (when all the eigenvalues of
$V$ are non-degenerate), by starting from $\{N^k_{1j}=\del_{kj}\}$, to
$\{N^k_{1j}, N^k_{2j}\}$, to $\{N^k_{1j}, N^k_{2j}, N^k_{3j}\}$, \etc.

After obtaining a list of fusion rules $N^{ij}_k$, we then, for each fusion
rule, use the Smith normal form of the integer matrix $\t M$ to find sets of
spins $\{s_i\}$ that satisfy \eqn{Ms}.  Last, we select the combination
$(N^{ij}_k,s_i)$ that satisfy all the conditions and compute the central charge
$c$ in the process.  This way we obtain a list of 2+1D topological orders.

\subsection{The stacking operation of topological order}

Before we present the result from the numerical calculation, let us discuss a
stacking operation,\cite{KW1458} denoted by $\boxtimes$.  We note that stacking
two rank $N$ and rank $N'$ topological orders described
$\cC=(N^{ij}_k,s_{i},c)$ and $\cC'=(N^{\prime\ k'}_{i'j'},s'_{i'},c')$ will
give us a third topological order $\cC''=\cC\boxtimes \cC'$ with rank $N''=N N'$
and
\begin{align}
\label{stack}
&
 (N^{\prime\prime})_{kk'}^{ii',jj'} = N^{ij}_k (N^{\prime})_{k'}^{i'j'}, \ s''_{ii'}=s_{i}+s'_{i'},
\  c''=c+c', 
\nonumber\\
&
 d''_{ii'} = d_{i}
d'_{i'}, \ S''_{ii',jj'}= S_{i,j} S'_{i',j'} ,\ S''_\text{top} = S_\text{top} +
S'_\text{top} . 
\end{align}
 where $S_\text{top}$ is the topological entanglement entropy
$S_\text{top}=\log_2 D$, $D=\sqrt{\sum_i d_i^2}$.

The stacking operation $\boxtimes$ will make the set of topological order into a
monoid. The trivial topological order $\cC_\text{tri}$ (the product state) is
the unit of the monoid.  However, in general, a topological order $\cC$ does not
have an inverse respect to the stacking operation (\ie there does not exist a
topological order $\bar \cC$ such that $\cC \boxtimes \bar \cC
=\cC_\text{tri}$). This is why the set of topological order only form a monoid
instead of a group.  However, some topological order does have an inverse
respect to the stacking $\boxtimes$ operation. Such kind of topological orders
are called invertible topological orders.\cite{FT1292,KW1458,F1478,K1467,K1459}  

In 2+1D, the  invertible topological orders form an Abelian group $Z$ under to
stacking $\boxtimes$ operation. The group is generated by the $E_8$ FQH state
described by the $K$-matrix
\begin{align} 
K_{E_8}={\footnotesize \begin{pmatrix}
2&1&0&0&0&0&0&0\\ 
1&2&1&0&0&0&0&0\\ 
0&1&2&1&0&0&0&0\\ 
0&0&1&2&1&0&0&0\\
0&0&0&1&2&1&0&1\\ 
0&0&0&0&1&2&1&0\\ 
0&0&0&0&0&1&2&0\\ 
0&0&0&0&1&0&0&2\\
\end{pmatrix} }  .
\end{align} 
The $E_8$ topological order $\cC_{E_8}$ is invertible\cite{KW1458} since it has
no topological excitations\cite{BW9045,WZ9290} (due to $\det(K_{E_8})=1$). It
is described by $(N^{ij}_k,s_i,c) =(1,0,8)$. Stacking an $E_8$ topological
order to an topological order $(N^{ij}_k,s_i,c)$ only shift the central charge
$c$ by 8: $(N^{ij}_k,s_i,c) \to (N^{ij}_k,s_i,c+8)$.  
Such an operation is invertible.

In our lists of 2+1D topological orders, we will only list topological orders
up to invertible topological orders, \ie we will only list the quotient 
\begin{align}
\text{\{Topological orders\}/\{Invertible topological orders\}}.
\end{align}
It turns out that modular tensor category only
describe topological orders up to invertible topological orders.  

\subsection{A list of 2+1D bosonic topological orders with rank
$N=1,2,\cdots,7$}

Table \ref{toplst} lists 2+1D bosonic topological orders with rank
$N=1,2,\cdots,5$ and with $N^{ij}_k=0,1,2,3$.  Here we have ignored the
invertible topological orders.\cite{KW1458}  So the term ``topological order''
really refers to topological order up to invertible topological orders.  

In the table, there is 1 rank $N=1$ topological order, which is actually a
trivial topological order (\ie corresponds to many-body states with no
topological order).  There are 4 non-trivial rank $N=2$ topological orders,
which correspond to $\nu=1/2$ bosonic Laughlin state with central charge $c=1$
and the Fibonacci state with central charge $c=\frac{14}{5}$, plus their time
reversal conjugates.  Those 4 topological orders orders are \emph{primitive} in
the sense that they cannot be obtained by stacking non-invertible topological
orders with lower rank.  

Our numeric calculation also produce 12 rank $N=3$ and 10 rank $N=5$
topological orders, which are all primitive since $N=3,5$ are prime numbers.

For rank $N=4$ topological orders, we find 18 of them.  Applying \eqn{stack},
we find that by stacking two of the rank $N=2$ topological orders, we can
obtain $3+3+4=10$ distinct rank $N=4$ topological orders.  (If  two
$(N^{ij}_k,s_{i},c)$'s are the same up to a permutation of the indices, we will
say they describe the same topological order.) Indeed, 10 of 18  rank $N=4$
topological orders are not primitive, corresponding to the stacking two of the
rank $N=2$ topological orders (see the blue entries in Table \ref{toplst}).  We
also see that 6 primitive topological orders are Abelian since their
topological excitations all have unit quantum dimensions $d_i=1$.  There are
only two non-Abelian rank $N=4$ topological orders, which are related by time
reversal transformation.

We like to pointed out the \Ref{RSW0777} gives a complete classification of all
70 modular tensor categories with rank $N\leq 4$. Compare with such a
classification result, we find that our list for 35 rank $N\leq 4$ topological
orders is complete. (The other 35  modular tensor categories have $S_{11}<0$
and do not correspond to unitary theory.)

We find 50 rank $N=6$ topological orders with $N^{ij}_k\leq 2$ (see Table
\ref{toplst6}).  Most of those 50 topological orders are not primitive and can
be obtained by stacking  rank $N=2$ and rank $N'=3$ topological orders (see the
last column of Table \ref{toplst6}), where we have denoted the topological
orders by their rank $N$ and their central charge $c$: $N^B_c$).  Only 10 among
the 50 are primitive. 
We also find 24 rank $N=7$ topological orders with $N^{ij}_k=0,1$ (see Table
\ref{toplst7}).  They are all primitive since $7$ is a prime number. 

\subsection{Understand the topological orders in the lists}

\subsubsection{Non-Abelian type of topological order}

In this section, we like to gain a better understanding of the topological
orders in the lists. 
Let us first use the stacking operation to introduce the notion of
non-Abelian type of topological order.  Two topological order $\cC_1$ and
$\cC_2$ have the same non-Abelian type iff there exist Abelian topological
orders $\cA_1$ and $\cA_2$ such that 
\begin{align} 
\cC_1\boxtimes \cA_1=\cC_2\boxtimes \cA_2 .  
\end{align} 
The quantum dimensions in Abelian
topological orders are all equal to 1, so topological orders with the same
non-Abelian type must have the same spectrum of the quantum dimensions
(disregard the degeneracy).

\subsubsection{Quantum dimensions as algebraic numbers}

We next note that the quantum dimensions are algebraic
numbers (the roots of polynomial with integer coefficients), since they are
eigenvalues of integer matrices.  So it is helpful to express those quantum
dimensions in terms of algebraic expressions, such as $\sqrt n$. But $\sqrt{n}$
is not enough. So here we introduce another set of algebraic numbers
\begin{align} 
\zeta_n^m=\frac{\sin[\pi(m+1)/(n+2)]}{\sin[\pi/(n+2)]}.
\end{align} 
It turns out that we can express all the quantum dimensions that we find in
terms of $\zeta_n^m$ and $\sqrt n$.

We note that the quantum dimensions that appear in $Z_n$-parafermion CFT
theory\citep{ZF8515} are all given by $\zeta_n^m$.
Also, the $Z_n$-parafermion theory has a central charge 
\begin{align} 
c_{Z_n}=\frac{2n-2}{n+2}.  
\end{align} 
This suggests that many topological orders that we obtain are related to
$Z_n$-parafermion theories.

We like to remark that \eqn{Ncnd} can be rewritten as
\begin{align}
 N_j N_k =
\sum_n N^{jk}_n N_n
\end{align}
 Since $N_i$ commute with each other, their
largest positive eigenvalues $d_i$ satisfy
\begin{align}
 d_j d_k =\sum_n N^{jk}_n d_n. 
\end{align}
 Thus, if we express the quantum dimension $d_i$ in
the basis of algebraic numbers, such as $\zeta_n^m$, with integer coefficients,
we can see the fusion rule $N^{ik}_n$ from the product of $d_i$'s.

\subsubsection{Topological orders of parafermion non-Abelian type}

Using the above concepts, we see that that the two $N=2$ non-Ableian
topological orders have the non-Abelian type of the $Z_3$-parafermion theory
since their quantum dimensions contain $\zeta^1_3$.  Similarly, the $N=3$
topological orders have non-Abelian types of the $Z_2$ and $Z_5$-parafermion
theories.  The primitive $N=4$ non-Abelian topological order has a non-Abelian
type of the $Z_7$-parafermion theory.  Among the $N=5$ topological orders, we
see the non-Abelian types of the $Z_4$- and $Z_9$-parafermion theories.  
Among the primitive $N=6$ topological orders, we see the non-Abelian types of
the $Z_{11}$-parafermion theories.  For $N=7$, we see that there are 16
topological orders with the non-Ableian type  of the $Z_6$-parafermion theory.

\subsubsection{Topological orders of $SO(k)_2$ non-Abelian type}

\begin{table}[tb]
\caption{ The fusion rule $j\otimes i$ for topological order $N^B_c=6^B_4$, which is same as the fusion rule of $SO(5)_2$ current algebra.
For example, $\al\otimes \al=\one \oplus a \oplus \bt$.
}
\label{N6orb}
\centering
\begin{tabular}{ |c|cccccc|}
 \hline 
$s_i$ & 0 & $ 0$ & $ \frac{2}{5}$ & $-\frac{2}{5}$ & $ 0$ & $ \frac{1}{2}$ \\
$d_i$ & 1 & $1$ & $2$ & $2$ & $\sqrt{5}$ & $\sqrt{5}$ \\
\hline
$j\backslash i$ & $\one$ & $a$ & $\al$ & $\bt$ & $\ga$ & $\chi$ \\
\hline
$\one$  & $\one$  & $a$  & $\al$  & $\bt$  & $\ga$  & $\chi$  \\
$a$  & $a$  & $\one$  & $\al$  & $\bt$  & $\chi$  & $\ga$  \\
$\al$  & $\al$  & $\al$  & $\one \oplus a \oplus \bt$  & $\al \oplus \bt$  & $\ga \oplus \chi$  & $\ga \oplus \chi$  \\
$\bt$  & $\bt$  & $\bt$  & $\al \oplus \bt$  & $\one \oplus a \oplus \al$  & $\ga \oplus \chi$  & $\ga \oplus \chi$  \\
$\ga$  & $\ga$  & $\chi$  & $\ga \oplus \chi$  & $\ga \oplus \chi$  & $\one \oplus \al \oplus \bt$  & $a \oplus \al \oplus \bt$  \\
$\chi$  & $\chi$  & $\ga$  & $\ga \oplus \chi$  & $\ga \oplus \chi$  & $a \oplus \al \oplus \bt$  & $\one \oplus \al \oplus \bt$  \\
\hline 
 \end{tabular}
\end{table}

\begin{table*}[tb]
\caption{ The fusion rule $j\otimes i$ for topological order $N^B_c=5^B_\frc{18}{7}$}
\label{N52} 
\centering
\begin{tabular}{ |c|ccccc|}
 \hline 
$s_i$ & 0 & $-\frac{1}{7}$ & $-\frac{1}{7}$ & $ \frac{1}{7}$ & $ \frac{3}{7}$ \\
$d_i$ & 1 & $\zeta_{5}^{2}$ & $\zeta_{5}^{2}$ & $\zeta_{12}^{2}$ & $\zeta_{12}^{4}$ \\
\hline
$j\backslash i$ & $\one$ & $\al$ & $\bt$ & $\ga$ & $\chi$ \\
\hline
$\one$  & $\one$  & $\al$  & $\bt$  & $\ga$  & $\chi$  \\
$\al$  & $\al$  & $\bt \oplus \ga$  & $\one \oplus \chi$  & $\bt \oplus \chi$  & $\al \oplus \ga \oplus \chi$  \\
$\bt$  & $\bt$  & $\one \oplus \chi$  & $\al \oplus \ga$  & $\al \oplus \chi$  & $\bt \oplus \ga \oplus \chi$  \\
$\ga$  & $\ga$  & $\bt \oplus \chi$  & $\al \oplus \chi$  & $\one \oplus \ga \oplus \chi$  & $\al \oplus \bt \oplus \ga \oplus \chi$  \\
$\chi$  & $\chi$  & $\al \oplus \ga \oplus \chi$  & $\bt \oplus \ga \oplus \chi$  & $\al \oplus \bt \oplus \ga \oplus \chi$  & $\one \oplus \al \oplus \bt \oplus \ga \oplus 2\chi$  \\
\hline 
 \end{tabular}
\end{table*}
\begin{table*}[tb]
\caption{ The fusion rule $j\otimes i$ for topological order $N^B_c=6^B_\frc{8}{3}$.}
\label{N62} 
\centering
\begin{tabular}{ |c|cccccc|}
 \hline 
$s_i$ & 0 & $ \frac{1}{9}$ & $ \frac{1}{9}$ & $ \frac{1}{9}$ & $ \frac{1}{3}$ & $-\frac{1}{3}$ \\
$d_i$ & 1 & $\zeta_{16}^{2}$ & $\zeta_{16}^{2}$ & $\zeta_{16}^{2}$ & $\zeta_{16}^{4}$ & $\zeta_{16}^{6}$ \\
\hline
$j\backslash i$ & $\one$ & $\al$ & $\bt$ & $\ga$ & $\chi$ & $\eta$ \\
\hline
$\one$  & $\one$  & $\al$  & $\bt$  & $\ga$  & $\chi$  & $\eta$  \\
$\al$  & $\al$  & $\one \oplus \al \oplus \chi$  & $\ga \oplus \eta$  & $\bt \oplus \eta$  & $\al \oplus \chi \oplus \eta$  & $\bt \oplus \ga \oplus \chi \oplus \eta$  \\
$\bt$  & $\bt$  & $\ga \oplus \eta$  & $\one \oplus \bt \oplus \chi$  & $\al \oplus \eta$  & $\bt \oplus \chi \oplus \eta$  & $\al \oplus \ga \oplus \chi \oplus \eta$  \\
$\ga$  & $\ga$  & $\bt \oplus \eta$  & $\al \oplus \eta$  & $\one \oplus \ga \oplus \chi$  & $\ga \oplus \chi \oplus \eta$  & $\al \oplus \bt \oplus \chi \oplus \eta$  \\
$\chi$  & $\chi$  & $\al \oplus \chi \oplus \eta$  & $\bt \oplus \chi \oplus \eta$  & $\ga \oplus \chi \oplus \eta$  & $\one \oplus \al \oplus \bt \oplus \ga \oplus \chi \oplus \eta$  & $\al \oplus \bt \oplus \ga \oplus \chi \oplus 2\eta$  \\
$\eta$  & $\eta$  & $\bt \oplus \ga \oplus \chi \oplus \eta$  & $\al \oplus \ga \oplus \chi \oplus \eta$  & $\al \oplus \bt \oplus \chi \oplus \eta$  & $\al \oplus \bt \oplus \ga \oplus \chi \oplus 2\eta$  & $\one \oplus \al \oplus \bt \oplus \ga \oplus 2\chi \oplus 2\eta$  \\
\hline 
 \end{tabular}
\end{table*}

However, there are four $N=6$ topological orders (see Table \ref{N6orb}) and
four $N=7$ topological orders that are not related to the parafermion theories.
They are the so called $TY(A,\chi,\tau)^{\Z_2}$ category studied in
\Ref{GN09053117}, with $A=\Z_5$ for $N=6$ cases and $A=\Z_7$ for $N=7$ cases.
They belong to \emph{metaplectic modular categories}, which are defined as any
modular category with the same fusion rules as $SO(k)_2$ for $k$ odd. They have
rank $N=(k +7)/2$ and dimension $D^2=4 N$. They have two 1-dimensional objects
and two $\sqrt{n}$-dimensional objects objects.  The remaining $\frac{N-1}{2}$
objects have dimension 2.\cite{BW14112313} We also like to point out that the
four $N=6$ topological orders and the four $N=7$ topological orders that are
closely related to $U(1)_k/Z_2$ orbifold CFT with
$k=5,7$.\cite{BW1023,BW1121}

\subsubsection{Other topological orders beyond parafermion non-Abelian type}

In addition to the $SO(k)_2$ non-Abelian topological orders, there are also a
few topological orders that are beyond parafermion non-Abelian type (see Tables
\ref{N52} and \ref{N62}).  Some of the fusion coefficient $N^{ij}_k = 2$
for those topological orders.

\section{Physical realization of the topologically ordered states}

In this section, we will discuss some physical realization of the topological
orders that we find through the classifying theory.  In this section, we will
refer different topological orders by their rank $N$ and central charge $c$, and
use $N^B_c$ to denote them.

\subsection{Abelian topological orders}

All the Abelian topological orders can be describe by the $K$-matrix and can be
realized by multilayer FQH states.  
\begin{enumerate} 
\item 
The topological
order $N^B_c= 2^B_1$ in Table \ref{toplst}, is described by a 1-by-1 $K$-matrix
$K=(2)$.  It realized by the Laughlin wave function for bosons
$\Psi_{2^B_1}=\prod(z_i-z_j)^2 \ee^{-\frac14 \sum |z_i|^2}$.  
\item
 The
topological order $4^B_1$  is described by another 1-by-1 $K$-matrix $K=(4)$, and
is realized by the Laughlin wave function for $\Psi_{4^B_1}=\prod(z_i-z_j)^4
\ee^{-\frac14 \sum |z_i|^2}$.  
\item
 The $3^B_2$ topological order is described
by a 2-by-2 $K$-matrix $K=\bpm 2&1\\
 1&2\\
 \epm$, and can be realized by a
double-layer bosonic FQH state $\Psi_{3^B_2}= \prod(z_i-z_j)^2 \prod(w_i-w_j)^2
\prod(z_i-w_j) \ee^{-\frac14 \sum (|z_i|^2+|w_i|^2)}$.  
\item
 Stacking two
$3^B_2$ topological orders give rise to a $9^B_4$ topological order described by
\begin{align} K=\bpm 2&1&0&0\\
 1&2&0&0\\
 0&0&2&1\\
 0&0&1&2\\
 \epm .
\end{align} Such a topological order has 9 different types of topological
excitations.  Their spins are given by
\begin{align}
 \{s_i\} =\{ 0,
\frac{1}{3}\times 4, -\frac{1}{3}\times 4, \}. 
\end{align}
 \ie there are 4
types of topological excitations with spin $\frac{1}{3}$, and 4 types of
topological excitations with spin $-\frac{1}{3}$.  
\item
 There are two Abelian
$4^B_0$ topological orders. The first one is the $Z_2$ topological order
described by $K=\bpm 0&2\\
 2&0\\
 \epm$, which can be realized by $Z_2$ spin
liquids \cite{RS9173,Wsrvb} or toric code model.\cite{K032} The other is the
double-semion topological order described by $K=\bpm 2&0\\
 0&-2\\
 \epm$, which
can be realized by a string-net model\cite{LWstrnet}.  
\item
 The $5^B_0$
topological order is described by $K=\bpm 2&3\\
 3&2\\
 \epm$, and can be
realized by a double-layer bosonic FQH state $\Psi_{5^B_0}= \prod(z_i-z_j)^2
\prod(w_i-w_j)^2 \prod(z_i-w_j)^3 \ee^{-\frac14 \sum (|z_i|^2+|w_i|^2)}$.
\item
 The Abelian $7^B_2$ topological order in Table \ref{toplst7} is
described by $K=\bpm 4&3\\
 3&4\\
 \epm$, and can be realized by a double-layer
bosonic FQH state $\Psi_{7^B_0}= \prod(z_i-z_j)^4 \prod(w_i-w_j)^4
\prod(z_i-w_j)^3 \ee^{-\frac14 \sum (|z_i|^2+|w_i|^2)}$.  
\item 
The $4^B_4$ and $5^B_4$ topological orders are described by 
\begin{align}
K_{4^B_4}= \begin{pmatrix} 
 2 & 1 &1 &1 \\
 1 & 2 &0 &0 \\
 1 & 0 &2 &0 \\
 1 & 0 &0 &2 \\
\end{pmatrix}, \ \ \ 
K_{5^B_4}= \begin{pmatrix} 
 2 & 1 &1 &1 \\
 1 & 2 &1 &0 \\
 1 & 1 &2 &0 \\
 1 & 0 &0 &2 \\
\end{pmatrix}.  
\end{align} 
They can be realized by a four-layer FQH
states.  
\end{enumerate}

\subsection{Non-Abelian topological orders of $Z_n$-parafermion type}

\begin{table}[tb]
\caption{The fusion rule $j\otimes i$ for a $Z_2$ parafermion topological order $N^B_c=3^B_\frc{5}{2}$.  Such a topological order can be realized\cite{Wnab,BW9215} by
wave function $\Psi_{3^B_{\frc{5}{2}}} = [\Psi^\text{LL}_2(\{z_i\})]^2 $.  
The edge states are described by $SU(2)_2\times U(1)$ Kac-Moody
algebra.  Note that $\zeta_{2}^{1}=\sqrt 2$.
}
\label{N3p5/2}
\centering
\begin{tabular}{ |c|ccc|}
 \hline 
$s_i$ & 0 & $ \frac{1}{2}$ & $ \frac{5}{16}$ \\
$d_i$ & 1 & $1$ & $\zeta_{2}^{1}$ \\
\hline
$j\backslash i$ & $\one$ & $\psi$ & $\si$ \\\hline
$\one$  & $\one$  & $\psi$  & $\si$  \\
$\psi$  & $\psi$  & $\one$  & $\si$  \\
$\si$  & $\si$  & $\si$  & $\one \oplus \psi$  \\
\hline 
 \end{tabular}
\end{table}

\begin{table}[tb]
\caption{The fusion rule $j\otimes i$ for a $Z_3$ parafermion (Fibonacci) topological order $N^B_c= {4^B_{-\frc{19}{5}}} \sim
4^B_\frc{21}{5} $.  Such a topological order can be realized\cite{Wnab,BW9215} by wave function $
\Psi_{4^B_{\frc{21}{5}}} = [\Psi^\text{LL}_3(\{z_i\})]^2 $.  
The edge states are described by $SU(3)_2\times U(1)$ Kac-Moody algebra. 
}
\label{N4p21/5}
\centering
\begin{tabular}{ |c|cccc|}
 \hline 
$s_i$ & 0 & $-\frac{1}{4}$ & $ \frac{7}{20}$ & $-\frac{2}{5}$ \\
$d_i$ & 1 & $1$ & $\zeta_{3}^{1}$ & $\zeta_{3}^{1}$ \\
\hline
$j\backslash i$ & $\one$ & $a$ & $\si$ & $\tau$ \\
\hline 
$\one$  & $\one$  & $a$  & $\si$  & $\tau$  \\
$a$  & $a$  & $\one$  & $\tau$  & $\si$  \\
$\si$  & $\si$  & $\tau$  & $\one \oplus \tau$  & $a \oplus \si$  \\
$\tau$  & $\tau$  & $\si$  & $a \oplus \si$  & $\one \oplus \tau$  \\
\hline 
 \end{tabular}
\end{table}

\begin{table}[tb]
\caption{The fusion rule $j\otimes i$ for the simplest $Z_3$ parafermion (Fibonacci) topological order
$N^B_c=2^B_\frc{14}{5}$.}
\label{N2p14/5}
\centering
\begin{tabular}{ |c|cc|}
 \hline 
$s_i$ & 0 & $ \frac{2}{5}$ \\
$d_i$ & 1 & $\zeta_{3}^{1}$ \\
\hline
$j\backslash i$ & $\one$ & $\si$ \\
\hline 
$\one$  & $\one$  & $\si$  \\
$\si$  & $\si$  & $\one \oplus \si$  \\
\hline 
 \end{tabular}
\end{table}

\begin{table}[tb]
\caption{ The fusion rule $j\otimes i$ for topological order $N^B_c={6^B_{-\frc{1}{7}}}\sim
6^B_\frc{55}{7} $.  It can be realized\cite{Wnab,BW9215} by wave function
$\Psi_{6^B_{\frc{55}{7}}} = [\Psi^\text{LL}_5(\{z_i\})]^2$.  The edge states are
described by $SU(5)_2\times U(1)$ Kac-Moody algebra. 
}
\label{N6p55/7}
\centering
\begin{tabular}{ |c|cccccc|}
 \hline 
$s_i$ & 0 & $ \frac{1}{4}$ & $ \frac{1}{7}$ & $ \frac{11}{28}$ & $-\frac{1}{28}$ & $-\frac{2}{7}$ \\
$d_i$ & 1 & $1$ & $\zeta_{5}^{1}$ & $\zeta_{5}^{1}$ & $\zeta_{5}^{2}$ & $\zeta_{5}^{2}$ \\
\hline
$j\backslash i$ & $\one$ & $a$ & $\si$ & $\si'$ & $\tau$ & $\tau'$ \\
 \hline 
$\one$  & $\one$  & $a$  & $\si$  & $\si'$  & $\tau$  & $\tau'$  \\
$a$  & $a$  & $\one$  & $\si'$  & $\si$  & $\tau'$  & $\tau$  \\
$\si$  & $\si$  & $\si'$  & $\one \oplus \tau'$  & $a \oplus \tau$  & $\si' \oplus \tau$  & $\si \oplus \tau'$  \\
$\si'$  & $\si'$  & $\si$  & $a \oplus \tau$  & $\one \oplus \tau'$  & $\si \oplus \tau'$  & $\si' \oplus \tau$  \\
$\tau$  & $\tau$  & $\tau'$  & $\si' \oplus \tau$  & $\si \oplus \tau'$  & $\one \oplus \si \oplus \tau'$  & $a \oplus \si' \oplus \tau$  \\
$\tau'$  & $\tau'$  & $\tau$  & $\si \oplus \tau'$  & $\si' \oplus \tau$  & $a \oplus \si' \oplus \tau$  & $\one \oplus \si \oplus \tau'$  \\
\hline 
 \end{tabular}
\end{table}

Most non-Abelian topological orders that we found are of the
$Z_n$-parafermion\citep{ZF8515} type.  For such kind of $Z_n$-parafermion-type
non-Abelian topological orders all the quantum dimensions are of the form
$\zeta_n^m$ for a set of $m$'s.  (Note that the quantum dimensions can be
$\zeta_n^0=1$.) In this section, we will discuss the physical realization of
some of the $Z_n$-parafermion-type non-Abelian topological orders.  
\begin{enumerate} 
\item The $3^B_{\frc{5}{2}}$ topological order in Table \ref{toplst} is of the
$Z_2$-parafermion type (see Table \ref{N3p5/2}).  It can be realized by the
following filling-fraction $\nu=1$ bosonic FQH wave function whose non-Abelian
properties was first revealed in \Ref{Wnab,BW9215} (Feb. 1991): 
\begin{align} 
\Psi_{3^B_{\frc{5}{2}}} = [\Psi^\text{LL}_2(\{z_i\})]^2 ,
\end{align} 
where $\Psi^\text{LL}_n(\{z_i\})$ is the fermionic wave function of $n$ filled
Landau levels.  The $\Psi_{3^B_{\frc{5}{2}}}$ state was shown to be a
non-Abelian FQH state described by $SU(2)_2\times U(1)$ Kac-Moody current
algebra, which is the same as $Z_2$-parafermion $\times U(1)\times U(1)$
non-Abelian FQH state.  \Ref{Wnab,W9139,BW9215} also studied the fermionic
version of the above $Z_2$-parafermion non-Abelian state 
\begin{align}
\Psi_{6_{\frc{5}{2}}} = \Psi^\text{LL}_1(\{z_i\})[\Psi^\text{LL}_2(\{z_i\})]^2 , 
\end{align} 
with rank $N=6$, central charge $c=5/2$, and filling-fraction $\nu=1/2$.  
\item The $3^B_{\frc{3}{2}}$ topological order in Table
\ref{toplst} is also of the $Z_2$-parafermion type.  It can be realized by the
following filling-fraction $\nu=1$ bosonic FQH wave function 
\begin{align}
\Psi_{3^B_{\frc{3}{2}}}=\cA( \frac{1}{z_1-z_2} \frac{1}{z_3-z_3}\cdots ) \prod
(z_i-z_j) \ee^{-\frac14 \sum |z_i|^2} . 
\end{align}
 It is closely related to
the $\nu=1/2$ fermionic Pfaffient state $6_{\frc{3}{2}}$ first proposed in
\Ref{MR9162} (Aug. 1991), which has a rank $N=6$ and a central charge $c=\frac
32$: 
\begin{align} 
\Psi_{6_{\frc{3}{2}}}=\cA( \frac{1}{z_1-z_2}
\frac{1}{z_3-z_3}\cdots ) \prod (z_i-z_j)^2 \ee^{-\frac14 \sum |z_i|^2} .
\end{align} 
The above two $Z_2$ parafermion states (one for bosonic electrons
and one for fermionic electrons) can also be described by \emph{patterns of
zeros} (or 1D occupation patterns)\cite{WW0808,WW0809,BW0932,SL0604,BKW0608,SY0802,BH0802,BH0802a,BH0882} $\{n_l\}=\{n_0,n_1,n_2,\cdots\}$:
\begin{align} 
\Psi_{3^B_{\frc{3}{2}}}:& \{n_l\}=20|20|20|\cdots , 
\nonumber\\
\Psi_{6_{\frc{3}{2}}}:& \{n_l\}=1100|1100|1100|\cdots .  
\end{align} 
\item 
The $4^B_{-\frc{19}{5}}\sim 4^B_{\frc{21}{5}}$ topological order in Table
\ref{toplst} is of the $Z_3$-parafermion (or Fibonacci) type (see Table
\ref{N4p21/5}).  It has the same $Z_3$-parafermion (Fibonacci) non-Abelian type
as the  $2^B_{\frc{14}{5}}$  topological order  (see Table \ref{N2p14/5}).  The
$4^B_{\frc{21}{5}}$ topological order can be realized by the following
filling-fraction $\nu=3/2$ bosonic FQH wave function with non-Abelian
properties:\cite{Wnab,BW9215} 
\begin{align}
\Psi_{4^B_{\frc{21}{5}}} = [\Psi^\text{LL}_3(\{z_i\})]^2 .  
\end{align}
The $\Psi_{4^B_{\frc{21}{5}}}$ state was shown to be a non-Abelian FQH state
whose  edge excitations are described by $SU(3)_2\times U(1)$ Kac-Moody current
algebra with central charge $c=\frac{21}{5}$.\cite{Wnab,W9139,BW9215} Due to
the level-rank duality, the $SU(3)_2$ non-Abelian type is the same as the
$SU(2)_3$ non-Abelian type, which is also the same as the $Z_3$-parafermion
non-Abelian type.  The fermionic version of the above $Z_3$-parafermion
non-Abelian state is given by\cite{Wnab,W9139,BW9215}
\begin{align} 
\Psi =\Psi^\text{LL}_1(\{z_i\})[\Psi^\text{LL}_3(\{z_i\})]^2 , 
\end{align} 
which has rank $N=10$, central charge $c=21/5$, and filling-fraction $\nu=3/5$.
\Ref{BW9215} also constructed/studied those type of non-Abelian FQH states
using parafermion CFTs in 1992.  The non-Abelian excitations from such
non-Abelian FQH states can perform universal topological quantum computations.
\item 
The $4^B_{\frc{9}{5}}$ topological order in Table \ref{toplst} is of the
$Z_3$-parafermion (Fibonacci) type.  It can be realized by the following
filling-fraction $\nu=3/2$ bosonic FQH wave function described by the following
pattern of zeros: $\{n_l\}=\{n_0,n_1,n_2,\cdots\}$:
\begin{align} 
\Psi_{4^B_{\frc{9}{5}}}: \{n_l\}=30|30|30|\cdots , 
\end{align} 
\ie
$n_\text{even}=3$ and $n_\text{odd}=0$.  It is closely related to the $\nu=3/5$
fermionic FQH state constructed using $Z_3$ parafermion CFT in
1998\cite{RR9984} with $N_c=10_\frc{9}{5}$ described by the pattern of zeros:
\begin{align} 
\Psi_{10_\frc{9}{5}}: \{n_l\}=11100|11100|11100|\cdots .
\end{align} 
\item 
The $6^B_{-\frc{1}{7}}\sim 6^B_{\frc{55}{7}}$ topological order in Table
\ref{toplst6} is of the $Z_5$-parafermion type (see Table \ref{N6p55/7}).  It
can be realized by the following filling-fraction $\nu=5/2$ bosonic FQH wave
function which is non-Abelian\cite{Wnab,BW9215}: 
\begin{align}
\Psi_{6^B_{\frc{55}{7}}} = [\Psi^\text{LL}_5(\{z_i\})]^2 .  
\end{align}
The fermionic version of the above
$Z_5$-parafermion non-Abelian state 
is given by\cite{Wnab,W9139,BW9215}
\begin{align} \Psi =
\Psi^\text{LL}_1(\{z_i\})[\Psi^\text{LL}_5(\{z_i\})]^2 , 
\end{align} 
which has rank $N=21$, central charge $c=55/7$,
and filling-fraction $\nu=5/7$.  The above non-Abelian FQH states and their
edge excitations are also described by $SU(5)_2\times U(1)$ Kac-Moody current
algebra.  
\item The $6^B_{\frc{15}{7}}$ topological order in Table \ref{toplst6}
is of the $Z_5$-parafermion type.  It can be realized by the following
filling-fraction $\nu=5/2$ bosonic FQH wave function
$\{n_l\}=\{n_0,n_1,n_2,\cdots\}$: 
\begin{align} 
\Psi_{6^B_{\frc{15}{7}}}: \{n_l\}=50|50|50|\cdots .  
\end{align} It is closely related to the $\nu=5/7$
fermionic $Z_5$-parafermion state\cite{RR9984} with $N_c=21_{\frc{15}{7}}$:
\begin{align} 
\Psi_{21_{\frc{15}{7}}}: \{n_l\}=1111100|1111100|1111100|\cdots .  
\end{align} 
\end{enumerate}

\subsection{Non-Abelian topological orders of $Z_n\times Z_{n'}$-parafermion
type}

Some non-Abelian topological orders that we found are of the $Z_n\times
Z_{n'}$-parafermion type.  For such kind of $Z_n\times Z_{n'}$-parafermion-type
non-Abelian topological orders, all the quantum dimensions are of the form
$\zeta_n^m \zeta_{n'}^{m'}$ for a set of $m,m'$'s.  Some of those topological
orders can be realized by stacking $Z_n$-parafermion  topological order with
$Z_{n'}$-parafermion  topological order.

For example, staking two $Z_3$-parafermion $2^B_{\frc{14}{5}}$ topological order
described by wave function $\Psi_{2^B_{\frc{14}{5}}}$ will give us a third
$Z_3\times Z_3$-parafermion $4^B_{\frc{28}{5}}=4^B_{-\frc{12}{5}}$ topological
order described by wave function 
\begin{align} 
\Psi_{4^B_{\frc{28}{5}}}(\{z_i\},
\{w_i\}) =\Psi_{2^B_{\frc{14}{5}}}(\{z_i\}) \Psi_{2^B_{\frc{14}{5}}}( \{w_i\}) .
\end{align}


Similarly, staking $Z_3$-parafermion $2^B_{\frc{14}{5}}$ topological order and
$Z_2$-parafermion $3^B_{\frc{1}{2}}$ topological order together produce a third
$Z_3\times Z_2$-parafermion $6^B_{\frc{33}{10}}$ topological order in the Table
\ref{toplst6}, which is described by
wave function 
\begin{align} 
\Psi_{6^B_{\frc{33}{10}}}(\{z_i\}, \{w_i\})
=\Psi_{2^B_{\frc{14}{5}}}(\{z_i\}) \Psi_{3^B_{\frc{1}{2}}}( \{w_i\}) .
\end{align} 

We may identify $z_i$ and $w_i$ in the above wave function, trying to obtain a new topologically ordered state.  If we are
lucky, the new wave function 
\begin{align}
\Psi_{6^B_{\frc{23}{10}}}(\{z_i\}) =\Psi_{2^B_{\frc{14}{5}}}(\{z_i\})
\Psi_{3^B_{\frc{1}{2}}}( \{z_i\}) .  
\end{align} 
will describe a gapped state, which will be a topological order with one less
central charge (for details, see \Ref{W9927}), \ie a $Z_3\times
Z_2$-parafermion $6^B_{\frc{23}{10}}$ topological order.  The
$6^B_{\frc{23}{10}}$ topological order does appear in our table \ref{toplst6},
which implies that identifying $z_i$ and $w_i$ will give us the $Z_3\times
Z_2$-parafermion topological order $6^B_{\frc{23}{10}}$.

\subsection{2+1D time-reversal symmetric topological orders}

We have found 6 topological orders with $c=0$ and $N\leq 7$: three $N^B_c=4^B_0$,
one $N^B_c=5^B_0$, and two $N^B_c=6^B_0$.  The spin spectrum has the $\{s_i\} \to
\{-s_i\}$ symmetry for all those topological orders.  It suggests that those
topological orders can be realized by time reversal symmetric systems.  In
contrast,  the spin spectrum does not have the $\{s_i\} \to \{-s_i\}$ symmetry
for most topological orders, suggesting that they cannot be realized by time
reversal symmetric systems.

\subsection{2+1D anomalous time-reversal symmetric topological orders}

We have found 4 topological orders with $c=4$ and $N\leq 7$: one $N^B_c=4^B_4$, one
$N^B_c=5^B_4$, and two $N^B_c=6^B_4$.  The spin spectrum has the $\{s_i\} \to \{-s_i\}$
symmetry for all those topological orders.  However, since $c\neq 0$ implies a
chiral edge state, those topological orders cannot be realized by time reversal
symmetric systems.  It was suggested in \Ref{VS1258,WS1334}, that the $N^B_c=4^B_4$
topological order can be realized as the time-reversal symmetric surface states
of a 3+1D time reversal symmetric symmetry-protected topological state.  We
believe all those topological orders can be realized as the time-reversal
symmetric surface states of the same 3+1D time reversal symmetric
symmetry-protected topological state.  In other words, those topological orders
have anomalous time-reversal symmetries, which have the same type of
anomaly.\cite{W1313}

\subsection{2+1D fermionic topological orders}

Although we have only discussed bosonic topological orders in this paper, we
can see fermionic topological orders\cite{GWW1017,LW150704673} from our classification of bosonic
topological orders. Let us illustrate this point through an example.

We start with the $N^B_c=4^B_0, s_i=(0,0,0,\frac12)$ topological order (\ie the
$Z_2$ topological order\cite{RS9173,Wsrvb,MS0181}) in Table \ref{toplst}.  We
know that the $Z_2$ topological order contain a fermionic excitation $f$.  If
we add the  fermionic excitations to the ground state and let the fermions to
form a product state, such an addition will not change the $Z_2$ topological
order.  However, if we let the fermions to form a $p+\ii p$ superconducting
state, then the $Z_2$ topological order will change to a different topological
order.  Since the $p+\ii p$ superconducting state has $c=1/2$ edge state, the
new topological order should also has $c=1/2$. This suggests that fermion
condensation into the $p+\ii p$ state will change the $N^B_c=4^B_0$ $Z_2$
topological order to the $N^B_c=3^B_\frc 12$ topological order in Table
\ref{toplst}.  We note that the $Z_2$-charge and the $Z_2$-vortex both behave
like the same $\pi$-flux to the fermion $f$.  In the $p+\ii p$ state, $\pi$-flux
will carry an Majorana zero mode and behave like a topological excitations of
quantum dimension $\sqrt 2=\zeta^1_2$.  Such a kind of topological excitations
appear in the $N^B_c=3^B_\frc 12$ state, confirming our identification.

Similarly, if we let the fermions to form $2n+1$ layers of  $p+\ii p$
superconducting states, then the $Z_2$ topological order will change to the
$N^B_c=3^B_\frc {2n+1} 2$ topological order in Table \ref{toplst}.  This is
because $2n+1$ layers of $p+\ii p$ states have chiral central charge
$c=(2n+1)/2$ edge state.  Also, if we let the fermions to form $2n$ layers of
$p+\ii p$ superconducting states (\ie a $\nu=n$ integer quantum Hall state),
then the $Z_2$ topological order will change to the $N^B_c=4^B_n$ topological order
in Table \ref{toplst}.  This is because $2n$ layers of $p+\ii p$ states have
chiral central charge $c=n$ edge state.

We also see that the $N^B_c=6^B_{\pm \frc{(3+10n)}{10}}$ states are related by the
fermion condensation into $\nu=\Del n$ integer quantum Hall state.
The $N^B_c=7^B_{\pm \frc{(1+2n)}{4}}$ states are related by the
fermion condensation into $\Del n$ layers of $p+\ii p$ states.

\section{A classification of 1+1D gravitational anomalies}

Since the 1+1D bosonic gravitational anomalies (both perturbative and global
gravitational anomalies of known or unknown types) are classified by the 2+1D
bosonic topological orders, $(S,T,c)$ or $(N^{ij}_k,s_i,c)$ give us a
classification of all 1+1D bosonic gravitational anomalies.  We may also view
the tables \ref{toplst}, \ref{toplst6}, and \ref{toplst7} as tables of simple
bosonic gravitational anomalies.  When $c\neq 0$, the 1+1D bosonic
gravitational anomaly contain perturbative gravitational anomaly. When $c=0$,
the 1+1D bosonic gravitational anomaly is a pure global gravitational anomaly.

Given a 1+1D low energy effective theory $\cL_\text{1+1D}$, how do we know if
the theory has gravitational anomaly or not?  According to \Ref{W1313,KW1458},
we first try to realize $\cL_\text{1+1D}$ by the edge of 2+1D gapped liquid
system described by $\cL_\text{2+1D}$.  We then use the non-Abelian geometric
phase\cite{Wrig} or wave function overlap\cite{MW1418} to compute $S,T$.  From
$(S,T,c)$, we learn the type of the  gravitational anomaly in the 1+1D theory
$\cL_\text{1+1D}$.

As an example, let us consider the following 1+1D bosonic system
\begin{align}
 \cL_\text{1+1D} = \frac{K_{IJ}}{4\pi} \prt_x \phi_I \prt_t \phi_J
-\frac{1}{4\pi} V_{IJ}  \prt_x \phi_I \prt_x \phi_J ,
\end{align}
where $\phi_I$ are compact real fields ($\phi_I\sim \phi_I+2\pi$).  Such 1+1D
effective theory can be realized by the edge of 2+1D $K$-matrix FQH
state.\cite{Wedgerev,Wtoprev}  We find that the 1+1D effective theory
$\cL_\text{1+1D}$ is anomaly free if det$(K)=\pm 1$ and $K$ has an equal number
of positive and negative eigenvalues.

If  $K$ has different numbers of positive and negative eigenvalues, then the
above 1+1D bosonic theory will have a perturbative gravitational anomaly.  If
$K=\bpm 2 & 0\\ 0 &-2 \\\epm$, $K=\bpm 2 & 3\\ 3 &2 \\\epm$ \etc, then the
above 1+1D bosonic theory will only have a global gravitational anomaly.

\section{Summary}

In this paper, we review the discovery and development of topological order --
a new kind of order  beyond Landau symmetry breaking theory in many-body
systems.  We stress that topological order can be defined/probed by measurable
quantities $(S,T,c)$ or $(N^{ij}_k, s_i, c)$.  

We know that symmetry breaking orders can be described and classified by group
theory. Using  group theory, we can obtain a list of symmetry breaking orders,
such as the 230 crystal orders in three dimensions.  Similarly, in this paper,
we present a simple theory of 2+1D bosonic topological order based on $(S,T,c)$
or $(N^{ij}_k, s_i, c)$.  This allows us to obtain a list of simple 2+1D
bosonic topological orders.  Although it is not clear if the theory presented
in this paper is a complete theory for topological order or not, it serves as
the first step in developing such a theory.  

We also discussed how to realize the some of the topological orders in the list
by concrete many-body wave functions. A more systematic way to realize those
topological orders is via simple current CFT, which will appear elsewhere.

From a mathematical point of view, we assumed that a unitary modular tensor
category (UMTC) can be uniquely characterized by its $(S,T)$ matrices. Under
this assumption, we found that there are 10 and only 10 rank-5 UMTC's with $D^2
\leq 120$ (see Table \ref{toplst2-6}).  It is very likely that there are only
10 rank-5 UMTC's.  We also found that there are 50 and only 50 rank-6 UMTC's
with $D^2 \leq 101$ (see Table \ref{toplst6}).  Most of those UTMC's are
stacking of rand-2 and rand-3 UMTC's.  Table \ref{toplst2-6} lists all 10
primitive rank-6 UMTC's with $D^2 \leq 101$.  Also Table \ref{toplst7-9} lists
all rank-7 UMTC's with $D^2 \leq 40$ (plus a few more with higher $D^2$).

This research is supported by NSF Grant No.  DMR-1005541, and NSFC 11274192. It
is also supported by the John Templeton Foundation No. 39901. Research at
Perimeter Institute is supported by the Government of Canada through Industry
Canada and by the Province of Ontario through the Ministry of Research.

\appendix

\section{A fusion category theory for the amplitudes of planar string
configurations}

In Section \ref{smptop}, we simply list many conditions on $(S,T,c)$ or
$(N^{ij}_k,s_i,c)$. We did not explain where do they come from, although they
are derived in various mathematical literature (for a review, see
\Ref{Wang10}). In the next a few sections, we will try to explain and
understand some of those conditions, in a simple and self-contained way.

\subsection{The string operators}

Although a topological excitation cannot be created alone, a pair of particle
and anti-particle $i,\bar i$ can be created by an open string operator $W_i$.
In some cases, the open string operator $W_i$ is a product of local operators
along the string
\begin{align}
 W_i = \prod_{i \in \text{string}} M_i(x_i).
\end{align} But more generally, the open string operator $W_i$ has a more
complicated structure.  We need to use local operators with two ``bond''
indices, $M_i^{ab}(x_i)$ to construct it:\cite{LWstrnet}
\begin{align}
 W_i =
\sum_{a_1a_2a_3\cdots} M^{aa_1}_i(i_1) M^{a_1a_2}_i(i_2)
M^{a_2a_3}_i(i_3)\cdots
\end{align}
 We see that the bond indices are traced
over and the above string operator is a \emph{matrix-product operator}.

We may choose the local operator $M_i^{ab}(x_i)$ properly such that the normal
of $W_i |\text{ground}\>$ does not depend on the length of the string operator,
and furthermore $W_i |\text{ground}\> \propto |\text{ground}\>$.  Such a string
operator obeys the so called ``zero law'' as described in \Ref{HW0541}.
\Ref{HW0541} pointed out that it is always possible to obtain such ``zero law''
string operator for each type of topological excitation.

Using the ``zero law'' closed string operator, we can have another way to
understand the simple type and composite type.  If $i$ is of a composite type,
$i=j\oplus k\oplus \cdots$, then the corresponding ``zero law'' closed string
operator can be decomposed into a sum of ``zero law'' closed string operators:
\begin{align} W_i = W_j+ W_k+\cdots. 
\end{align}
 If a ``zero law'' closed
string operator cannot be decomposed, then the corresponding particle is of a
simple type.  The correspondence between the composite type and the sum of the
string operators, as well as the correspondence between the fusion of
topological excitations and the product of string operators, allow us to see
that the ``zero law'' closed string operators for simple types satisfy an
algebra described by the fusion coefficients $N^{ij}_k$
\begin{align}
\label{WWNW} W_i W_j=\sum_k N^{ij}_k W_k .  
\end{align}
 (We will derive this
relation later in Section \ref{strFA}.)

\begin{figure}[tb] 
\centerline{ \includegraphics[scale=0.5]{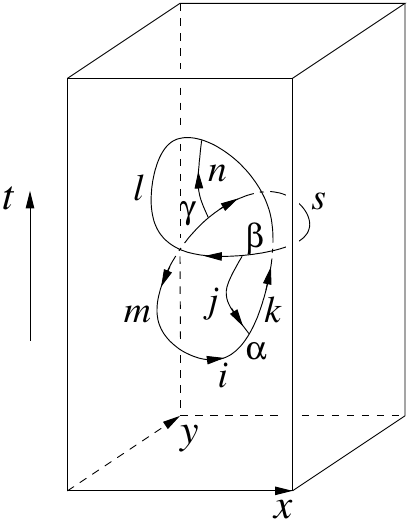} } 
\caption{
World-lines in a local region represent a local  tunneling process, where the
topological excitations are created in pairs, and then braided and fused, and
at last annihilated in pairs.  The picture is also a 2D projection of a 3D
string configuration.  } 
\label{qptung} 
\end{figure}

\subsection{The space-time world lines and quasiparticle tunneling process}

In the space-time path integral picture, a ``zero law'' string operator
correspond to a string in a time slice of a fixed time.  We can consider more
general ``zero law'' strings in space-time that can go through different times.
Those more general strings in space-time correspond to the world-lines of the
topological excitations.  If the all the world-lines are confined in a local
region, then they will represent a local  tunneling process, where the
topological excitations are created in pairs, and then braided and fused, and
at last annihilated in pairs (see Fig. \ref{qptung}).  Since the degenerate
ground state are locally indistinguishable, such a local tunneling process
causes the same amplitude for different degenerate ground states.  Thus
world-line confined in a local region correspond the a complex number (the
amplitude) in the space-time path integral picture.  Since the world-line
correspond to ``zero law'' strings, the above complex number (the amplitude)
does not depend on the shape and length of the world-line.  It only depend on
the linking and the fusion of the world lines.

\begin{figure}[tb] 
\centerline{ \includegraphics[scale=0.5]{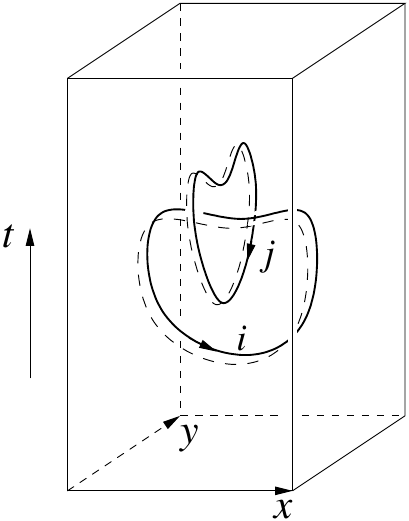} } 
\caption{
A local tunneling process of two linked world-lines.  The dash lines are the
framing of the world-lines.  In such a tunneling process the framing dash lines
do not link with the original strings, indicating that the particles do not
twist (or rotate) during the tunneling process.  The amplitude of the above
linked loops is a complex number denoted as $S^\text{Lnk}_{ij}$.  }
\label{lnklpr} 
\end{figure}

Clearly, number of types of the strings is given by the rank $N$, and the
strings are labeled by the simple type $i$ of the topological excitations.
Also, the strings are oriented if the particle $i$ and anti-particle $\bar i$
are different (see  Fig. \ref{lnklpr}).

The fusion of particles is represented by the branching point of the strings.
If $N^{ij}_k=0$, then the amplitude of the string configuration that contain a
branching point of $i,j,k$ strings will be zero.  When $N^{ij}_k>1$, it means that
the space $i\otimes j$ contain several copies of the space $k$. We will label
the  copies of the space $k$ by $\al=1,2,\cdots,N^{ij}_k$.  There will be a
tunneling amplitude into each copy of the space $k$, and the tunneling
amplitude will depend on $\al$.  So we will include the index $\al\in
[1,\cdots,N^{ij}_k]$ on each branching point.  In this case, each such labeled
graph of strings gives rise to an amplitude.  We will use $A(X)$ to represent
such an amplitude for a labeled string configuration $X$, such as the one in
Fig. \ref{qptung}.

We also like to mention that the world-lines have a finite cross section which
is not circular. So more precisely, the world-lines are represented by framed
strings in \ref{lnklpr}.  The framing represents the  finite cross section,
which do not have rotation symmetry.

\subsection{Planar string configurations} 

In this paper, we will mostly draw the string in 3-dimensional space-time in
terms of their projection on a particular plane.  In the 2D projected
representation, we will always choose a canonical framing, by displacing all
the 2D strings a little bit in a direction perpendicular to the 2D plane to
obtain the framing dash-lines.  So when we draw such 2D string configurations,
we will assume the above canonical framing and will not draw the framing
dash-lines.  

In the rest of this section, we will consider all the amplitudes for planar
string configurations, where the strings in the 2D projection do not cross each
other.  It turns out that the amplitudes for different planar string
configurations have a lot of relations, so that we can determine the amplitudes
for all the planar string configurations from a set a tensors that satisfy a
certain relations.  This turns out to be a fusion category theory of the
amplitudes for planar string configurations. \Ref{GWW1017} presented such a
fusion category theory for more general fermionic case.  Here, we will present
the simplified case for bosons.

Since the planar string configurations can be viewed as the world line of
particles in 1+1D space-time, the fusion category theory described here can
also be viewed as the classifying theory for anomalous 1+1D topological
orders,\cite{LW1384,KW1458,FV14095723} which can be described by the particles
tunneling process in 1+1D space-time.

\subsection{The first type of linear relations: the F-move} 

Let us consider a local region in the 2D projected string configuration. We fix
all strings cutting across the boundary of the region, and consider all the
different ways that the strings connect to each other in the region.  Those
different string configurations describe different local tunneling process.  If
a subset of string configurations already describe all the channel of the
tunneling processes, then the amplitude of every other local string
configuration can be expressed as a linear combination of the amplitudes for
the  subset of string configurations.

In fact, the graph $\bmm \includegraphics[scale=.40]{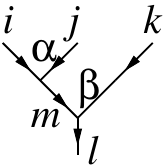} \emm$ with fixed
$ijkl$ but different $m\al\bt$ is a subset of string configurations that
describe all the channel of the local tunneling processes with fixed $ijkl$.
The graph $\bmm \includegraphics[scale=.40]{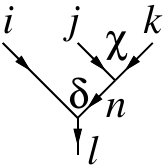} \emm$ with fixed $ijkl$ but
different $n\chi\del$ is another subset of string configurations that describe
all the channel of the local tunneling processes with fixed $ijkl$.  So we can
express the amplitudes for string configurations in one subset in terms of the
amplitudes for string configurations in another subset: 
\begin{align}
\label{IHwave} 
A \bpm \includegraphics[scale=.40]{F1g} \epm = \sum_{n\chi\del}
F^{ijm,i\bt}_{kln,\chi\del} A \bpm \includegraphics[scale=.40]{F2g} \epm .
\end{align} 
We note that
\begin{align}
 \label{Feq0} &
F^{ijm,\al\bt}_{kln,\chi\del} = 0 \text{ when} \\
 & N_{ij}^{m}<1 \text{ or }
N_{mk}^{l}<1 \text{ or } N_{jk}^{n}<1 \text{ or } N_{in}^{l}<1 . 
\nonumber
\end{align} When $N^{ij}_{m}<1$ or $ N^{mk}_{l}<1$, the left-hand-side of
\eqn{IHwave} is always zero.  Thus $F^{ijm,\al\bt}_{kln,\chi\del} = 0$ when
$N^{ij}_{m}<1$ or $ N^{mk}_{l}<1$.  When $N^{jk}_{n}<1$ or  $N^{in}_{l}<1$,
amplitude on the right-hand-side of \eqn{IHwave} is always zero.  So we can
choose $F^{ijm,\al\bt}_{kln,\chi\del} = 0$ when $N^{jk}_{n}<1$ or
$N^{in}_{l}<1$.

For fixed $i$, $j$, $k$, and $l$, the matrix $F^{ij}_{kl}$ with matrix elements
$(F^{ij}_{kl})^{m,\al\bt}_{n,\chi\del} = F^{ijm,\al\bt}_{kln,\chi\del} $ is a
matrix of dimension $\sum_m N^{ij}_m N^{mk}_l \times \sum_n N^{in}_l N^{jk}_n$.
The  matrix describe the relation of the tunneling amplitude through one set of
channels described by basis $m\al\bt$ and through another set of channels
described by basis $n\chi\del$.  We note that  the tunneling maps $i,j,k$ to
$l$ with degeneracy.  The first tunneling path gives rise to basis $m\al\bt$ of
the degenerate subspace.  The second tunneling path gives rise to basis
$n\chi\del$ of the degenerate subspace.
The degenerate subspace of $l$ should to the same, regardless the
tunneling paths.
So we require $N^{ij}_k$ to satisfy
\begin{align} 
\sum_m N^{ij}_m N^{mk}_l = \sum_n N^{in}_l N^{jk}_n 
\end{align}
and the matrix $F^{ij}_{kl}$ to be unitary:  
\begin{align} 
\label{2FFstar}
\sum_{n\chi\del} F^{ijm',\al'\bt'}_{kln,\chi\del}
(F^{ijm,\al\bt}_{kln,\chi\del})^* =\del_{m,m'}\del_{\al,\al'}\del_{\bt,\bt'}.
\end{align} 
(But here we do not require $N^{ij}_k=N^{ji}_k$).  It is easy to
see that the unitary condition implies: 
\begin{align} 
\label{inverseF} 
A \bpm
\includegraphics[scale=.40]{F2g} \epm = \sum_{m\al\bt} \left(
F^{ijm,\al\bt}_{kln,\chi\del}\right)^\dagger A \bpm
\includegraphics[scale=.40]{F1g} \epm .  
\end{align}

Similarly, we have a dual F-move
\begin{align}
 \label{dFmv} A \bpm
\includegraphics[scale=.40]{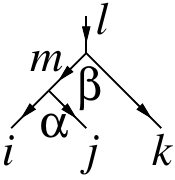} \epm = \sum_{n\chi\del} \t
F^{ijm,i\bt}_{kln,\chi\del} A \bpm \includegraphics[scale=.40]{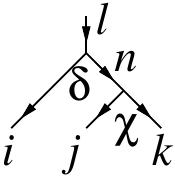} \epm ,
\end{align} where $\t  F^{ijm,i\bt}_{kln,\chi\del}$ also satisfies a unitary
condition.

The F-move \eq{IHwave} can be viewed as a relationship between amplitudes for
different graphs that are only differ by a local transformation.  Since we can
transform one graph to another graph through different paths (\ie different sets of local F-moves), the
F-move \eq{IHwave} must satisfy certain self consistent conditions.  For
example the graph $ \bmm \includegraphics[scale=.35]{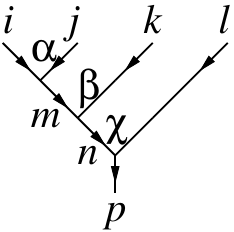} \emm $ can be
transformed to $ \bmm \includegraphics[scale=.35]{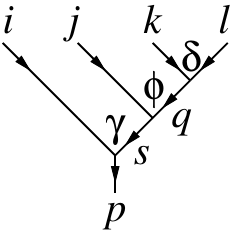} \emm $ through two
different paths; one contains two steps of F-moves and another
contains three steps of F-moves as described by \eqn{IHwave}.
The two paths lead to the following relations between the wave functions:
\begin{widetext} 
\begin{align} 
\label{FFFrelG} 
A \bpm
\includegraphics[scale=.40]{pent1g} \epm & = \sum_{t\eta\vphi}
F^{ijm,\al\bt}_{knt,\eta\vphi} A \bpm \includegraphics[scale=.40]{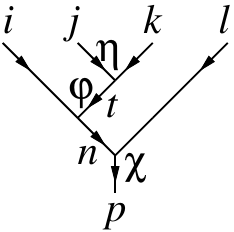} \epm
= \sum_{t\eta\vphi;s\ka\ga} F^{ijm,\al\bt}_{knt,\eta\vphi}
F^{itn,\vphi\chi}_{lps,\ka\ga} A \bpm \includegraphics[scale=.40]{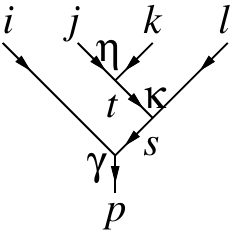} \epm
\nonumber\\
 & = \sum_{t\eta\ka;\vphi;s\ka\ga;q\del\phi}
F^{ijm,\al\bt}_{knt,\eta\vphi} F^{itn,\vphi\chi}_{lps,\ka\ga}
F^{jkt,\eta\ka}_{lsq,\del\phi} A \bpm \includegraphics[scale=.40]{pent3g} \epm
.  
\end{align}
\begin{align} 
\label{FFrelG} 
A \bpm \includegraphics[scale=.40]{pent1g} \epm &
= \sum_{q\del\eps} F^{mkn,\bt\chi}_{lpq,\del\eps} A \bpm
\includegraphics[scale=.40]{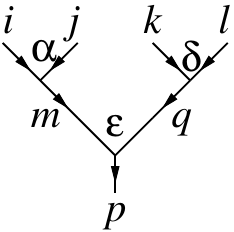} \epm = \sum_{q\del\eps;s\phi\ga}
F^{mkn,\bt\chi}_{lpq,\del\eps} F^{ijm,\al\eps}_{qps,\phi\ga} A \bpm
\includegraphics[scale=.40]{pent3g} \epm , 
\end{align}
\end{widetext} The consistence of the above two relations leads a condition on
the $F$-tensor:
\begin{align}
 \label{penid} &\ \ \ \ \sum_{t}
\sum_{\eta=1}^{N^{jk}_{t}} \sum_{\vphi=1}^{N^{it}_{n}}
\sum_{\ka=1}^{N^{tl}_{s}} F^{ijm,\al\bt}_{knt,\eta\vphi}
F^{itn,\vphi\chi}_{lps,\ka\ga} F^{jkt,\eta\ka}_{lsq,\del\phi}
\nonumber\\
 &  =
\sum_{\eps=1}^{N^{mq}_{p}} F^{mkn,\bt\chi}_{lpq,\del\eps}
F^{ijm,\al\eps}_{qps,\phi\ga} . 
\end{align}
 which is the famous pentagon
identity.  The above pentagon identity \eq{penid} is a set of nonlinear
equations satisfied by the rank-10 tensor $F^{ijm,\al\bt}_{kln,\chi\del}$.  The
above consistency relations \eq{penid} are equivalent to the requirement that
the local unitary transformations described by \eqn{IHwave} on different paths
all commute with each other.

\subsection{The second type of linear relations: the O-move}

The second type of linear relations re-express the amplitude for $\bmm
\includegraphics[scale=.35]{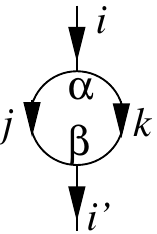} \emm$ in terms of the amplitude for $\bmm
\includegraphics[scale=.35]{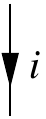} \emm$:
\begin{align}
 \label{PhiO} A \bpm
\includegraphics[scale=.40]{iOip} \epm =  O^{jk,\al\bt}_i \del_{ii'} A \bpm
\includegraphics[scale=.40]{iline} \epm  . 
\end{align}
 We call such local
change of graph an O-move.  Here $O_{i}^{jk,\al\bt} $ satisfies
\begin{align}
\label{Onorm} \sum_{k,j} \sum_{\al=1}^{N^{jk}_i} \sum_{\bt=1}^{N^{jk}_i}
O^{jk,\al\bt}_i (O^{jk,\al\bt}_i)^*=1 
\end{align} 
and
\begin{align}
 \label{Oz}
O^{jk,\al\bt}_i=0 \text{ if }& N^{jk}_i<1 . 
\end{align}
 We note that the
number of choices for the four indices $(j,k,\al,\bt)$ in $O_{i}^{jk,\al\bt}$
must be equal or greater than $1$:
\begin{align}
 \label{Di1} D_i=\sum_{jk}
(N^{jk}_i)^2 \geq 1 . 
\end{align}

\subsection{The third type of linear relations: the Y-move}

For fixed $i,j$,  $\bmm \includegraphics[scale=.35]{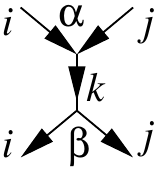} \emm$ for different
$k\al\bt$ describe all the possible tunneling channels.  So the amplitude for
$\bmm \includegraphics[scale=.35]{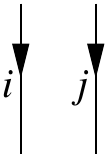} \emm$ can be expressed in terms of the
amplitudes of  $\bmm \includegraphics[scale=.35]{ijkY} \emm$:
\begin{align}
\label{PhiY} \sum_{k,\al\bt} Y^{ij}_{k,\al\bt} A \bpm
\includegraphics[scale=.40]{ijkY} \epm = A \bpm \includegraphics[scale=.40]{ij}
\epm
\end{align}
 We will call such a local change as a Y-move.  We can choose
\begin{align} \label{Yz} Y^{ij}_{k,\al\bt}=0, \text{ if }& N^{ij}_k<1.
\end{align}

\subsection{A relation between $O_{i}^{jk,{\al\bt}}$ and $Y_{k}^{ij,{\al\bt}}$}

We find that the following tunneling amplitude has two ways of reduction:
\begin{align} 
\label{iOiOOi} 
\sum_{\bt\ga}Y_{i,{\bt\ga}}^{jk} A \bpm
\includegraphics[scale=.40]{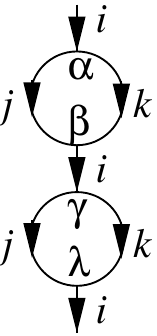} \epm &  = A \bpm
\includegraphics[scale=.40]{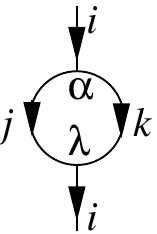} \epm
\nonumber \\
 &  = O_{i}^{jk,{\al\la}} A
\bpm \includegraphics[scale=.40]{iline} \epm ,
\end{align}
\begin{align} &\ \ \ \ \sum_{\bt\ga}Y_{i,{\bt\ga}}^{jk} A \bpm
\includegraphics[scale=.40]{iOOi} \epm
\nonumber\\
 &  =
\sum_{\bt\ga}Y_{i,{\bt\ga}}^{jk} O_{i}^{jk,{\ga\la}} A \bpm
\includegraphics[scale=.40]{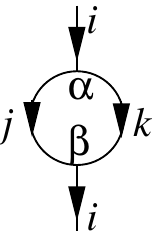} \epm
\nonumber \\
 &  =
\sum_{\bt\ga}Y_{i,{\bt\ga}}^{jk} O_{i}^{jk,{\ga\la}} O_{i}^{jk,{\al\bt}} A \bpm
\includegraphics[scale=.40]{iline} \epm
\end{align}
 The two reductions should
agree, which leads to the condition
\begin{align}
 \label{YO1}
O_{i}^{jk,{\al\la}} = \sum_{\bt\ga}Y_{i,{\bt\ga}}^{jk} O_{i}^{jk,{\ga\la}}
O_{i}^{jk,{\al\bt}}
\end{align}

\subsection{A freedom of changing basis at each vertex}

We note that the following transformation changes the basis at the branching
point labeled by $\al$
\begin{align}
 A \bpm \includegraphics[scale=.40]{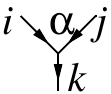}
\epm \to \sum_\bt f^{ij,\al}_{k,\bt} A \bpm \includegraphics[scale=.40]{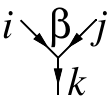}
\epm ,
\end{align}
 where $f^{ij}_k$ is a unitary matrix
\begin{align}
 \sum_\bt
f^{ij,\al}_{k,\bt} (f^{ij,\al'}_{k,\bt})^*=\del_{\al\la'} . 
\end{align}

Similarly, we have  unitary transformation $f^{k,\al}_{ij,\bt}$ for vertices
with one incoming edges and two outgoing edge.  Such transformations
correspond to a choice of basis and should be regarded as an equivalent
relation.

The above transformation induce the following transformation on
$(F^{ijm,\al\bt}_{kln,\ga\la}$,$O^{jk,\al\bt}_i$,$Y_{k,\al\bt}^{ij})$:
\begin{align} 
\label{ftrans} 
O^{jk,\al\bt}_{i}&\to f^{i,\al}_{jk,\al'}
f^{jk,\bt}_{i,\bt'} O^{jk,\al'\bt'}_{i} ,
\\
 Y^{ij}_{k,\al\bt}&\to
(f^{ij,\al'}_{k,\al})^* (f^{k,\bt'}_{ij,\bt})^* Y^{ij}_{k,\al'\bt'} ,
\nonumber\\
 F^{ijm,\al\bt}_{kln,\chi\del} &\to f^{ij,\al}_{m,\al'}
f^{mk,\bt}_{l,\bt'} (f^{jk,\chi'}_{n,\chi})^* (f^{in,\del'}_{l,\del})^*
F^{ijm,\al'\bt'}_{kln,\chi'\del'} . 
\nonumber 
\end{align}
We note that the first line of the above equation is a singular value
decomposition, since $f^{i,\al}_{jk,\al'}$ and $f^{jk,\bt}_{i,\bt'} $, for
fixed $i,j,k$, are independent unitary matrices.  Thus, we can use the above
basis-changing freedom to choose
\begin{align}
 O^{jk,\al\bt}_{i} =
O^{jk,\al}_{i} \del_{\al\bt}, \ \ \ \ O^{jk,\al}_{i} \geq 0.\label{gauge}
\end{align} 
We see that $O^{jk,\al}_{i}$, as the singular values, can be chosen to be
positive real numbers.
Then \eqn{YO1} implies that 
\begin{align}
 \label{YO}
Y^{ij}_{k,\al\bt}=Y^{ij}_{k,\al} \del_{\al\bt},\ \ \ \ 
Y^{ij}_{k,\al} = 1/O^{ij,\al}_k. 
\end{align}

\subsection{A relation between $O_{i}^{jk,\al}$ and
$F^{ijm,\al\bt}_{kln,\del\chi}$}

We also find another graph that can have two ways of reduction as well:
\begin{align} &A \bpm \includegraphics[scale=.45]{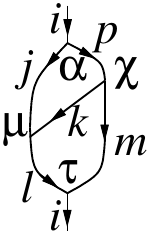} \epm = \sum_{s}
F^{jkl,\mu\tau}_{mis,\chi\al} A \bpm \includegraphics[scale=.45]{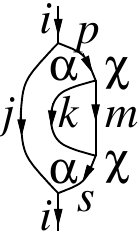} \epm
\nonumber\\
 & = F^{jkl,\mu\tau}_{mip,\chi\al} O_{p}^{km,\chi} O_{i}^{jp,\al} A
\bpm \includegraphics[scale=.45]{iline} \epm
\end{align}

\begin{align} A \bpm \includegraphics[scale=.45]{iQQi3} \epm & = \t
F^{jkl,\mu\tau}_{mip,\chi\al} A \bpm \includegraphics[scale=.45]{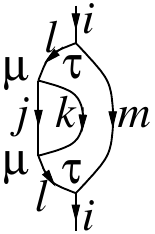} \epm
\nonumber\\
 & = \t F^{jkl,\mu\tau}_{mip,\chi\al} O_l^{jk,\mu} O_i^{lm,\tau} A
\bpm \includegraphics[scale=.45]{iline} \epm
\end{align}
 This allows us to
obtain another condition
\begin{align}
 \t F^{jkl,\mu\tau}_{mip,\chi\al} =
F^{jkl,\mu\tau}_{mip,\chi\al} O_{p}^{km,\chi}
O_{i}^{jp,\al}(O_i^{lm,\tau})^{-1}(O_l^{jk,\mu})^{-1}
\end{align}
 We require
$\t F^{jkl,\mu\tau}_{mip,\chi\al}$ to be unitary, which leads to
\begin{align}
&\ \ \ \sum_{l\mu\tau} (F^{jkl,\mu\tau}_{mip',\chi'\al'})^*
\frac{O_{p'}^{km,\chi'} O_{i}^{jp',\al'}}{ O_i^{lm,\tau} O_l^{jk,\mu}}
F^{jkl,\mu\tau}_{mip,\chi\al} \frac{O_{p}^{km,\chi} O_{i}^{jp,\al}}{
O_i^{lm,\tau} O_l^{jk,\mu} }
\nonumber\\
 &=\sum_{l\mu\tau}
\left(F^{jkl,\mu\tau}_{mip',\chi'\al'}\right)^* F^{jkl,\mu\tau}_{mip,\chi\al}
\frac{O_{p'}^{km,\chi'} O_{i}^{jp',\al'}O_{p}^{km,\chi} O_{i}^{jp,\al} }{
(O_i^{lm,\tau} O_l^{jk,\mu})^2}
\nonumber\\
 &= \del_{pp'} \del_{\chi\chi'}
\del_{\al\al'} ,
\end{align}
 or
\begin{align}
 &\sum_{l\mu\tau} \frac{
(F^{jkl,\mu\tau}_{mip',\chi'\al'})^* F^{jkl,\mu\tau}_{mip,\chi\al} }{
(O_i^{lm,\tau} O_l^{jk,\mu})^2} = \frac{ \del_{pp'} \del_{\chi\chi'}
\del_{\al\al'} }{ (O_{p}^{km,\chi} O_{i}^{jp,\al})^2 } , \label{FOcondition}
\end{align}

The above condition can be satisfied by the following ansatz (note that
$O^{ij,\al}_{k}$ is real and positive)
\begin{align}
O^{ij,\al}_{k}=\sqrt{\frac{w_iw_j}{ w_k}}\del^{ij}_k, \ \ \ 
\ \ \ w_i>0, \label{DefO}
\end{align}
 where $\delta^{jk}_i=1$ for
$N^{jk}_i>0$ and $\delta^{jk}_i=0$ for $N^{jk}_i=0$.  From \eqn{Onorm}, we find
that $w_i$ satisfy
\begin{align}
 \label{Nddd} \sum_{ij} w_iw_j N^{ij}_k =
w_kD^2,\ \ \  D=\sqrt{\sum_l w_l^2}. 
\end{align}
 The solution of such an
equation gives us $w_i$.  

Let us consider the fusion of $n$ type-$i$ particles. The dimension of the
fusion space is
\begin{align}
 D^i(n)=\sum_{k_1,\cdots,k_{n-1}}
N^{ik_{n-2}}_{k_{n-1}} \cdots N^{ik_2}_{k_3} N^{ik_1}_{k_2} N^{ii}_{k_1}
\end{align} Let $d_i$ be the eigenvalue of matrix $N_i$ (defined as
$(N_i)_{kj}=N^{ij}_k$) with largest absolute value.  $d_i$ will called quantum
dimension of type-$i$ particle. Since all the entry of $N^{ij}_k$ are
non-negative, one can show that $d_i$ is real and positive.  We see that the
dimension of the fusion space is given by
\begin{align}
 D^i(n) = O(1) d_i^n .
\end{align} Now, consider $n^2$ type-$i$ particles and $n^2$ type-$j$
particles.  We first fuse $n$ type-$i$ particles, then fuse the result with $n$
type-$j$ particles, and then fuse the result with $n$ type-$i$ particles, \etc.
The dimension of the fusion space is
\begin{align}
 & \sum_{k_1,\cdots,k_{2
n^2-1}} \cdots N^{ik_{3n-2}}_{k_{3n-1}} \cdots  N^{ik_{2n}}_{k_{2n+1}}
N^{ik_{2n-1}}_{k_{2n}}
\nonumber\\
 & \ \ \  \ \ N^{jk_{2n-2}}_{k_{2n-1}} \cdots
N^{jk_n}_{k_{n+1}} N^{jk_{n-1}}_{k_n} N^{ik_{n-2}}_{k_{n-1}} \cdots
N^{ik_1}_{k_2} N^{ii}_{k_1}. 
\end{align}
 The above dimension of the fusion
space should be $O(1) d_i^{n^2} d_j^{n^2}$.  But if the largest-eigenvalue
eigenvectors of $N_i$ and $N_j$ are different, we will get $O(1) d_i^{n^2}
d_j^{n^2} f^n$ as the dimension of the fusion space.  The fact that $f=1$
implies that the largest-eigenvalue eigenvectors of $N_i$ and $N_j$ must be the
same (as implied by the condition \eqn{NN=NN}).  Let $(v_1,v_2, \cdots)$ be the
common largest-eigenvalue eigenvector for $N_i$'s:
\begin{align}
 N_i v  = d_i
v. 
\end{align}
 Since all the entry of $N_i$ are non-negative, one can show
that $v_i$ is real and positive.  Using \eqn{NN=NN}, we find
\begin{align}
 &
\sum_{m,k} N^{ij}_m N^{mk}_l v_k=\sum_{m,k} N^{im}_l N^{jk}_m v_k,
\nonumber\\
& \sum_m N^{ij}_m d_m v_l= d_id_j v_l ,\ \ \ d_id_j = \sum_m N^{ij}_m d_m.
\end{align} We see that $d^T$ is the left eigenvector of $N_i$ with eigenvalue
$d_i$:
\begin{align}
 \label{dNdd} d^T N_i =d^T d_i .  
\end{align} 
In other
words, the left eigenvector of $N_i$ with the largest-eigenvalue is independent
of $i$.  Such a common left eigenvector is given by $d^T$, and the
corresponding largest-eigenvalue is the quantum dimension $d_i$.

%
%

\subsection{The fourth type of linear relations: the H-move }

Let us consider a new type of move -- $H$-move.  First, for fixed $i,j,k,l$
$\bpm \includegraphics[scale=.25]{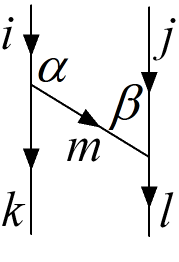} \epm$ and $\bpm
\includegraphics[scale=.25]{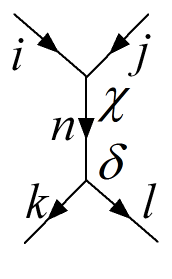} \epm$ describe all the possible tunneling
channels and their can express each other via unitary linear relations:
\begin{align} \label{Hwave1} A \bpm \includegraphics[scale=.25]{H1.png} \epm =
\sum_{n\chi\del} H^{kim,\al\bt}_{jln,\chi\del} A \bpm
\includegraphics[scale=.25]{H2.png} \epm . 
\end{align}
 In the following, we
will show how to compute the coefficients $H^{kim,\al\bt}_{jln,\chi\del}$ from
$F^{ijm,\al\bt}_{kln,\chi\del}$ and $w_i$.

First, by applying the Y-move, we have:
\begin{align}
 \label{Hwave2} A \bpm
\includegraphics[scale=.25]{H1.png} \epm = \sum_{n,\chi^\prime \del}
Y^{kl}_{n,\chi^\prime \del} A \bpm
\includegraphics[scale=.25]{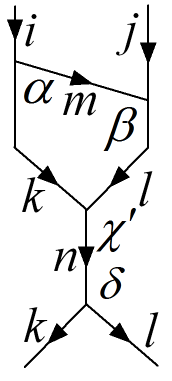} \epm . 
\end{align}
 Next, by applying an
inverse F-move, we obtain:
\begin{align}
 \label{Hwave3} A \bpm
\includegraphics[scale=.25]{H3.png} \epm = \sum_{i^\prime,\beta^\prime\chi}
{\left(F^{kmi^\prime,\bt^\prime \chi}_{jnl,\bt\chi^\prime}\right)}^\dagger A
\bpm \includegraphics[scale=.25]{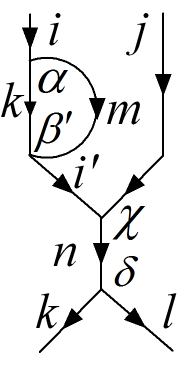} \epm . 
\end{align}
 Finally, by
applying the O-move, we end up with:
\begin{align}
 \label{Hwave4} A \bpm
\includegraphics[scale=.25]{H4.png} \epm =
O^{km,\alpha\beta^\prime}_i\delta_{ii'} A \bpm
\includegraphics[scale=.25]{H2.png} \epm . 
\end{align}
 All together, we find:
\begin{align} H^{kim,\al\bt}_{jln,\chi\del}=\sum_{\chi^\prime
\beta^\prime}Y^{kl}_{n,\chi^\prime \del}{\left(F^{kmi,\bt^\prime
\chi}_{jnl,\bt\chi^\prime}\right)}^\dagger O^{km,\alpha\beta^\prime}_i
\end{align}

Under the proper basis choice eqn. (\ref{gauge}), we can further express the
coefficients $H^{ijm,\al\bt}_{kln,\chi\del}$ as:
\begin{align}
H^{kim,\al\bt}_{jln,\chi\del}&= Y^{kl}_{n,\del}{(F^{kmi,\alpha
\chi}_{jnl,\bt\del})}^* O^{km,\alpha }_i\nonumber\\
 &={(F^{kmi,\alpha
\chi}_{jnl,\bt\del})}^* {(O^{kl,\del}_{n})}^{-1}O^{km,\alpha }_i
\end{align}

The unitarity condition for H-move requires that:
\begin{align}
\sum_{n\chi\del} \frac{F^{km^\prime i,\alpha
\chi}_{jnl,\bt^\prime\del^\prime}{(F^{kmi,\alpha \chi}_{jnl,\bt\del})}^*}
{(O^{kl,\del}_{n})^2}=\frac{\delta_{mm^\prime}\delta_{\al\al^\prime}\delta_{\bt\bt^\prime}}{(O^{km,\alpha}_i)^2},
\end{align} With the special ansatz eqn. (\ref{DefO}), we can further simplify
the above two expressions as:
\begin{align}
 H^{kim,\al\bt}_{jln,\chi\del}=
\sqrt{\frac{w_mw_n}{w_iw_l}} {\left(F^{kmi,\alpha \chi}_{jnl,\bt\del}\right)}^*
\end{align} and
\begin{align}
 \sum_{n\chi\del} w_n F^{km^\prime i,\alpha
\chi}_{jnl,\bt^\prime\del^\prime}{(F^{kmi,\alpha \chi}_{jnl,\bt\del})}^*
=\frac{w_iw_l}{w_m}
\delta_{mm^\prime}\delta_{\al\al^\prime}\delta_{\bt\bt^\prime},
\end{align}

Similarly, we can also construct the dual-H move:
\begin{align}
 A \bpm
\includegraphics[scale=.25]{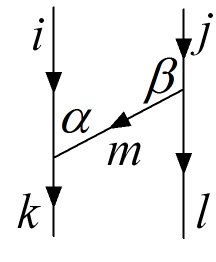} \epm = \sum_{n\chi\del} \t
H^{kim,\al\bt}_{jln,\chi\del} A \bpm \includegraphics[scale=.25]{H2.png} \epm .
\end{align} and we can express $\t H^{kim,\al\bt}_{jln,\chi\del}$ as:
\begin{align} \t H^{kim,\al\bt}_{jln,\chi\del} = \t
H^{kim,\al\bt}_{jln,\chi\del},\label{DefHdual}
\end{align}
 where the
coefficients $\t H^{kim,\al\bt}_{jln,\chi\del}$ can be expressed as:
\begin{align} \t H^{kim,\al\bt}_{jln,\chi\del}&= Y^{kl}_{n,\del}F^{imk,\alpha
\del}_{lnj,\bt\chi} O^{ml,\bt }_j\nonumber\\
 &= F^{imk,\alpha
\chi}_{lnj,\bt\del} {(O^{kl,\del}_{n})}^{-1}O^{ml,\bt }_j
\end{align}
 Again,
with the special ansatz eqn. (\ref{DefO}), we have:
\begin{align}
 \t
H^{kim,\al\bt}_{jln,\chi\del}=\sqrt{\frac{w_mw_n}{w_jw_k}}F^{imk,\alpha
\del}_{lnj,\bt\chi} 
\end{align}
 It is easy to see that the unitarity condition
for dual H-move is automatically satisfied if the H-move is unitary.

\subsection{Summary of the conditions on the linear relations}

We see that valid tunneling amplitudes $A(X)$ can be characterized by tensor
data $(N^{ij}_{k}, F^{ijm,\al\bt}_{kln,\ga\la})$.  However, only  certain
tensor data $(N^{ij}_{k}, F^{ijm,\al\bt}_{kln,\ga\la})$, that satisfy the
conditions eqns.  (\ref{2FFstar}, \ref{Feq0}, \ref{penid}, \ref{Nddd}), can
self-consistently describe valid tunneling amplitudes $A(X)$.  

Those conditions form a set of non-linear equations whose variables are
$N^{ij}_{k}$, $F^{ijm,\al\bt}_{kln,\ga\la}$, $w_i$ (where $w_i$ can be
determined by $N^{ij}_k$ alone). Let us collect those conditions and list them
below
\begin{align}
 \label{Neq} &\bullet\ \sum_{m=0}^N N^{ij}_{m} N^{mk}_{l}
=\sum_{n=0}^N N^{jk}_{n} N^{in}_l
\nonumber\\
 &\bullet\ \sum_{jk} (N^{jk}_i)^2
\geq 1 ;
\nonumber\\
\end{align}
\begin{align} \label{Feq} & \bullet\ \sum_{n\chi\del}
F^{ijm',\al'\bt'}_{kln,\chi\del} (F^{ijm,\al\bt}_{kln,\chi\del})^*
=\del_{m,m'}\del_{\al,\al'}\del_{\bt,\bt'},
\nonumber\\
 &\bullet\
F^{ijm,\al\bt}_{kln,\chi\del} = 0 \text{ when}
\nonumber \\
 & \ \ \ \
N^{ij}_{m}<1 \text{ or } N^{mk}_{l}<1 \text{ or } N^{jk}_{n}<1 \text{ or }
N^{in}_{l}<1 ,
\nonumber\\
 &\bullet\ \sum_{t} \sum_{\eta=1}^{N^{jk}_{t}}
\sum_{\vphi=1}^{N^{it}_{n}} \sum_{\ka=1}^{N^{tl}_{s}}
F^{ijm,\al\bt}_{knt,\eta\vphi} F^{itn,\vphi\chi}_{lps,\ka\ga}
F^{jkt,\eta\ka}_{lsq,\del\phi}
\nonumber\\
 &= \sum_{\eps=1}^{N^{mq}_{p}}
F^{mkn,\bt\chi}_{lpq,\del\eps} F^{ijm,\al\eps}_{qps,\phi\ga} . 
\end{align}
\begin{align} \label{Oeq} \bullet\  \sum_{i,j} w_iw_j N^{ij}_k =w_kD^2,\ \ \
D=\sqrt{\sum_l w_l^2}. 
\end{align}
\begin{align} \label{dFeq} \bullet\  \sum_{n\chi\del} w_n F^{km^\prime i,\alpha
\chi}_{jnl,\bt^\prime\del^\prime}{(F^{kmi,\alpha \chi}_{jnl,\bt\del})}^*
=\frac{w_iw_l}{w_m}
\delta_{mm^\prime}\delta_{\al\al^\prime}\delta_{\bt\bt^\prime},
\end{align}

\subsection{A derivation of string fusion algebra}

\label{strFA}

\begin{figure}[tb] 
\centerline{ \includegraphics[scale=0.42]{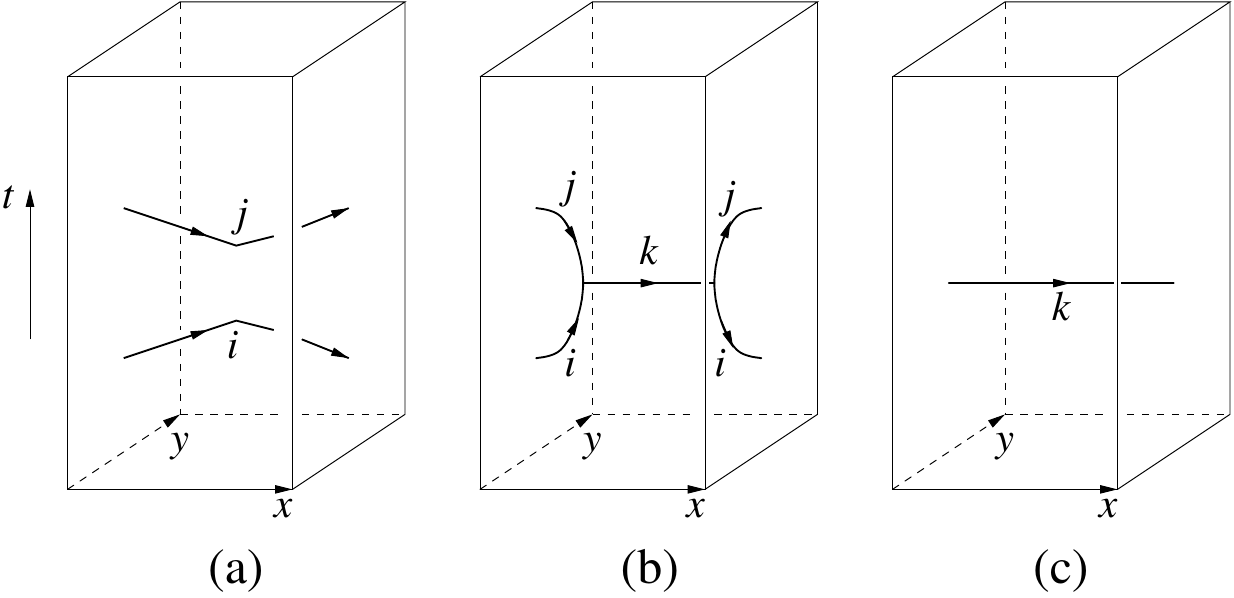} } 
\caption{
(a): Two tunneling processes: $W^x_i$ and $W^x_j$. (b): The tunneling path of
the above two tunneling processes can be deformed according to the Y-move (c):
The O-move can reduce (b) to (c).  } \label{falgT} 
\end{figure}

As an application of the above algebraic structure, let us consider the
``zero-law'' string operators on a torus $S^1_x\times S^1_y$, wrapping around
the $x$-direction $W^x_i$. Viewing those string operators as world-lines in
space-time, applying the Y-move and then the O-move, and using \eqn{YO} (see
Fig.  \ref{falgT}), we find that 
\begin{align}
 \label{AAA} W^x_i
W^x_j&=\sum_{k=1}^N \sum_{\al=1}^{N^{ij}_k} Y^{ij}_{k,\al} O^{ij,\al}_k W^x_k
\nonumber\\
 &= \sum_{k}
N^{ij}_{k} W^x_k . 
\end{align}
 We see that the algebra of the loop operator
$W^x_i$ forms a representation of fusion algebra $i\otimes j=\sum_{k}
N^{ij}_{k} k$.  The operators $W^y_i =SW^x_i S^{-1}$, where $S$ is the
$90^\circ$ rotation, satisfy the same fusion algebra
\begin{align} \label{BBB} W^y_i W^y_j &= \sum_{k} N^{ij}_{k} W^y_k .
\end{align} This way, we derived \eqn{WWNW}.

\section{Unitary m-fusion category}

The tensor data $(N^{ij}_{k}, F^{ijm,\al\bt}_{kln,\ga\la})$ satisfying the
conditions eqns.  \eq{Neq}, \eq{Feq}, \eq{Oeq}, and \eq{dFeq} form a so call
\emph{unitary m-fusion category}.  In fact, we can view the graph $\bmm
\includegraphics[scale=.40]{F1g} \emm$ and the graph $\bmm
\includegraphics[scale=.40]{F2g} \emm$ as the different ways to fusion three
particle types $i,j,k$ to $N^{ijk}_l\equiv \sum_m N^{ij}_m N^{mk}_l = \sum_m
N^{in}_l N^{jk}_n$ copies of particles $l$. But the two ways of fusions lead to
different basis of the space of $N^{ijk}_l$ copies of $l$. The
$F^{ijm,\al\bt}_{kln,\ga\la}$ tensor is nothing but the unitary transformation
that relates the two basis.  

However, here we do not require the existence of trivial particles type.  Thus
the structure we described is not a unitary fusion category. So we call it an
unitary m-fusion category (UmFC).

We like to stress that the fusion discussed here is not symmetric (\ie we do
not require $N^{ij}_k=N^{ji}_k$).  Thus fusion that we are talking about is the
fusion of 1D particles, where their order cannot be changed.  Therefore UmFC is
a classifying theory of 1+1D anomalous topological orders
$\cC_{1+1}$.\cite{KW1458,LW1384} Such anomalous topological orders cannot be
realized by any well defined 1D lattice models, but they can realized as
boundary of 2D lattice models with non-trivial 2+1D topological orders.  Those
2+1D topological orders $\cC_{2+1}$ are described by modular tensor categories,
which are uniquely determined by the $1+1D$ anomalous topological order on the
boundary. In fact, the 2+1D bulk topological order is the Drinfeld center of
the 1+1D anomalous boundary topological order: $C_{2+1}=Z(\cC_{1+1})$.  One
concrete way to compute the Drinfeld center is described in \Ref{LW1384}.

\section{Unitary fusion category and the trivial particle type}

\subsection{Trivial particle type and rule of adding trivial strings}

In the above discussion, we did not assume the existence of a trivial particle
type. Here we will assume such a trivial particle type to exist, and
denoted it by $1$, which satisfies the following fusion rule 
\begin{align}
1\otimes i=i\otimes 1=i .  
\end{align} 
Thus $N^{ij}_k$ satisfies
\begin{align}
N^{1i}_j=N^{i1}_j=\del_{ij}. 
\end{align}
 We also requires that for every $i$
there exists a unique $\bar i$ such that
\begin{align}
 \bar{\bar i}=i,\ \ \
\bar 1=1,\ \ \ N^{ij}_1=\del_{i\bar j}
\end{align}
 By setting $l=1$ in
\eqn{NN=NN}, we find the following symmetry condition on $N^{ij}_k$:
\begin{align} 
N_{\bar k}^{ij}= N_{\bar i}^{jk} .  
\end{align} 
Using the above, we can rewrite the condition \eqn{Oeq} as 
\begin{align} 
\sum_k w_{\bar i}w_j N^{jk}_i =w_{\bar k} D^2.  
\end{align} 
We see that $ w_{\bar i}$ is the left eigenvector of $\sum_j w_j N_j$ with
eigenvalue $D^2$. $D^2$ is the largest eigenvalue of $\sum_j w_j N_j$, since
the eigenvector has positive elements.  As a result  $ w_{\bar i}$ is common
left eigenvector of $N_j$ for all $j$'s, with eigenvalue $w_j$.  Since $
w_{\bar i}$ is non-negative, $w_j$ is the largest eigenvalue of $N_j$.
Therefore
\begin{align} \label{wd} w_j=d_j
\end{align}
 is the quantum dimension of type-$j$ particle (see \eqn{dNdd}).  The largest
left eigenvalue of $N_1$ is 1.  Thus $d_1=w_1=1$.

We can represent a type-$1$ string by a dash line.  By examine the O-move with
$k=1$:
\begin{align}
 A \bpm \includegraphics[scale=.40]{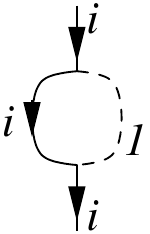} \epm =  A
\bpm \includegraphics[scale=.40]{iline} \epm  . 
\end{align}
 we see that we can remove or add any vertex with dash line without affecting
the amplitude.  In other words, a vertex with dash line can be added/removed
freely.  The unitary m-fusion categories with the trivial particle type will be
called \emph{unitary fusion categories}.

\subsection{Amplitudes for loops}
\begin{figure}[tb] 
\centerline{ \includegraphics[scale=0.5]{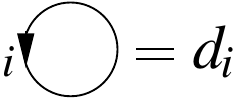} } 
\caption{A loop of type-$i$ string has an amplitude $d_i$.} 
\label{lpdi} 
\end{figure}

The tensors $N^{ij}_{k}$, $F^{ijm,\al\bt}_{kln,\ga\la}$ characterize the four
types of  linear relations between graphs with some local differences.  Those
local changes are almost complete, in the sense that any graphs of strings can
be reduce to graphs that contain only isolated loops.  Since the amplitude of a
graph that contain disconnect parts is given by the product of the amplitudes
for those parts, therefore, if we know the amplitude for single loops of
string, then the amplitude of any string configuration can be computed from the
tensor data $(N^{ij}_{k}, F^{ijm,\al\bt}_{kln,\ga\la})$.  

With the presence of trivial particle type in the unitary fusion category, we
can determine the amplitude for a loop of $i$-string.  Using the rule of adding
dash lines (the trivial strings) and $O$-move \eqn{DefO}, we find
\begin{align}
A \bpm \includegraphics[scale=.40]{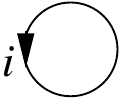} \epm &= A \bpm
\includegraphics[scale=.40]{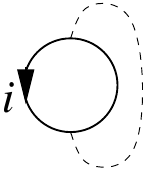} \epm = O^{i\bar i}_1 A \bpm
\includegraphics[scale=.40]{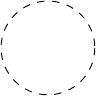} \epm
\nonumber\\
 &= d_i A \bpm
\includegraphics[scale=.40]{1lp} \epm
\end{align}
 We may choose $ A \bpm
\includegraphics[scale=.40]{1lp} \epm=1$.  This allows we to determine
(see Fig. \ref{lpdi})
\begin{align} 
\label{loopw} 
A \bpm \includegraphics[scale=.40]{ilp} \epm =d_i.
\end{align}

\section{Modular tensor category for the amplitudes of non-planar string
configurations} \label{MTC}

\subsection{Commutative unitary fusion category}

We have being considering planar graphs and the related fusion category theory.
In this section we will consider non-planar graphs.  Since the particles now
live in 2D space, the fusion of the particles satisfies
\begin{align}
 i\otimes
j =j\otimes i,
\end{align}
 and thus
\begin{align}
 N^{ij}_k=N^{ji}_k.
\end{align} So the fusion of 2D particles are commutative (while the fusion of 1D
particles may not be commutative).  The fusion with $N^{ij}_k=N^{ji}_k$ is called
commutative.  Also, we assume the existence of trivial particle type.  Thus, in
this section, the fusion of the particles is described by a commutative unitary
fusion category.

The commutative unitary fusion category for planar graphs plus the extra
structure for non-planar graphs and their amplitudes will give us a modular
tensor category theory.  In this section, we will derive many conditions that
involve amplitudes of non-planar graphs.  

\subsection{Amplitude for linked loops and Verlinde formula}

\begin{figure}[tb] 
\centerline{ \includegraphics[scale=0.5]{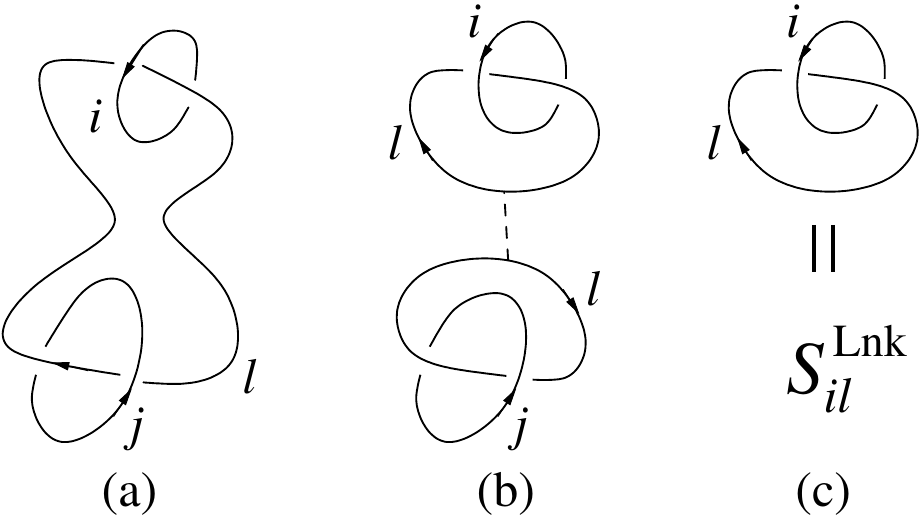} } 
\caption{ A Y-move can change a linking of three loops in (a) to two linkings
of two loops in (b).  The amplitude for two linked $i$-loop and $l$-loop in (c)
is denoted as $S^\text{Lnk}_{il}$.  } 
\label{ijkLnk} 
\end{figure}

As the first application of non-planar graphs, consider a three linked loops in
Fig. \ref{ijkLnk}a.  We can evaluate the graph in two ways: (a) we fuse the
$i$-loop and $j$-loop using \eqn{WWNW} to produce a single $k$-loop; (b) we use
a  Y-move to change the linking of three loops to two linkings of two loops.
We defined the amplitude for two linked loops $i$ and $j$ as
$S^\text{Lnk}_{ij}$ (see Fig. \ref{lnklpr}).  $S^\text{Lnk}_{ij}$ satisfies
\begin{equation} \label{STCsymm} S^\text{Lnk}_{ij} =S^\text{Lnk}_{ji} .
\end{equation} This allows us to obtain
\begin{align}
 \label{VerTC} \sum_k
N^{ij}_k S^\text{Lnk}_{kl} = Y^{\bar ll}_1 
S^\text{Lnk}_{il}S^\text{Lnk}_{jl} =
\frac{S^\text{Lnk}_{il}S^\text{Lnk}_{jl}}{d_l},
\end{align}
 which is the tensor
category version of Verlinde formula.

\subsection{Degenerate ground states on torus and excitation basis}

\begin{figure}[tb] \centerline{ \includegraphics[scale=0.5]{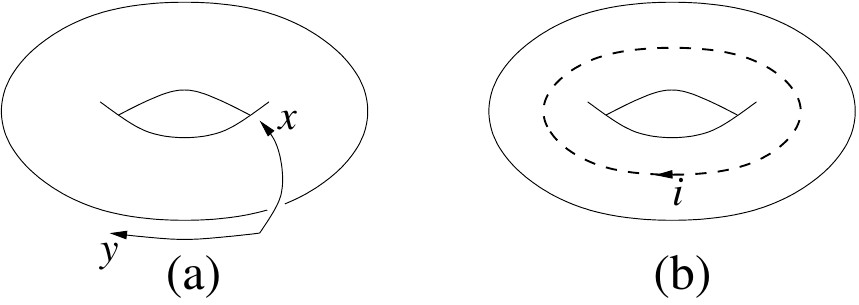} } \caption{
(a): The ground state $|1\>$ on a torus that corresponds to the trivial
quasiparticle can be represented by an empty solid torus.  (b): The other
ground state $|i\>$ that corresponds to a type $i$ quasiparticle can be
represented by an solid torus with a loop of type $i$ in the center.  }
\label{sttT} \end{figure}
\begin{figure}[tb] 
\centerline{ \includegraphics[scale=0.5]{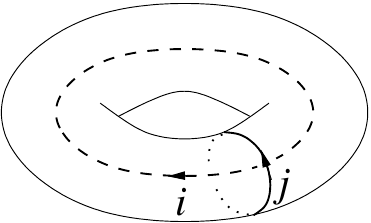} } 
\caption{The graphic representation of $W^x_j|i\>$.  } 
\label{sttTA} 
\end{figure}
\begin{figure}[tb] \centerline{ \includegraphics[scale=0.35]{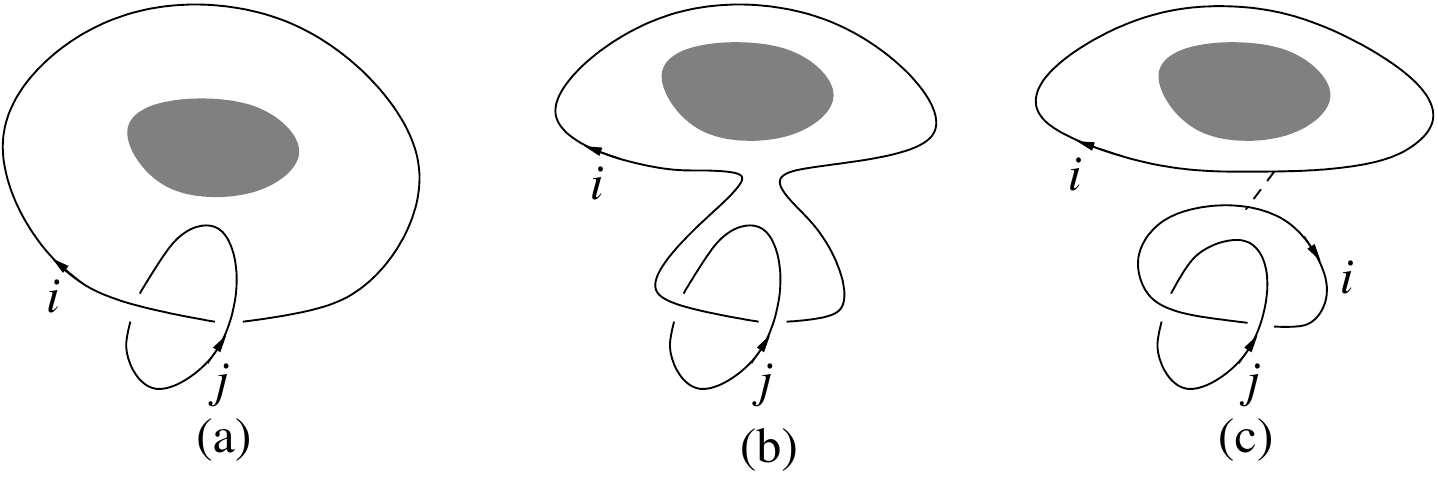} }
\caption{ (a): The graphic representation of $W^x_j|i\>$.  (b):  The graphic
representation of $W^x_j|i\>$.  (c): The Y-move can deformed the graph in (b)
to the graph in (c).  The shaded area represents the hole of the torus.  }
\label{sttTAlnk} \end{figure}

To obtain the  algebraic structure for  non-planar graphs, let us first try to
represent the degenerate ground states on torus graphically.  One of the
degenerate ground state that corresponds to the trivial quasiparticle $i=1$ can
be represented by an empty solid torus $S^1_x\times D^2_{yt}$ (see Fig.
\ref{sttT}a), where the circle in $y$-direction $S^1_y$ is a boundary of the
disk $D^2_{yt}$.  In other words, the path integral on the space-time
$S^1_x\times D^2_{yt}$ give rise to the state $|1\>$ on the surface
$S^1_x\times S^1_y$.  We denote such a state as $|1\>$.  Other degenerated
ground states can be obtained by the action of the $W^y_i$ operators
\begin{equation} \label{Bga0} |i\>\equiv W^y_i |1\> .  \end{equation} $|i\>$'s
form a orthonormal basis if we assume $(W^y_i)^\dag =W^y_{\bar i}$ and
$\<i|1\>=\del_{i1}$. This is because
\begin{align}
 \<j|i\>&=\<1| =\<1|W^y_{\bar
j} W^y_i |1\> =\<1|\sum_k N^{\bar j i}_k W^y_k |1\>
\nonumber\\
 &=\<1|N^{\bar j
i}_1 |1\>=\del_{ij}
\end{align}
 We will call such a basis of the degenerate
ground state an \emph{excitation basis}.

Since $|i\>$ is created by the tunneling operator $W^y_i$, $|i\>$ can be
represented by adding a $i$-loop that corresponds to the $W^y_i$ operator to
the center of the solid torus (see Fig.  \ref{sttT}b).

$|i\>$ is a natural basis, where the matrix elements of $W^x_l$ and $W^y_l$
have simple forms.  From \eqn{BBB}, we see that
\begin{align}
 W^y_j W^y_i|1\>
&= W^y_j |i\>  = \sum_{k} N^{ji}_{k} |k\> .  
\end{align}
The action of $W^x_{j}$ on $|i\>$ is represented by Fig. \ref{sttTA}.  From
Fig. \ref{sttTAlnk}, we find that 
\begin{equation} 
\label{AgaS}
W^x_j|i\>=Y^{\bar i i}_1 S^\text{Lnk}_{ij} |i\>
=\frac{S^\text{Lnk}_{ij}}{d_i}|i\> .  \end{equation} We see that $|i\>$'s are
common eigenstates of the commuting set of operators $W^x_j$.  The
corresponding eigenvalue for $W^x_j$ is $\frac{S^\text{Lnk}_{ij}}{d_i}$. We see
that different $|i\>$'s have different set of eigenvalues, which support our
assumption $\<i|1\>=\del_{i1}$.


\begin{table*}[tb] 
\caption{A list of primitive bosonic topological orders (up to invertible ones)
in 2+1D  with rank $N=2,3,\cdots,6$.  The list contain all topological orders
with rank $N=2,3,4$.  The list may not be complete for rank $N=5,6$.  However,
it contains all rank $N=5$ primitive topological orders (\ie UMTC's) with $D^2
\leq 120$, and  all rank $N=6$ primitive topological orders (\ie UMTC's) with
$D^2 \leq 101$.
} 
\label{toplst2-6} 
\centering
\begin{tabular}{ |c|c|c|l|l|c| } 
\hline 
$N^B_c$ & $S_\text{top}$ & $D^2$ & $d_1,d_2,\cdots$ & $s_1,s_2,\cdots$ & $K$-matrix/SCA\\
 \hline 
$2_{ 1}^B$ & $0.5$ & $2$ & $1,1$ & $0,\frac{ 1}{4}$ & $(2)$ \\
$2_{-1}^B$ & $0.5$ & $2$ & $1,1$ & $0,-\frac{1}{4}$ & $(-2)$ \\
$2_{ 14/5}^B$ & $0.9276$ & $3.6180$ & $1,\zeta_{3}^{1}$ & $0,\frac{ 2}{5}$ & $(A_1,3)_{1/2},\ (G_2,1)$ \\
$2_{-14/5}^B$ & $0.9276$ & $3.6180$ & $1,\zeta_{3}^{1}$ & $0,-\frac{2}{5}$ & $(A_1,-3)_{1/2},\ (G_2,-1)$ \\
 \hline 
$3_{ 2}^B$ & $0.7924$ & $3$ & $1,1,1$ & $0,\frac{ 1}{3},\frac{ 1}{3}$ & $(2\ 2;1)$ \\
$3_{-2}^B$ & $0.7924$ & $3$ & $1,1,1$ & $0,-\frac{1}{3},-\frac{1}{3}$ & $-(2\ 2;1)$ \\
$3_{ 3/2}^B$ & $1$ & $4$ & $1,1,\sqrt{2}$ & $0,\frac{ 1}{2},\frac{ 3}{16}$ & $(A_1,2), (B_9,1)$\\
$3_{ 5/2}^B$ & $1$ & $4$ & $1,1,\sqrt{2}$ & $0,\frac{ 1}{2},\frac{ 5}{16}$ & $(B_2,1)$ \\
$3_{ 7/2}^B$ & $1$ & $4$ & $1,1,\sqrt{2}$ & $0,\frac{ 1}{2},\frac{ 7}{16}$ & $(B_3,1)$ \\
$3_{-7/2}^B$ & $1$ & $4$ & $1,1,\sqrt{2}$ & $0,\frac{ 1}{2},-\frac{7}{16}$ & $(B_4,1)$ \\
$3_{-5/2}^B$ & $1$ & $4$ & $1,1,\sqrt{2}$ & $0,\frac{ 1}{2},-\frac{5}{16}$ & $(B_5,1)$ \\
$3_{-3/2}^B$ & $1$ & $4$ & $1,1,\sqrt{2}$ & $0,\frac{ 1}{2},-\frac{3}{16}$ & $(B_6,1)$ \\
$3_{-1/2}^B$ & $1$ & $4$ & $1,1,\sqrt{2}$ & $0,\frac{ 1}{2},-\frac{1}{16}$ & $(B_7,1)$ \\
$3_{ 1/2}^B$ & $1$ & $4$ & $1,1,\sqrt{2}$ & $0,\frac{ 1}{2},\frac{ 1}{16}$ & $(B_8,1)$ \\
$3_{ 8/7}^B$ & $1.6082$ & $9.2958$ & $1,\zeta_{5}^{1},\zeta_{5}^{2}$ & $0,-\frac{1}{7},\frac{ 2}{7}$ & $(A_1,5)_{1/2}$\\
$3_{-8/7}^B$ & $1.6082$ & $9.2958$ & $1,\zeta_{5}^{1},\zeta_{5}^{2}$ & $0,\frac{ 1}{7},-\frac{2}{7}$ & $(A_1,-5)_{1/2}$\\
 \hline 
$4_{ 0}^B$ & $1$ & $4$ & $1,1,1,1$ & $0, 0, 0,\frac{ 1}{2}$ & $(0,0;2)$ \\
$4_{ 1}^B$ & $1$ & $4$ & $1,1,1,1$ & $0,\frac{ 1}{8},\frac{ 1}{8},\frac{ 1}{2}$ & $(4)$ \\
$4_{-1}^B$ & $1$ & $4$ & $1,1,1,1$ & $0,-\frac{1}{8},-\frac{1}{8},\frac{ 1}{2}$ & $(-4)$ \\
$4_{ 3}^B$ & $1$ & $4$ & $1,1,1,1$ & $0,\frac{ 3}{8},\frac{ 3}{8},\frac{ 1}{2}$ & $(2\ 2\ 2;1\ 1;1)$ \\
$4_{-3}^B$ & $1$ & $4$ & $1,1,1,1$ & $0,-\frac{3}{8},-\frac{3}{8},\frac{ 1}{2}$ & $-(2\ 2\ 2;1\ 1;1)$ \\
$4_{ 4}^B$ & $1$ & $4$ & $1,1,1,1$ & $0,\frac{ 1}{2},\frac{ 1}{2},\frac{ 1}{2}$ & $(2\ 2\ 2\ 2;1\ 0\ 0;1\ 0;1)$ \\
$4_{ 10/3}^B$ & $2.1328$ & $19.234$ & $1,\zeta_{7}^{1},\zeta_{7}^{2},\zeta_{7}^{3}$ & $0,\frac{ 1}{3},\frac{ 2}{9},-\frac{1}{3}$ & $(A_1,7)_{1/2}$ \\
$4_{-10/3}^B$ & $2.1328$ & $19.234$ & $1,\zeta_{7}^{1},\zeta_{7}^{2},\zeta_{7}^{3}$ & $0,-\frac{1}{3},-\frac{2}{9},\frac{ 1}{3}$ & $(A_1,-7)_{1/2}, (G_2,2)$ \\
 \hline 
$5_{ 0}^B$ & $1.1609$ & $5$ & $1,1,1,1,1$ & $0,\frac{ 1}{5},\frac{ 1}{5},-\frac{1}{5},-\frac{1}{5}$ & $(2\ 2;3)$ \\
$5_{ 4}^B$ & $1.1609$ & $5$ & $1,1,1,1,1$ & $0,\frac{ 2}{5},\frac{ 2}{5},-\frac{2}{5},-\frac{2}{5}$ &  $(2\ 2\ 2\ 2;1\ 1\ 0;1\ 0;1)$  \\
$5_{ 2}^{B,a}$ & $1.7924$ & $12$ & $1,1,\sqrt{3},\sqrt{3},2$ & $0, 0,\frac{ 1}{8},-\frac{3}{8},\frac{ 1}{3}$ & $(A_1,4)$\\
$5_{ 2}^{B,b}$ & $1.7924$ & $12$ & $1,1,\sqrt{3},\sqrt{3},2$ & $0, 0,-\frac{1}{8},\frac{ 3}{8},\frac{ 1}{3}$ & $(5_2^{B,a}\boxtimes 2_1^B \boxtimes 2_{-1}^B)_{1/4}$\\
$5_{-2}^{B,a}$ & $1.7924$ & $12$ & $1,1,\sqrt{3},\sqrt{3},2$ & $0, 0,-\frac{1}{8},\frac{ 3}{8},-\frac{1}{3}$ & $(A_1,-4)$ \\
$5_{-2}^{B,b}$ & $1.7924$ & $12$ & $1,1,\sqrt{3},\sqrt{3},2$ & $0, 0,\frac{ 1}{8},-\frac{3}{8},-\frac{1}{3}$ &  $(5_{-2}^{B,a}\boxtimes 2_1^B \boxtimes 2_{-1}^B)_{1/4}$ \\
$5_{ 16/11}^B$ & $2.5573$ & $34.646$ & $1,\zeta_{9}^{1},\zeta_{9}^{2},\zeta_{9}^{3},\zeta_{9}^{4}$ & $0,-\frac{2}{11},\frac{ 2}{11},\frac{ 1}{11},-\frac{5}{11}$ & $(A_1,9)_{1/2},\ (F_4,2)$\\
$5_{-16/11}^B$ & $2.5573$ & $34.646$ & $1,\zeta_{9}^{1},\zeta_{9}^{2},\zeta_{9}^{3},\zeta_{9}^{4}$ & $0,\frac{ 2}{11},-\frac{2}{11},-\frac{1}{11},\frac{ 5}{11}$ & $(A_1,-9)_{1/2},\ (E_8,3)$\\
$5_{ 18/7}^B$ & $2.5716$ & $35.342$ & $1,\frac12\zeta_{12}^{6},\frac12\zeta_{12}^{6},\zeta_{12}^{2},\zeta_{12}^{4}$ & $0,-\frac{1}{7},-\frac{1}{7},\frac{ 1}{7},\frac{ 3}{7}$ & $(A_1,12)_{1/4},\ (A_2,4)_{1/3}$\\
$5_{-18/7}^B$ & $2.5716$ & $35.342$ & $1,\frac12\zeta_{12}^{6},\frac12\zeta_{12}^{2},\zeta_{12}^{2},\zeta_{12}^{4}$ & $0,\frac{ 1}{7},\frac{ 1}{7},-\frac{1}{7},-\frac{3}{7}$ & $(A_3,3)_{1/4}$ \\
 \hline 
$6_{ 0}^{B,a}$ & $2.1609$ & $20$ & $1,1,2,2,\sqrt{5},\sqrt{5}$ & $0, 0,\frac{ 1}{5},-\frac{1}{5}, 0,\frac{ 1}{2}$ & $(D_5,2)_{1/4},\ (U(1)_5/\Z_2)_{1/2}$\\
$6_{ 0}^{B,b}$ & $2.1609$ & $20$ & $1,1,2,2,\sqrt{5},\sqrt{5}$ & $0, 0,\frac{ 1}{5},-\frac{1}{5},\frac{ 1}{4},-\frac{1}{4}$ & $(6_0^{B,a}\boxtimes 2_1^B\boxtimes 2_{-1}^B)_{1/4}$\\
$6_{ 4}^{B,a}$ & $2.1609$ & $20$ & $1,1,2,2,\sqrt{5},\sqrt{5}$ & $0, 0,\frac{ 2}{5},-\frac{2}{5},\frac{ 1}{4},-\frac{1}{4}$ & $(B_2,2)$ \\
$6_{ 4}^{B,b}$ & $2.1609$ & $20$ & $1,1,2,2,\sqrt{5},\sqrt{5}$ & $0, 0,\frac{ 2}{5},-\frac{2}{5}, 0,\frac{ 1}{2}$ & $(6_4^{B,a}\boxtimes 2_1^B\boxtimes 2_{-1}^B)_{1/4}$ \\
$6_{ 46/13}^B$ & $2.9132$ & $56.746$ & $1,\zeta_{11}^{1},\zeta_{11}^{2},\zeta_{11}^{3},\zeta_{11}^{4},\zeta_{11}^{5}$ & $0,\frac{ 4}{13},\frac{ 2}{13},-\frac{6}{13},\frac{ 6}{13},-\frac{1}{13}$ & $(A_1,11)_{1/2}$ \\
$6_{-46/13}^B$ & $2.9132$ & $56.746$ & $1,\zeta_{11}^{1},\zeta_{11}^{2},\zeta_{11}^{3},\zeta_{11}^{4},\zeta_{11}^{5}$ & $0,-\frac{4}{13},-\frac{2}{13},\frac{ 6}{13},-\frac{6}{13},\frac{ 1}{13}$ & $(A_1,-11)_{1/2}$ \\
$6_{ 8/3}^B$ & $3.1107$ & $74.617$ & $1,\frac12\zeta_{16}^{8},\frac12\zeta_{16}^{8},\zeta_{16}^{2},\zeta_{16}^{4},\zeta_{16}^{6}$ & $0,\frac{ 1}{9},\frac{ 1}{9},\frac{ 1}{9},\frac{ 1}{3},-\frac{1}{3}$ & $(A_1,16)_{1/4}$ \\
$6_{-8/3}^B$ & $3.1107$ & $74.617$ & $1,\frac12\zeta_{16}^{8},\frac12\zeta_{16}^{8},\zeta_{16}^{2},\zeta_{16}^{4},\zeta_{16}^{6}$ & $0,-\frac{1}{9},-\frac{1}{9},-\frac{1}{9},-\frac{1}{3},\frac{ 1}{3}$ & $(A_1,-16)_{1/4},\ (A_2,6)_{1/9}$ \\
$6_{ 2}^B$ & $3.3263$ & $100.61$ & $1,\frac{3+\sqrt{21}}{2},\frac{3+\sqrt{21}}{2},\frac{3+\sqrt{21}}{2},\frac{5+\sqrt{21}}{2},\frac{7+\sqrt{21}}{2}$ 
& $0,-\frac{1}{7},-\frac{2}{7},\frac{ 3}{7}, 0,\frac{ 1}{3}$ & $(G_2,-3)$\\
$6_{-2}^B$ & $3.3263$ & $100.61$ & 
$1,\frac{3+\sqrt{21}}{2},\frac{3+\sqrt{21}}{2},\frac{3+\sqrt{21}}{2},\frac{5+\sqrt{21}}{2},\frac{7+\sqrt{21}}{2}$
& $0,\frac{ 1}{7},\frac{ 2}{7},-\frac{3}{7}, 0,-\frac{1}{3}$ & $(G_2,3)$\\
 \hline 
\end{tabular} 
\end{table*}
\begin{table*}[tb] 
\caption{A list of primitive bosonic topological orders (up to invertible ones)
in 2+1D  with rank $N=7,8,9$.  The list may not be complete.  However, it
contains all rank $N=7$ primitive topological orders (\ie UMTC's) with $D^2
\leq 40$, all rank $N=8$ primitive topological orders with $D^2 \leq 25$, and
all rank $N=9$ primitive topological orders with $D^2 \leq 20$.  
} 
\label{toplst7-9} 
\centering
\begin{tabular}{ |c|c|c|l|l|c| } 
\hline 
$N^B_c$ & $S_\text{top}$ & $D^2$ & $d_1,d_2,\cdots$ & $s_1,s_2,\cdots$ & $K$-matrix/SCA \\
 \hline 
$7_{ 2}^{B,a}$ & $1.4036$ & $7$ & $1,1,1,1,1,1,1$ & $0, \frac{1}{7}, \frac{1}{7}, \frac{2}{7}, \frac{2}{7},-\frac{3}{7},-\frac{3}{7}$ & $(4\ 4;3)$ \\
$7_{-2}^{B,a}$ & $1.4036$ & $7$ & $1,1,1,1,1,1,1$ & $0,-\frac{1}{7},-\frac{1}{7},-\frac{2}{7},-\frac{2}{7}, \frac{3}{7}, \frac{3}{7}$ & $-(4\ 4;3),\ (A_6,1)$ \\
$7_{ 9/4}^B$ & $2.3857$ & $27.313$ & $1,1,\zeta_{6}^{1},\zeta_{6}^{1},\zeta_{6}^{2},\zeta_{6}^{2},\zeta_{6}^{3}$ & $0, \frac{1}{2}, \frac{3}{32}, \frac{3}{32}, \frac{1}{4},-\frac{1}{4},\frac{ 15}{32}$ & $(A_1,6)$ \\
$7_{ 13/4}^B$ & $2.3857$ & $27.313$ & $1,1,\zeta_{6}^{1},\zeta_{6}^{1},\zeta_{6}^{2},\zeta_{6}^{2},\zeta_{6}^{3}$ & $0,\frac{ 1}{2},\frac{ 7}{32},\frac{ 7}{32},\frac{ 1}{4},-\frac{1}{4},-\frac{13}{32}$ & $(7_{9/4}^B\boxtimes 4_1^B)_{1/4}$ \\
$7_{-15/4}^B$ & $2.3857$ & $27.313$ & $1,1,\zeta_{6}^{1},\zeta_{6}^{1},\zeta_{6}^{2},\zeta_{6}^{2},\zeta_{6}^{3}$ & $0,\frac{ 1}{2},\frac{ 11}{32},\frac{ 11}{32},\frac{ 1}{4},-\frac{1}{4},-\frac{9}{32}$ & $(7_{13/4}^B\boxtimes 4_1^B)_{1/4}$ \\
$7_{-11/4}^B$ & $2.3857$ & $27.313$ & $1,1,\zeta_{6}^{1},\zeta_{6}^{1},\zeta_{6}^{2},\zeta_{6}^{2},\zeta_{6}^{3}$ & $0,\frac{ 1}{2},\frac{ 15}{32},\frac{ 15}{32},\frac{ 1}{4},-\frac{1}{4},-\frac{5}{32}$ & $(7_{-15/4}^B\boxtimes 4_1^B)_{1/4}$ \\
$7_{-7/4}^B$ & $2.3857$ & $27.313$ & $1,1,\zeta_{6}^{1},\zeta_{6}^{1},\zeta_{6}^{2},\zeta_{6}^{2},\zeta_{6}^{3}$ & $0,\frac{ 1}{2},-\frac{13}{32},-\frac{13}{32},\frac{ 1}{4},-\frac{1}{4},-\frac{1}{32}$ & $(7_{-11/4}^B\boxtimes 4_1^B)_{1/4}$ \\
$7_{-3/4}^B$ & $2.3857$ & $27.313$ & $1,1,\zeta_{6}^{1},\zeta_{6}^{1},\zeta_{6}^{2},\zeta_{6}^{2},\zeta_{6}^{3}$ & $0,\frac{ 1}{2},-\frac{9}{32},-\frac{9}{32},\frac{ 1}{4},-\frac{1}{4},\frac{ 3}{32}$ & $(7_{-7/4}^B\boxtimes 4_1^B)_{1/4}$ \\
$7_{ 1/4}^B$ & $2.3857$ & $27.313$ & $1,1,\zeta_{6}^{1},\zeta_{6}^{1},\zeta_{6}^{2},\zeta_{6}^{2},\zeta_{6}^{3}$ & $0,\frac{ 1}{2},-\frac{5}{32},-\frac{5}{32},\frac{ 1}{4},-\frac{1}{4},\frac{ 7}{32}$ & $(7_{-3/4}^B\boxtimes 4_1^B)_{1/4}$ \\
$7_{ 5/4}^B$ & $2.3857$ & $27.313$ & $1,1,\zeta_{6}^{1},\zeta_{6}^{1},\zeta_{6}^{2},\zeta_{6}^{2},\zeta_{6}^{3}$ & $0,\frac{ 1}{2},-\frac{1}{32},-\frac{1}{32},\frac{ 1}{4},-\frac{1}{4},\frac{ 11}{32}$ & $(7_{1/4}^B\boxtimes 4_1^B)_{1/4}$ \\
$7_{ 7/4}^B$ & $2.3857$ & $27.313$ & $1,1,\zeta_{6}^{1},\zeta_{6}^{1},\zeta_{6}^{2},\zeta_{6}^{2},\zeta_{6}^{3}$ & $0,\frac{ 1}{2},\frac{ 13}{32},\frac{ 13}{32},\frac{ 1}{4},-\frac{1}{4},\frac{ 1}{32}$ & $(C_6,1)$ \\
$7_{ 11/4}^B$ & $2.3857$ & $27.313$ & $1,1,\zeta_{6}^{1},\zeta_{6}^{1},\zeta_{6}^{2},\zeta_{6}^{2},\zeta_{6}^{3}$ & $0,\frac{ 1}{2},-\frac{15}{32},-\frac{15}{32},\frac{ 1}{4},-\frac{1}{4},\frac{ 5}{32}$ & $(7_{7/4}^B\boxtimes 4_1^B)_{1/4}$ \\
$7_{ 15/4}^B$ & $2.3857$ & $27.313$ & $1,1,\zeta_{6}^{1},\zeta_{6}^{1},\zeta_{6}^{2},\zeta_{6}^{2},\zeta_{6}^{3}$ & $0,\frac{ 1}{2},-\frac{11}{32},-\frac{11}{32},\frac{ 1}{4},-\frac{1}{4},\frac{ 9}{32}$ & $(7_{11/4}^B\boxtimes 4_1^B)_{1/4}$ \\
$7_{-13/4}^B$ & $2.3857$ & $27.313$ & $1,1,\zeta_{6}^{1},\zeta_{6}^{1},\zeta_{6}^{2},\zeta_{6}^{2},\zeta_{6}^{3}$ & $0,\frac{ 1}{2},-\frac{7}{32},-\frac{7}{32},\frac{ 1}{4},-\frac{1}{4},\frac{ 13}{32}$ & $(7_{15/4}^B\boxtimes 4_1^B)_{1/4}$ \\
$7_{-9/4}^B$ & $2.3857$ & $27.313$ & $1,1,\zeta_{6}^{1},\zeta_{6}^{1},\zeta_{6}^{2},\zeta_{6}^{2},\zeta_{6}^{3}$ & $0,\frac{ 1}{2},-\frac{3}{32},-\frac{3}{32},\frac{ 1}{4},-\frac{1}{4},-\frac{15}{32}$ & $(7_{-13/4}^B\boxtimes 4_1^B)_{1/4}$ \\
$7_{-5/4}^B$ & $2.3857$ & $27.313$ & $1,1,\zeta_{6}^{1},\zeta_{6}^{1},\zeta_{6}^{2},\zeta_{6}^{2},\zeta_{6}^{3}$ & $0,\frac{ 1}{2},\frac{ 1}{32},\frac{ 1}{32},\frac{ 1}{4},-\frac{1}{4},-\frac{11}{32}$ & $(7_{-9/4}^B\boxtimes 4_1^B)_{1/4}$ \\
$7_{-1/4}^B$ & $2.3857$ & $27.313$ & $1,1,\zeta_{6}^{1},\zeta_{6}^{1},\zeta_{6}^{2},\zeta_{6}^{2},\zeta_{6}^{3}$ & $0,\frac{ 1}{2},\frac{ 5}{32},\frac{ 5}{32},\frac{ 1}{4},-\frac{1}{4},-\frac{7}{32}$ & $(7_{-5/4}^B\boxtimes 4_1^B)_{1/4}$ \\
$7_{ 3/4}^B$ & $2.3857$ & $27.313$ & $1,1,\zeta_{6}^{1},\zeta_{6}^{1},\zeta_{6}^{2},\zeta_{6}^{2},\zeta_{6}^{3}$ & $0,\frac{ 1}{2},\frac{ 9}{32},\frac{ 9}{32},\frac{ 1}{4},-\frac{1}{4},-\frac{3}{32}$ & $(7_{-1/4}^B\boxtimes 4_1^B)_{1/4}$ \\
$7_{ 2}^{B,b}$ & $2.4036$ & $28$ & $1,1,2,2,2,\sqrt{7},\sqrt{7}$ & $0, 0,\frac{ 1}{7},\frac{ 2}{7},-\frac{3}{7},\frac{ 1}{8},-\frac{3}{8}$ & $(U(1)_7/\Z_2)_{1/2}$ \\
$7_{ 2}^{B,c}$ & $2.4036$ & $28$ & $1,1,2,2,2,\sqrt{7},\sqrt{7}$ & $0, 0,\frac{ 1}{7},\frac{ 2}{7},-\frac{3}{7},-\frac{1}{8},\frac{ 3}{8}$ & $(7_2^{B,b}\boxtimes 2_1^B\boxtimes 2_{-1}^B)_{1/4}$ \\
$7_{-2}^{B,b}$ & $2.4036$ & $28$ & $1,1,2,2,2,\sqrt{7},\sqrt{7}$ & $0, 0,-\frac{1}{7},-\frac{2}{7},\frac{ 3}{7},-\frac{1}{8},\frac{ 3}{8}$ & $(B_3,2),\ (D_7,2)_{1/2}$ \\
$7_{-2}^{B,c}$ & $2.4036$ & $28$ & $1,1,2,2,2,\sqrt{7},\sqrt{7}$ & $0, 0,-\frac{1}{7},-\frac{2}{7},\frac{ 3}{7},\frac{ 1}{8},-\frac{3}{8}$ & $(7_{-2}^{B,b}\boxtimes 2_1^B\boxtimes 2_{-1}^B)_{1/4}$ \\
$7_{ 8/5}^B$ & $3.2194$ & $86.750$ & $1,\zeta_{13}^{1},\zeta_{13}^{2},\zeta_{13}^{3},\zeta_{13}^{4},\zeta_{13}^{5},\zeta_{13}^{6}$ & $0,-\frac{1}{5},\frac{ 2}{15}, 0,\frac{ 2}{5},\frac{ 1}{3},-\frac{1}{5}$ & $(A_1,13)_{1/2}$ \\
$7_{-8/5}^B$ & $3.2194$ & $86.750$ & $1,\zeta_{13}^{1},\zeta_{13}^{2},\zeta_{13}^{3},\zeta_{13}^{4},\zeta_{13}^{5},\zeta_{13}^{6}$ & $0,\frac{ 1}{5},-\frac{2}{15}, 0,-\frac{2}{5},-\frac{1}{3},\frac{ 1}{5}$ & $(A_1,-13)_{1/2}$ \\
$7_{ 1}^B$ & $3.2715$ & $93.254$ & {\footnotesize $1,\zeta_{6}^{2},\zeta_{6}^{2},1+\zeta_6^2,1+\zeta_6^2,2\zeta_6^2,1+2\zeta_6^2$} & $0,\frac{ 1}{2},\frac{ 1}{2},\frac{ 1}{4},\frac{ 1}{4},-\frac{3}{8}, 0$ & $(A_2,5)_{1/3}$ \\
$7_{-1}^B$ & $3.2715$ & $93.254$ & {\footnotesize $1,\zeta_{6}^{2},\zeta_{6}^{2},1+\zeta_6^2,1+\zeta_6^2,2\zeta_6^2,1+2\zeta_6^2$} & $0,\frac{ 1}{2},\frac{ 1}{2},-\frac{1}{4},-\frac{1}{4},\frac{ 3}{8}, 0$ & $(A_2,-5)_{1/3}$ \\
$7_{ 30/11}^B$ & $3.5425$ & $135.77$ & $1,\zeta_{20}^{2},\frac12\zeta_{20}^{10},\frac12\zeta_{20}^{10},\zeta_{20}^{4},\zeta_{20}^{6},\zeta_{20}^{8}$ & $0, \frac{1}{11}, \frac{4}{11}, \frac{4}{11}, \frac{3}{11},-\frac{5}{11},-\frac{1}{11}$ & $(A_1,20)_{1/4}$ \\
$7_{-30/11}^B$ & $3.5425$ & $135.77$ & $1,\zeta_{20}^{2},\frac12\zeta_{20}^{10},\frac12\zeta_{20}^{10},\zeta_{20}^{4},\zeta_{20}^{6},\zeta_{20}^{8}$ & $0,-\frac{1}{11},-\frac{4}{11},-\frac{4}{11},-\frac{3}{11}, \frac{5}{11}, \frac{1}{11}$ & $(A_1,-20)_{1/4}$ \\
 \hline 
$8_{ 1}^{B,a}$ & $1.5$ & $8$ & $1,1,1,1,1,1,1,1$ & $0, 0,\frac{ 1}{16},\frac{ 1}{16},\frac{ 1}{4},\frac{ 1}{4},-\frac{7}{16},-\frac{7}{16}$ & $(8)$ \\
$8_{ 1}^{B,b}$ & $1.5$ & $8$ & $1,1,1,1,1,1,1,1$ & $0, 0,-\frac{3}{16},-\frac{3}{16},\frac{ 1}{4},\frac{ 1}{4},\frac{ 5}{16},\frac{ 5}{16}$ & $(2\ 2\ 2;1\ 2;3)$ \\
$8_{-1}^{B,a}$ & $1.5$ & $8$ & $1,1,1,1,1,1,1,1$ & $0, 0,-\frac{1}{16},-\frac{1}{16},-\frac{1}{4},-\frac{1}{4},\frac{ 7}{16},\frac{ 7}{16}$ & $(-8)$ \\
$8_{-1}^{B,b}$ & $1.5$ & $8$ & $1,1,1,1,1,1,1,1$ & $0, 0,\frac{ 3}{16},\frac{ 3}{16},-\frac{1}{4},-\frac{1}{4},-\frac{5}{16},-\frac{5}{16}$ & $-(2\ 2\ 2;1\ 2;3)$ \\
$8^{ B}_{ 0}$ & $2.5849$ & $36$ & $1, 1, 2, 2, 2, 2, 3, 3$ & $0, 0, 0, 0, \frac{1}{3}, \frac{2}{3}, 0, \frac{1}{2}$ & $S_3$ gauge theory\\
$8^{ B}_{ 0}$ & $2.5849$ & $36$ & $1, 1, 2, 2, 2, 2, 3, 3$ & $0, 0, 0, 0, \frac{1}{3}, \frac{2}{3}, \frac{1}{4}, \frac{3}{4}$ & \\
$8_{ 0}^{B,a}$ & $2.5849$ & $36$ & $1,1,2,2,2,2,3,3$ & $0, 0, 0,\frac{ 1}{9},-\frac{2}{9},\frac{ 4}{9}, 0,\frac{ 1}{2}$ & $(B_4,2)$ \\
$8_{ 0}^{B,b}$ & $2.5849$ & $36$ & $1,1,2,2,2,2,3,3$ & $0, 0, 0,-\frac{1}{9},\frac{ 2}{9},-\frac{4}{9}, 0,\frac{ 1}{2}$ & $(B_4,-2)$ \\
$8_{ 0}^{B,c}$ & $2.5849$ & $36$ & $1,1,2,2,2,2,3,3$ & $0, 0, 0,\frac{ 1}{9},-\frac{2}{9},\frac{ 4}{9},\frac{ 1}{4},-\frac{1}{4}$ & 
\\
$8_{ 0}^{B,d}$ & $2.5849$ & $36$ & $1,1,2,2,2,2,3,3$ & $0, 0, 0,-\frac{1}{9},\frac{ 2}{9},-\frac{4}{9},\frac{ 1}{4},-\frac{1}{4}$ & 
\\
$8_{ 62/17}^B$ & $3.4879$ & $125.87$ & $1,\zeta_{15}^{1},\zeta_{15}^{2},\zeta_{15}^{3},\zeta_{15}^{4},\zeta_{15}^{5},\zeta_{15}^{6},\zeta_{15}^{7}$ & $0,\frac{ 5}{17},\frac{ 2}{17},\frac{ 8}{17},\frac{ 6}{17},-\frac{4}{17},-\frac{5}{17},\frac{ 3}{17}$ & $(A_1,15)_{1/2}$ \\
$8_{-62/17}^B$ & $3.4879$ & $125.87$ & $1,\zeta_{15}^{1},\zeta_{15}^{2},\zeta_{15}^{3},\zeta_{15}^{4},\zeta_{15}^{5},\zeta_{15}^{6},\zeta_{15}^{7}$ & {\footnotesize $0,-\frac{5}{17},-\frac{2}{17},-\frac{8}{17},-\frac{6}{17},\frac{ 4}{17},\frac{ 5}{17},-\frac{3}{17}$} & $(A_1,-15)_{1/2}$ \\
 \hline 
$9_{ 0}^B$ & $1.5849$ & $9$ & $1,1,1,1,1,1,1,1,1$ & $0, 0, 0,\frac{ 1}{9},\frac{ 1}{9},-\frac{2}{9},-\frac{2}{9},\frac{ 4}{9},\frac{ 4}{9}$ & $(4\ 4;5)$ \\
$9_{ 0}^B$ & $1.5849$ & $9$ & $1,1,1,1,1,1,1,1,1$ & $0, 0, 0,-\frac{1}{9},-\frac{1}{9},\frac{ 2}{9},\frac{ 2}{9},-\frac{4}{9},-\frac{4}{9}$ & $-(4\ 4;5)$ \\
 \hline 
\end{tabular} 
\end{table*}

\subsection{The relation between $S^\text{Lnk}$ and $S$}

\begin{figure}[tb] 
\centerline{ \includegraphics[scale=0.45]{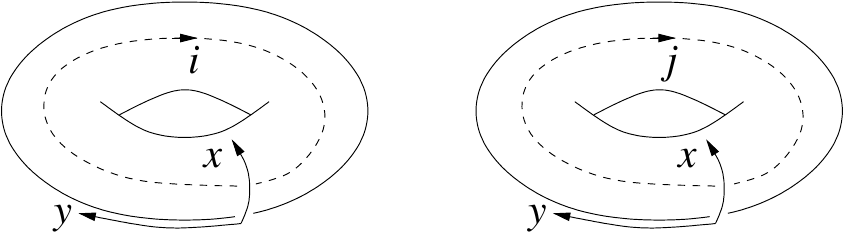} } 
\caption{
Gluing two solid tori $D^2_{xt} \times S^1_y$ without twist forms a $S^2\times
S^1$.  The gluing is done by identifying the $(x,y)$ point on the surface of the
first torus with the $(x,-y)$ point on the surface of the second torus.  If we
add an additional $\cS$ twist, \ie if we identify $(x,y)$ with $(-y,-x)$, the
gluing will produce a $S^3$.  } 
\label{glue3d} 
\end{figure}

\begin{figure}[tb] \centerline{ \includegraphics[scale=0.5]{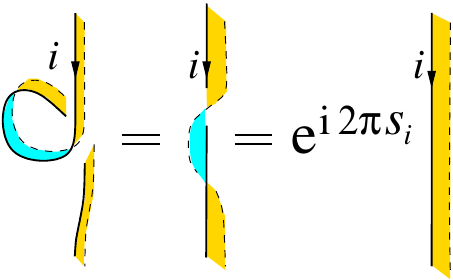} } 
\caption{ (Color online) 
A ``self-loop" with canonical framing corresponds to a twist by
$2\pi$.  A twist by $2\pi$ induces a phase $\ee^{\ii 2\pi s_i}$ that defines
the spin $s_i$ of the particle.}
\label{twist} 
\end{figure}
\begin{figure}[tb] \centerline{ \includegraphics[scale=0.5]{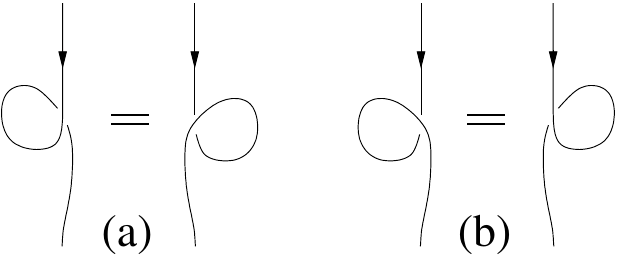} } 
\caption{
The two ``self-loops'' in (a) are ``right-handed'' and correspond to the same
twist.  The two ``self-loops'' in (b) are ``left-handed'' and also correspond
to the same twist that is opposite to that in (a).  } 
\label{twist2}
\end{figure}

The amplitude of two linked loops, $S^\text{Lnk}_{ij}$, and the representation
of the modular group in the excitations basis, $S$, are closely related.  From
Fig. \ref{glue3d}, we see that 
\begin{align} 
S_{ij}= \<i|\hat S|j\>= (\text{two linked loops in } S^3) = S^\text{Lnk}_{ij} Z_\text{inv}(S^3), 
\end{align} 
where $Z_\text{inv}(S^3)$ is the volume independent part of the partition
function on three-sphere $S^3$.\cite{KW1458} Since $S$ is unitary and
$S^\text{Lnk}_{i1}=d_i$, we see that 
\begin{align}
Z_\text{inv}(S^3)=1/D.
\end{align}
This way, we obtain an important relation:
\begin{align}
S_{ij} =S^\text{Lnk}_{ij}/D,  
\end{align} 
that connects the amplitude of two linked loops to a modular transformation of
the degenerate ground state on torus.  This allows us to rewrite \eqn{VerTC} as 
\begin{align} 
\label{Ver1} 
\sum_k N^{ij}_k S_{kl} = \frac{S_{il}S_{jl}}{S_{1l}}, 
\end{align} 
which is the Verlinde formula.

\subsection{The spin $s_i$ of topological excitations and $T$}

\begin{figure}[tb] 
\centerline{ \includegraphics[scale=0.6]{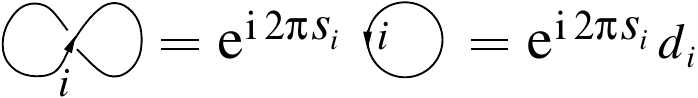} } 
\caption{A figure ``8'' of type-$i$ string has an amplitude $\ee^{2\pi \ii s_i} d_i$.} 
\label{twsidi} 
\end{figure}

\begin{figure}[tb]
\centerline{
\includegraphics[scale=0.45]{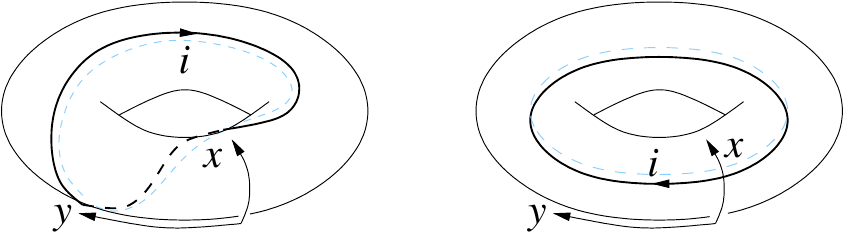}
}
\caption{
Gluing two solid tori $S^1_x\times D^2$ with an additional $\hat T$ twist,
\ie identifying $(x,y)$ with $(x+y,-y)$, will produce a $S^2\times S^1$.  The
$i$-loop in $y$-direction in the second solid torus at right can be
deformed into a $i$-loop in the first solid torus at left. We see that
the loop is twisted by $2\pi$ in the anti-clockwise direction.
}
\label{TTrns}
\end{figure}
\begin{figure}[tb] \centerline{ \includegraphics[scale=0.45]{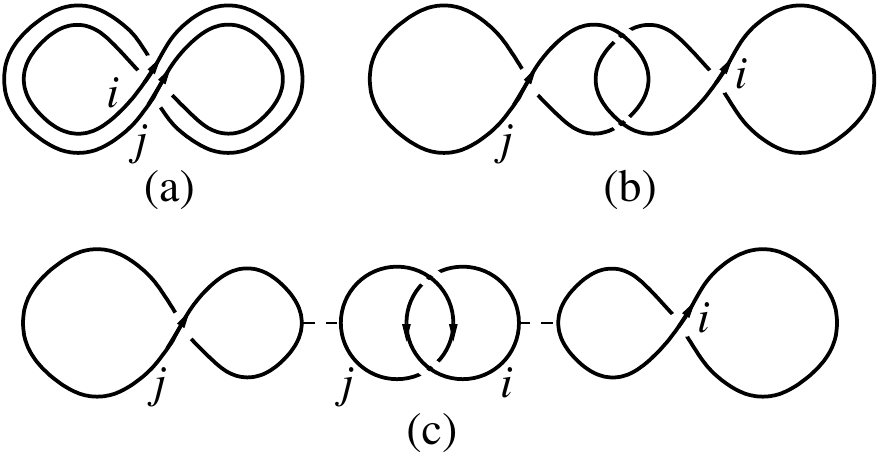} } 
\caption{A double figure ``8'' in (a) is equal to two linked  figure ``8'' in
(b) after sliding one of the  figure ``8''. Applying two Y-moves to (b), we obtain
(c)} 
\label{dtwist} 
\end{figure}
\begin{figure}[tb] \centerline{ \includegraphics[scale=0.45]{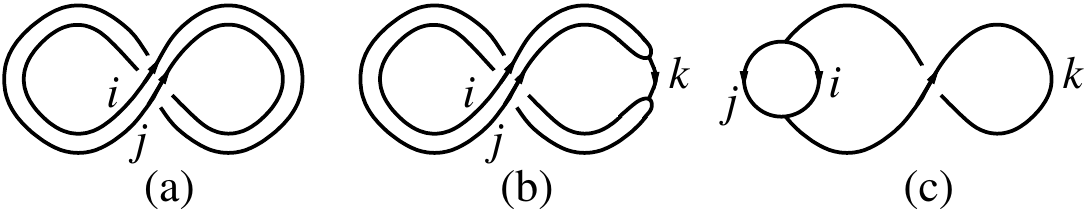} } 
\caption{A double figure ``8'' in (a) is 
changed to (b) after a Y-move.
(b) is equal to (c) after sliding the vertices.
} 
\label{dtwist1} 
\end{figure}

We have studied two linked loops, which is a string configuration with
crossing.  Another important string configuration with crossing is a
``self-loop'' (see Fig. \ref{twist}).  Such a ``self-loop'' corresponds to a
twist by $2\pi$, which equal to a straight line with a phase $\ee^{\ii 2\pi
s_i}$.  Here $s_i$ is the \emph{spin} of the type-$i$ topological excitation,
which is defined mod 1.  We also note that the handness of the ``self-loop''
determines the direction of the twist (see Fig. \ref{twist2}).
As a result, a figure ``8'' of type-$i$ string has an amplitude $\ee^{2\pi \ii s_i} d_i$ (see Fig. \ref{twsidi}).

It is clear that the Dehn twist $\hat T$, when
acting on $|i\>=W^y_i|1\>$, will twist the string $i$ by $2\pi$
and induces a phase $\ee^{\ii 2\pi s_i}$ (see Fig. \ref{TTrns}).
The  Dehn twist $\hat T$ also change the space-time metrics
which may causes an additional $i$-independent phase, which is denoted as
$\ee^{-\ii 2\pi \frac{c}{24}}$.
This way, we ontain another important relation:
\begin{align}
 T |i\> =\ee^{- 2\pi\ii \frac{c}{24}} \ee^{2\pi \ii s_i} |i\>.
\end{align}

\subsection{Relation between $N^{ij}_k$ and $S_{ij}$}

To understand the relation
\eq{SNsss}, let us compute the amplitude of a double figure ``8''
in two ways, as shown in Figs. \ref{dtwist} and  \ref{dtwist1}.
This allows us to show
\begin{align}
&\ \ \ \ Y^{\bar j j}_1
 Y^{\bar i i}_1 D S_{i\bar j} \ee^{2\pi \ii (s_i+s_j)}d_id_j
\nonumber\\
&=
\sum_{\al,k} Y^{ji}_{k,\al} O^{ji,\al}_k \ee^{2\pi \ii s_k}d_k,
\end{align}
which can be simplified to (noting $\al=1,\cdots,N^{ij}_k$)
\begin{align}
\frac{1}{d_j}\frac{1}{d_i}
 D S_{i\bar j} \ee^{2\pi \ii (s_i+s_j)}d_id_j
=
\sum_k N^{ij}_k \ee^{2\pi \ii s_k}d_k,
\end{align}
or
\begin{align}
D S_{i\bar j} \ee^{2\pi \ii (s_i+s_j)}
=
\sum_k N^{ij}_k \ee^{2\pi \ii s_k}d_k.
\end{align}
Using $S_{i\bar j}^*=S_{ij}$, the above becomes \eqn{SNsss}.

In this paper, we have derived most of the $(N^{ij}_k,s_i,c)$ conditions,
expect the condition \eqn{nuga}.
Here, we would like mention
that the condition \eqn{nuga} can be found in \Ref{Wang10}.

\section{List of primitive topological orders}

In this section, we give lists that contain more topological orders (see Tables
\ref{toplst2-6} and \ref{toplst7-9}, where only the primitive topological
orders are listed). The lists are generated using a different numerical code. 

The abelian states with $d_i=1$ are described by $K$-matrices. We use the
notation $(K_{11} K_{22} \cdots;K_{12} K_{23} \cdots;K_{13} K_{24}
\cdots;\cdots)$ to denote the $K$-matrices.  

In \Ref{SW150801111}, we show that the non-abelian states can be generated by
simple current algebra (SCA)\cite{BW9215,WW9455,LWW1024}.  (See also
https://www.math.ksu.edu/$\sim$gerald/voas/ ) The SCA's are denoted by $(R,\pm
k)_\al$ (see \Ref{RSW0777,SW150801111}), where $R=A_n,B_n,C_n,D_n,$ \etc, and
$(R,- k)_\al$ is the time-reversal conjugate of $(R,+ k)_\al$.  The last column
of Tables \ref{toplst2-6} and \ref{toplst7-9} indicates how the corresponding
topological order is realized by the $K$-matrix state or the SCA state.

\vfill
\break

\bibliography{../../bib/wencross,../../bib/all,../../bib/publst,./local}

\end{document}